\documentclass[10pt,a4paper]{article}
\usepackage{style}
\usepackage{amsmath,amssymb,epsfig,bm,amsfonts,yfonts,bbm,mathrsfs,hyperref,slashed}
\usepackage{pgfplots}
\usepackage[compat=1.1.0]{tikz-feynman}
\pgfplotsset{compat=newest}

\usepackage[scr=boondox,  % heavily sloped
            cal=esstix]   % slightly sloped
           {mathalpha}

\usepackage{pdfpages}
\usepackage{caption}

\definecolor{rot}{rgb}{0.7,0,0}
\definecolor{gruen}{rgb}{0,0.7,0}
\definecolor{blau}{rgb}{0,0,0.6}
\definecolor{hellblau}{rgb}{0.0, 0.7, 1.0}

\title{Response theory for quantum fields in isolation}

\author{Stefan Floerchinger}
\emailAdd{stefan.floerchinger@uni-jena.de}
\affiliation{Theoretisch-Physikalisches Institut, Max-Wien-Platz 1, 07743 Jena, Germany}

\abstract{Response theory describes the reaction of observales to perturbations in external fields. We review this formalism for quantum fiels in isolation that have unitary time evolution. An emphasis is put on consequences of causality and the resulting spectral representations for linear and nonlinear response functions, on functional techniques and generating functionals, including the description of the initial state, the evolution, and measurements. We review consequences of time reversal symmetry and relations for the statistics of work, and discuss a large class of quantum correlation functions, and their relation to response functions through fluctuation-dissipation relations. Consequences of conservation laws and gauge symmetries are mentioned briefly.}

\begin{document}

\maketitle

\section{Introduction}

Response theory deals with an answer, or response, of a physical system to an impulse or perturbation of some kind. The main appeal of the theory lies in its generality. Response theory is valid for weakly or strongly interacting theories, exists in a quantum as well as in a classical version, and can be used for systems that evolve in isolation, but also be formulated for open situations where an exchange of (quantum) information, energy, or particle number with a surrounding is possible.

Most of the time one is interested in the response of a macroscopic observable, one that can actually be observed, to a perturbation induced in another macroscopic quantity. In this setup response theory provides a very general framework for describing and investigating macroscopic properties of a theory. This is non-trivial for theories that are fundamentally anchored in the microscopic regime. Response theory helps to formulate here a very valuable bridge between the microscopic theory, which has typically very many degrees of freedom, and a macroscopic regime where only some collective degrees of freedom might be accessible. For example, one might have, in addition to a microscopic or fundamental description an effective theory, whose properties are emergent and \textit{a priori} not known.\footnote{Interestingly, approximations like linear response might work well for macroscopic observables while they fail for microscopic quantities, see e.\ g.\ \cite{10.1063/1.5122740}.}

A prime example for a macroscopic theory is fluid dynamics, a theory describing the large-scale motion of liquids of all kind that is widely independent of their microscopic composition. Only some fluid properties, like the thermodynamic equation of state, or viscosities and conductivities carry information from the microscopic domain. Response theory can be used to find expressions, known as Kubo relations, for example for viscosities, that can be evaluated within a microscopic calculation \cite{DeGrootMazur1961,Kadanoff:1963axw, Hohenberg:1977ym}.

Historically, response theory was developed starting with the beginning of the 20th century at several places in the world. Time dependent observables become first amendable to thermodynamical considerations through Albert Einsteins theoretical treatment of Brownian motion in 1905 (in Bern, Switzerland), with parallel developments by William Sutherland (Melbourne, Australia) and Marian Smoluchowski (Lviv, Ukraine), and later accounts by Paul Langevin (Paris, France), Norbert Wiener (Boston, USA), and many others.

Thermal noise in electric circuits was investigated theoretically by Geertuida de Haas-Lorentz in Leiden (Netherlands) in 1912, and around 1927 discovered experimentally by Frits Zernike in Groningen (Netherlands) and independently by John B. Johnson at the Bell laboratories in New York (USA). These findings were subsequently described theoretically, in 1928, also in New York, by Harry Nyquist.

Lars Onsager, working at Brown university (USA), published in 1929 and 1931 a substantially more general theory of thermal fluctuations and dissipation, based on consequences of microscopic time-reflection symmetry, including his famous reciprocal relations. These were generalized in 1945 by Hendrik Brugt Gerhard Casimir (Eindhoven, Netherlands).

Another big step was taken in 1951 by Herbert B.~Callen in Philadelphia (USA) and Theodore Welton at Oak Ridge (USA), who extended Nyquists theory of noise in electric conductors to a more general relation between correlation functions, quantifying fluctuations in thermal equilibrium, and response functions, that amongst other quantify dissipation -- their famous fluctuation-dissipation relation \cite{Callen:1951vq}. This can also be seen as an extension and generalization of the Einsteins relation for the diffusion coefficient found from his treatment of Brownian motion, and they managed to establish the relation in quantum theory. 

The 1950's were particularly productive in Japan, with contributions from (alphabetical order) Natsuki Hashitsume, Ryogo Kubo, Hazime Mori, Sadao Nakajima, Huzio Nakano, Hidetosi Takahasi, Morikazu Toda, Kazuhisa Tomita, Tsunenobu Yamamoto, Mario Yokota, and others. The history of this period is reviewed in ref.\ \cite{doi:10.1142/S0217979293002900}. Contributions in that period were also made by Melville S. Green, and Lars Onsager with Stefan Machlup in the USA. The works of Ryogo Kubo on the fluctuation--dissipation theorem and response theory were particularly influential. They established expressions for response functions in terms of expectation values of commutators of quantum mechanical operators. Non-linear response theory and fluctuation-dissipation relations on that basis were investigated also by William Bernhard and Herbert B.~Callan (Philadelphia, USA), as well as Ruslan L.~Stratonovich (Moscow, Russia), and German F.~Efremov (Nizhny Novgorod, Russia), see \cite{Stratonovich1992, Stratonovich1994} for a review. See also for more recent work on this topic \cite{Wang:1998wg}.

The focus of attention moved during the 1970's towards non-equilibrium states, and early version of what are now called fluctuation relations, developed by German N.~Bochkov and Yuriy E.~Kuzovlev in (Nizhny Novgorod, Russia), see their review \cite{Bochkov_2013}, and Peter Hänggi and collaborators (Basel, Switzerland, later Augsburg, Germany), reviewed in \cite{Campisi:2011wqf}.

In the 1990's, fluctuation theorems aimed at non-equilibrium states of classical systems were developed by Denis James Evans, Debra J. Searles, Garry P. Morris and others in Australia, together with Ezechiel Godert David Cohen (USA) and Giovanni Gallavotti (Italy), see \cite{Evans01112002, Evans_Morriss_2008, MARCONI2008111, Seifert_2012, Cuneo_2026} for reviews. These address fluctuations in entropy, or entropy production, out-of-equilibrium.

Christopher Jarzynski (then Seattle, USA) derived an influential equation for the statistics of work in 1997 \cite{Jarzynski1997}, see \cite{Jarzynski2008} and \cite{Jarzynski2011} for reviews. A closely related identity was found by Gavin E.~Crooks (Berkeley, USA) 1998 \cite{PhysRevE.60.2721}. These identities come in different variants, one of which addresses the work done on a driven, but otherwise isolated quantum systems, and we will discuss them in their quantum variants below.

The quantum versions of fluctuation theorems needed more time to be developed and understood properly, essentially because observations on a quantum system cannot be done without back-action. Work done on a quantum driven system is not a standard observable \cite{Talkner_2007} but can only be determined through \textit{two} measurements. For the history of these interesting developments we refer to recent dedicated reviews \cite{Esposito:2008dpw, Campisi:2011wqf,Hänggi2015}. See also \cite{PhysRevLett.133.070405} for recent developments concerning the linear response regime.

Also related with the peculiarity of quantum theory is the fact that there are very many ways to define correlation functions despite them having the same classical limit. This arises because operators for observables may not commute with the density matrix. In a series of works this was investigated by Dénes Petz and collaborators in Budapest (Hungary), see \cite{petz2010quantum} for a review. This relates to the interesting topic of quantum information geometry, reviewed in \cite{bengtsson2017}. A relation between linear response theory and the quantum correlation functions in the form of quantum Fisher information was recently established by Tomohiro Shitara and Masahito Ueda (Tokyo, Japan) \cite{shitara2016}.

The main point of attention in the current work are quantum field theories in isolation, i.\ e.\ with unitary time evolution. Typical applications we have in mind are for high-energy nuclear collisions, astrophysics and cosmology, or condensed matter experiments \cite{Kapusta:2007xjq, CalzertaHu2009, Altland:2010lla, Bellac:2011kqa, Laine:2016hma}. We will not restrict the nature of the field theory further, and it could be relativistic or non-relativistic. 

We will develop response theory at linear and non-linear order, and draw connections to closely related topics. In order to simplify the notation we typically write linear and quadratic orders explicitly, but in such a way that generalizations to cubic and higher orders should be clear. The aim is to set the stage for further developments in non-perturbative quantum field theory, using functional techniques and methods, and we will therefore also put an emphasis on the use of functional techniques like generating functionals.

Due to the long history of the subject sketched above, most of the material is not new, but some aspects are, for example we introduce a waek measurement scheme for intermediate and final times, a corresponding generating observable and related partition function in section \ref{sec:MoreOnMeasurements}, and we refine the connection between response functions and quantum correlation functions in sections \ref{sec:TimedependentPerturbationTheory} and \ref{sec:QuantumCorrelationFunctions}.

Much of the formalism is applicable to any quantum theory, and for those topics we use a general notation, while we specialize to a more field theoretic notation in later sections.

Starting point is accordingly a quantum theory with a Lagrangian or Hamiltonian that depends (possibly in a nonlinear way) on external fields in addition to the actual dynamical degrees of freedom. We are interested in the response of expectation values, or other observables, in the quantum theory, to a time-dependent change in the external fields. The expansion is done here around global thermal equilibrium states. We describe the evolution in a Hamiltonian picture and a Lagrangian or action based formalism in parallel.

From a relativistic point of view, we start on some initial Cauchy surface $\Sigma_\text{i}$, with a global thermal equilibrium quantum state. In order for such a state to exist in a generic spacetime, one must assume in the region before $\Sigma_\text{i}$ the existence of a timelike Killing vector field $\beta^\mu(x)$, i.\ e.\ a vector field $\beta^\mu(x)$ such that
\begin{equation}
  \mathscr{L}_\beta g_{\mu\nu}(x) = \nabla_\mu \beta_\nu(x) + \nabla_\nu \beta_\mu(x) = 0.
\end{equation}
The flow along this field can be used to define a time coordinate $t$, at least in the vicinity of $\Sigma_\text{i}$, such that $t=t_\text{i}$ corresponds to $\Sigma_\text{i}$. We will also assume that this time coordinate can be extended towards the future of $\Sigma_\text{i}$ and we will assume the existence of a Hamiltonian $H(t)$ that describes the evolution along this time.

For times $t>t_\text{i}$ we will allow time-dependent perturbations in this Hamiltonian that bring the state away from global equilibrium. This can be done in many ways. A time-dependent change in the spacetime metric is one possibility, but the time-dependence can also be induced by other fields, that we assume to be provided externally. 

We work most oft the time in natural units where $c=\hbar=k_\text{B}=1$.

\section{Quantum states and quantum correlations}
The general stage for quantum response theory is set by introducing Hamiltonians and Lagrangians that depend on time through external fields.

\subsection{Hamiltonians, Lagrangians, actions and their dependence on source fields}\label{sec:HamiltoniansLagrangiansActions}
The general idea of response theory is that deviations from a given situation -- oftentimes thermal equilibrium -- occur because of a change in the Hamiltonian and one is interested in the reaction of an observable to this change. The perturbation arises usually from an external parameter or field that can be manipulated in an experiment, such as the electromagnetic field in a condensed matter experiment. Or maybe one wants to study a fluid and one could take the spacetime metric as an external field that can be varied, at least theoretically.

Microscopically, quantum mechanics or quantum field theory can either be described in a Hamiltonian picture with operators, or in terms of a Lagrangian picture in terms or path or functional integrals. Sometimes one of the formalisms has advantages over the other, and sometimes it is tradition or personal preference that turns the balance. In the following, wherever possible, we collect the information and formulas for both formalisms.

We first consider a time-dependent Hamiltonian that we take to be of the form
\begin{equation}
  H(t) = H(j(t)) = H(j^1(t), \ldots, j^N(t)).
\end{equation}
Here $j^n(t)$ are externally provided real functions of time $t$, that we refer to as \textit{sources}. Besides time, they depend on an index $n$ that could be discrete, as appropriate for quantum mechanics, or also continuous. For quantum field theory, spatial position $\mathbf{x}$ can be seen as such a continuous index. For simplicity we use here mainly the notation for the discrete case, but a generalization is straight forward.\footnote{Some care is needed when a Fourier wave number $\mathbf{p}$ is taken as a continuous index, because fields that are real or hermitian as a function of $\mathbf{x}$ are complex in Fourier space.}

The source field $j^n(t)$ could couple linearly to an elementary field $\chi$ or to some composite operator.  It could also be an external gauge field, spacetime metric or similar. We also allow a non-linear dependence of the Hamiltonian $H$ or the corresponding Lagrangian or action on this source term.

Response theory also depends on the initial quantum state. One might be tempted to keep it arbitrary, but that would severely restrict the statements that can be made, in particular if one wants to take the microscopic theory as a starting point. One therefore usually starts with a thermal equilibrium state at some early time, and assumes that more general non-equilibrium states can be produced through the means of response theory itself, i.\ e.\ with time-dependent external sources. To have a global thermal equilibrium state at early time, one needs a stationary Hamiltonian in that period. In order for the Hamiltonian to be independent of time $t$ at early times, $t<t_\text{i}$, we assume
\begin{equation}
  j^n(t) = j^n_\text{i} \quad\quad\quad(t<t_\text{i}),
\end{equation}
so that $H(t) = H_\text{i} = H(j_\text{i})$ in that region. Similarly, we shall sometimes assume for late times $j^n(t) = j^n_\text{f}$ for $t>t_\text{f}$, with again a time-independent Hamiltonian $H(t) = H_\text{f} = H(j_\text{f})$.

For the Hamiltonian we assume an expansion
\begin{equation}
\begin{split}
  H(j(t)) = & H(j_\text{i}) - \sum_m  (j^m(t)-j^m_\text{i}) \phi_m(j_\text{i}) - \frac{1}{2}\sum_{mn} (j^m(t)-j^m_\text{i}) (j^n(t)-j^n_\text{i}) \phi_{mn}(j_\text{i}) \\
  & - \frac{1}{3!} \sum_{mnk} (j^m(t)-j^m_\text{i}) (j^n(t)-j^n_\text{i}) (j^k(t)-j^k_\text{i}) \phi_{mnk}(j_\text{i}) - \ldots,
\end{split}
\label{eq:ExpansionHamiltonianResponseOperators}
\end{equation}
where the $\phi_n(j_\text{i})$, $\phi_{mn}(j_\text{i})$ etc.\ are hermitian operators. We also write the derivative as
\begin{equation}
  - \frac{\partial}{\partial j^m(t)} H(j(t)) =  \phi_m(j(t)),
\end{equation}
with the source-dependent operator
\begin{equation}
  \phi_m(j(t)) = \phi_m(j_\text{i}) + \sum_n  (j^n(t) - j^n_\text{i}) \phi_{mn}(j_\text{i}) + \frac{1}{2} \sum_{nk}  (j^n(t)-j^n_\text{i}) (j^k(t)-j^k_\text{i}) \phi_{mnk}(j_\text{i}) + \ldots.
\label{eq:SourceDependentOperator}
\end{equation}

Alternatively, in a Lagrangian formulation of quantum theory using the path or functional integral one works with an action $S[\chi, j]$ that depends in a functional way on (a set of) fundamental fields $\chi(t)$, as well as the sources $j^n(t)$. It has the form of an integral over the Lagrangian,
\begin{equation}
  S[\chi, j] = \int dt \, L(\chi(t), \dot\chi(t), j(t)).
\end{equation}
Working directly with the action, we can assume a functional Taylor expansion in the source fields that is of the form
\begin{equation}
\begin{split}
  S[\chi, j] = & S[\chi, j_\text{i}] + 
  \int d t \left(j^n(t) - j^n_\text{i}\right) \tilde \phi_n(\chi(t), \dot \chi(t), j_\text{i}) \\ & + \frac{1}{2}\int dt \left(j^m(t) - j^m_\text{i} \right) \left( j^n(t) - j^n_\text{i} \right) \tilde \phi_{mn}(\chi(t), \dot \chi(t), j_\text{i}) + \ldots
\end{split}\label{eq:expansionMicroscopicActionResponsePerturbation}
\end{equation}
%The second term would drop out if the source couples linearly but we shall not assume that. 
We assume $j^n(t)$ as well as $\tilde \phi_n(\chi(t), \dot \chi(t), j_\text{i})$ and $\tilde \phi_{nm}(\chi(t), \dot \chi(t), j_\text{i})$ and so on to be real. We note that the terms $\tilde \phi_n(\chi(t), \dot \chi(t), j_\text{i})$ etc.\ that enter eq.\ \eqref{eq:expansionMicroscopicActionResponsePerturbation} differ from the operator $\phi_n(j_\text{i})$ etc.\ that enter the expansion of the Hamiltonian in eq.\ \eqref{eq:ExpansionHamiltonianResponseOperators}. Both are related through the Legendre transform that relates Lagrangian and Hamiltonian.

\subsection{Example complex scalar field}
As an example, we consider the action for a complex scalar field $\varphi(x)$ where $x=(t,\mathbf{x})$, with external gauge field $A_\mu(x)$, metric $g_{\mu\nu}(x)$ and metric determinant $g(x) = - \det(g_{\mu\nu}(x))$,
\begin{equation}
\begin{split}
  & S[\varphi, A, g] = \int d^d x \mathscr{L} \\
  & = \int d^{d}x \sqrt{g(x)} \left\{ - g^{\mu\nu}(x) \left[ (\partial_\mu +i A_\mu(x)) \varphi^*(x) \right] \left[ (\partial_\nu - i A_\nu(x)) \varphi(x) \right] - V(\varphi^*(x) \varphi(x)) \right\}.
\end{split}\label{eq:ScalarFieldCoupledToGaugeFields}
\end{equation}
Here one can see the external gauge field $A_\mu(x)$ and the metric $g_{\mu\nu}(x)$ as source fields.

We also determine the Hamiltonian. To that end we first determine the canonical momentum fields associated to $\varphi$ and $\varphi^*$,
\begin{equation}
\begin{split}
  \pi(t, \mathbf{x}) = & \frac{\partial \mathscr{L}}{\partial \dot \varphi(x)} = - \sqrt{g(x)} g^{0\mu}(x) (\partial_\mu + i A_\mu(x)) \varphi^*(x),\\  
  \pi^*(t, \mathbf{x}) = & \frac{\partial \mathscr{L}}{\partial \dot \varphi^*(x)} = - \sqrt{g(x)} g^{0\mu}(x) (\partial_\mu - i A_\mu(x)) \varphi(x),
\end{split}
\end{equation}
which can be used to express $\dot \varphi(x)$ and $\dot \varphi^*(x)$ in terms of $\pi(x)$ and $\pi^*(x)$. One finds eventually the Hamiltonian
\begin{equation}
\begin{split}
  H = & \int d^{d-1}x \left\{ \pi(x) \dot \varphi(x) + \pi^*(x) \dot \varphi^*(x) - \mathscr{L}\right\} \\
  = & \int d^{d-1}x \left\{ - \frac{1}{\sqrt{g} g^{00}} \pi^* \pi - \frac{g^{0j}}{g^{00}} \left[ \pi (\partial_j -i A_j) \varphi + \pi^* (\partial_j + i A_j) \varphi^* \right] + i A_0 \left[ \pi \varphi - \pi^* \varphi^* \right] \right. \\
  & \quad\quad\quad\quad\quad \left. + \sqrt{g} \left[ g^{jk} - \frac{g^{0j}g^{0k}}{g^{00}} \right] (\partial_j + i A_j) \varphi^* (\partial_k - i A_k) \varphi + \sqrt{g} U(\varphi^* \varphi)  \right\}.
\end{split}\label{eq:HamiltonianComplexScalarWithExternalGaugeFields}
\end{equation}
We note that this Hamiltonian is nonlinear in the components of the metric, and there is a term that is quadratic in the spatial part of the gauge field. The dependence on $A_0(x)$ is linear, though. We note also that the dependence of $H$ on the metric $g_{\mu\nu}(x)$ and gauge field $A_\mu(x)$ is local, i. e. there is no dependence on their derivatives.

We also note that in contrast to the Lagrangian density $\mathscr{L}$, the Hamiltonian $H$ is not gauge invariant. For $\varphi(x)\to \exp(i\alpha(x))\varphi(x)$, $\varphi^*(x)\to \exp(-i\alpha(x))\varphi^*(x)$, $A_\mu(x) \to A_\mu(x)+ \partial_\mu \alpha(x)$ one has $\pi(x)\to \exp(-i\alpha(x))\pi(x)$, $\pi^*(x)\to \exp(i\alpha(x))\pi^*(x)$ and
\begin{equation}
  H \to H + \int d^{d-1} x  \left[ \pi(x) \varphi(x) - \pi^*(x) \varphi^*(x) \right] i \partial_0 \alpha(x). 
\label{eq:ChangeHamiltonianGaugeTransform}
\end{equation}
For spatially constant $\partial_0 \alpha(x)$ this becomes $H\to H - N \partial_0 \alpha$. The Hamiltonian changes by a term proportional to the particle number or charge $N$ associated to the U$(1)$ symmetry.

\subsection{Initial density matrix}
We now characterize the initial quantum state, before the time-dependent perturbation of the Hamiltonian sets in. At the initial time $t_\text{i}$, and any time before that, we assume a grand-canonical equilibrium state with the density matrix
\begin{equation}
  \rho_\text{i} = \frac{1}{Z(\beta, \mu, j_\text{i})} \exp\left( - \beta (H(j_\text{i}) - \mu N)\right),
  \label{eq:initialThermalDensityMatrix}
\end{equation}
and the partition function
\begin{equation}
  Z(\beta, \mu, j_\text{i}) = \text{Tr} \left\{ \exp\left(-\beta (H(j_\text{i}) - \mu N) \right) \right\} = \int D\chi \exp\left( - S_\text{E}[\chi] \right).
\label{eq:thermalPartitionFunctionHamiltonianAndAction}
\end{equation}
Here we use the usual abbreviations $\beta=1/T$ for the inverse temperature and $\mu$ is the chemical potential. %Chemical potentials have been taken to be part of the parameters $j_\text{i}$. This can be introduced like an external gauge field $A_\mu(x)$ coupling to the conserved particle currents, the time component of which is $A^0 = \mu$. While this modifies the Hamiltonian by subtracting $\mu N$, the new Hamiltonian nevertheless leads to the correct time-dependent Heisenberg operator fields later on (see [H. A. Weldon, PRD 76, 125029 (2007)] for a detailed discussion). 
The last equation introduces a functional integral representation where the Euclidean action depends also on temperature, chemical potential and the source parameters $j_\text{i}$ (see also below).

Being an equilibrium state of the time-independent Hamiltonian $H(j_\text{i})$ the density matrix $\rho(t) = \rho_\text{i}$ is independent of time for $t<t_\text{i}$. In analogy to the position space representation of non-relativistic quantum mechanics, the matrix elements are $\langle \chi_{\text{i}+} | \rho_\text{i} | \chi_{\text{i}-} \rangle$, where $\chi$ stand for the microscopic degrees of freedom, for example field configurations at fixed time in quantum field theory.

\subsection{Complex time evolution}\label{sec:ComplexTimeEvolution}
To find a functional integral representation for $\rho_\text{i}$ and the partition function, one uses a generalization of the time evolution operator,
\begin{equation}
  \langle \chi_2 | U(\Delta t) | \chi_1 \rangle = \langle \chi_2 | \exp\left(-i \Delta t (H-\mu N)\right) | \chi_1 \rangle = \sum_n \langle \chi_2 | n \rangle \langle n | \chi_1 \rangle \exp\left(-i \Delta t (E_n-\mu N_n)\right) .
\end{equation} 
In the second step we inserted a factor of unity in the form of a sum of simultaneous eigenstates of energy and particle number $|n\rangle$, so that $H|n\rangle = E_n | n\rangle$ and $N|n\rangle = N_n | n\rangle$.

For real time difference $\Delta t$, the terms $\exp(-i \Delta t (E_n-\mu N_n))$ are pure phases, and the evolution operator is unitary. There is an additional time-dependent phase $\sim \mu$ which can be seen as an additive change of the energy that happens to depend on the conserved particle number. This does not change the physical dynamics as long as there are no interference terms between sectors with different particle numbers. Alternatively, one may consider the change in the Hamiltonian $H\to H-\mu N$ as a result of a gauge transformation, see the discussion around eq.\ \eqref{eq:ChangeHamiltonianGaugeTransform}. For a detailed discussion of the role of chemical potentials for real time evolution see ref.\ \cite{Weldon:2007zz}.

One can also make $\Delta t$ complex, and the exponential terms become $\exp(-i \text{Re}(\Delta t) (E_n-\mu N_n) +  \text{Im}(\Delta t) (E_n-\mu N_n))$. For $\text{Im}(\Delta t)<0$ the evolution operator remains well defined because high energy states are suppressed. It is not unitary any more, though. In fact it could be decomposed into a unitary operator $\exp(-i \text{Re}(\Delta t) (H-\mu N))$ and a hermitian operator $\exp(\text{Im}(\Delta t) (H-\mu N))$. We will use the symbol $\mathscr{U}(t+\Delta t, t)$ for evolution operators in this generalized sense, where $\Delta t$ can have a negative imaginary part. Let us stress again the importance of $\text{Im}(\Delta t)$ being negative. Complex time evolution with positive imaginary part of $\Delta t$ would have infinitely large contributions from high energy states, and therefore be ill defined.

Using the above construction, the density matrix can be expressed as an evolution operator in negative imaginary time direction, for example
\begin{equation}
 \langle \chi_{\text{i}+} | \rho_\text{i} | \chi_{\text{i}-} \rangle = \frac{1}{Z(\beta, \mu, j_\text{i})}\langle \chi_{\text{i}+} | \mathscr{U}(t_\text{i}-i \beta, t_\text{i}, j_\text{i}) | \chi_{\text{i}-} \rangle,
\label{eq:rhoThermalMatrixElement}
\end{equation}
with
\begin{equation}
  \langle \chi_{\text{i}+} | \mathscr{U}(t_\text{i}-i \beta, t_\text{i}, j_\text{i}) | \chi_{\text{i}-} \rangle = \int_{\chi_{\text{i}+}, \chi_{\text{i}-}} D \chi \exp\left(-S_\text{E}[\chi, j_\text{i}]\right).
\label{eq:rhoThermalPartitionFunction}
\end{equation}
We stress again that $\mathscr{U}$ is \textit{not} unitary, but here in fact hermitian.

This depends on the fields $\chi_{\text{i}+}, \chi_{\text{i}-}$, on the initial Cauchy surface $\Sigma_\text{i}$ as boundary values for the Euclidean functional integral at the times $t_\text{i} - i \beta$ and $t_\text{i}$. The Euclidean action $S_\text{E}[\chi, j]$ is an integral along $t=t_\text{i}-i \tau $ where $0 \leq \tau \leq \beta$. %Equivalently one could also work with a symmetric version of the imaginary time contour from $t_\text{i}+i\beta/2$ down to $t_\text{i}-i\beta/2$.

One can extend the sources $j$ to being functions of the time coordinate $\tau$, and this will be needed for example for analytic continuation from imaginary to real times. One has
\begin{equation}
  - S_\text{E}[\chi, j] = i S[\chi, j] = i \int_{t_\text{i}}^{t_\text{i}-i\beta} dt L(\chi(t), \tfrac{\partial}{\partial t} \chi(t), j(t)) = \int_0^{\beta} d\tau L(\chi(t_\text{i}-i\tau), i\tfrac{\partial}{\partial \tau} \chi(t_\text{i}-i\tau), j(t_\text{i}-i\tau)).
\label{eq:EuclideanActionThermalState}
\end{equation}
The imaginary unit in front of the time derivative can be absorbed in a relativistic theory into a rotation of the spacetime metric to the Euclidean domain (given there is a timelike Killing vector field available).

The partition function is then a functional of $j^n(t_\text{i}-i\tau)$,
\begin{equation}
  Z(\beta, \alpha)[j] = \int D\chi \exp\left(-S_\text{E}[\chi, j] \right),
\end{equation}
where bosonic (fermionic) fields $\chi$ are periodic (anti-periodic) in imaginary time $\tau$.
One must be careful, however, because arbitrary functional forms $j^n(t_\text{i}-i\tau)$ would lead to unphysical situations, either because they violate thermal equilibrium conditions or even make $\rho_\text{i}$ non-hermitian. For thermal equilibrium states one can consider the $\tau$-dependent sources on the Matsubara circle as a calculational device, but for equilibrium states one must eventually evaluate functional derivatives at constant sources $j^n_\text{i}$.

\section{Static response and correlation functions}
Before we come to dynamical response, it is beneficial to discuss static response. The basic idea it to change a parameter that characterizes a thermal equilibrium state in a static way, or very very slowly. Such a static or equilibrium response is closely connected to thermal equilibrium correlation functions and it can be seen as a thermodynamic property itself. As usual in thermodynamics one must be careful how the variation is done, in particular which quantities are kept constant while others are varied.

As a warmup, we first characterize thermal equilibrium states through their expectation values, response and correlation functions. We start with a global thermal equilibrium state in the grand canonical ensemble. The density matrix is as in eq.\ \eqref{eq:initialThermalDensityMatrix} with the partition function in eq.\ \eqref{eq:thermalPartitionFunctionHamiltonianAndAction}. It is customary to also introduce the grand-canonical potential as
\begin{equation}
  \Omega(T,\mu, j_\text{i}) = E - TS - \mu N = - T \ln Z(\beta, \alpha, j_\text{i}).
\end{equation}
One has the differentials (we keep the volume fixed) 
\begin{equation}
\begin{split}
  d \ln Z = & - E_\text{i}  d\beta + N d\alpha + \frac{1}{T} \sum_m \Phi_m^\text{i} dj^m_\text{i}, \\
  d\Omega = & - S dT - N d\mu - \sum_m \Phi_m^\text{i} dj^m_\text{i},     
\end{split}
\end{equation}
where we employ the entropy
\begin{equation}
  S = - \text{Tr}\{ \rho_\text{i} \ln \rho_\text{i} \} = \ln Z + \beta E_\text{i} - \alpha N = - \frac{\partial}{\partial T} \Omega(T, \mu, j_\text{i}),
\end{equation}
and the expectation values
\begin{equation}
\begin{split}
E_\text{i} = & \text{Tr}\{  \rho_\text{i} H_\text{i} \} =  - \frac{\partial}{\partial \beta} \ln Z(\beta, \alpha, j_\text{i}) = \Omega(T,\mu, j_\text{i}) + TS+\mu N, \\
N = & \text{Tr} \{ \rho_\text{i} N \} = \frac{\partial}{\partial \alpha} \ln Z(\beta, \alpha, j_\text{i}) = - \frac{\partial}{\partial \mu} \Omega(T, \mu, j_\text{i}), \\
\Phi_m^\text{i} = & \text{Tr}\{ \rho_\text{i} \phi_m(j_\text{i}) \} = T \frac{\partial}{\partial j^m_\text{i}} \ln Z(\beta, \alpha, j_\text{i}) = - \frac{\partial}{\partial j^m_\text{i}} \Omega(T,\mu, j_\text{i}).
\end{split}\label{eq:ExpValuesInitialThermalState}
\end{equation}
These relations generalize the well known thermal equilibrium expressions to a situation where also the sources $j_\text{i}^n$ can be varied. 

For a source that is taken to depend also on the imaginary time $\tau$ the thermal expectation value is
\begin{equation}
  \Phi^\text{i}_m = T \int_{0}^{\beta} d\tau \frac{\delta}{\delta j^m(t_\text{i}-i\tau)} \ln Z[j]{\big |}_{j=j_\text{i}}.
\end{equation}
Note that the right hand side can be seen as an uniform average over different positions on the periodic imaginary Matsubara time. By translational symmetry with respect to the imaginary time direction, the functional derivatives at different imaginary times $\tau$ all have to agree, so that one could also write
\begin{equation}
  \Phi^\text{i}_m = \frac{\delta}{\delta j^m(t_\text{i}-i\tau)} \ln Z[j]{\big |}_{j=j_\text{i}},
\end{equation}
where the evaluation point $\tau$ with $0\leq \tau\leq \beta$ is arbitrary.

\subsection{Static response at constant temperature and chemical potential}
We can ask for the initial thermal equilibrium state: how does the expectation value $\Phi_m^\text{i}$ change when the source $j^n$ changes? If this is done with a coupling to an external heat bath that keeps temperature $T$ and chemical potential $\mu$ constant, the response is quantified by the derivative
\begin{equation}
  \chi^{(T, \mu)}_{mn} = \left(\frac{\partial \Phi_m^\text{i}}{\partial j^n} \right)_{T,\mu}  = T \frac{\partial^2}{\partial j^n \partial j^m} \ln Z(\beta, \alpha, j) {\big |}_{j=j_\text{i}} = - \frac{\partial^2}{\partial j^n \partial j^m} \Omega(T, \mu, j) {\big |}_{j=j_\text{i}}.
\label{eq:defStaticSusceptibility}
\end{equation}
This is an \textit{isothermal static susceptibility}. In a similar way one can define higher order derivatives
\begin{equation}
  \chi^{(T, \mu)}_{mnk} = \left(\frac{\partial^2 \Phi_m^\text{i}}{\partial j^n \partial j^k} \right)_{T, \mu} = T \frac{\partial^3}{\partial j^m \partial j^n \partial j^k} \ln Z(\beta, \alpha, j) {\big |}_{j=j_\text{i}} = - \frac{\partial^3}{\partial j^m \partial j^n \partial j^k} \Omega(T, \mu, j) {\big |}_{j=j_\text{i}}.
\label{eq:defEqQuadSuscept}
\end{equation}
As partial derivatives, $\chi_{mn}^{(T, \mu)}$, $\chi_{mnk}^{(T, \mu)}$ and higher orders are symmetric with respect to the interchange of the indices $m$, $n$ and $k$.

The static susceptibilities can be determined using the functional integral representation of the thermal partition function $Z(\beta, j)$ in eq.\ \eqref{eq:thermalPartitionFunctionHamiltonianAndAction} with the Euclidean action \eqref{eq:EuclideanActionThermalState}. 

With sources that are depending on the imaginary time $\tau$, susceptibilities become averages over the Matsubara torus, e.\ g.\
\begin{equation}
\begin{split}
  \chi^{(T, \mu)}_{mn}  = & T \int_0^{\beta} d\tau_1 d\tau_2 \frac{\delta^2}{\delta j^m(t_\text{i}-i\tau_1) \delta j^n(t_\text{i}-i\tau_2)} \ln Z[j]{\big |}_{j=j_\text{i}} \\
  = &  \int_0^{\beta} d\tau \frac{\delta^2}{\delta j^m(t_\text{i}) \delta j^n(t_\text{i}-i\tau)} \ln Z[j]{\big |}_{j=j_\text{i}}.
\end{split}
\end{equation}
In the second line we used the translational symmetry in Matsubara time and performed one imaginary time integral.

One could also employ a Fourier decomposition of $j^n(t_\text{i}-i\tau)$ with Matsubara frequencies, and the static susceptibility would involve derivatives only with respect to the zero frequency component.

\subsection{Static response in terms of operators}
It might be useful to give also operator expressions for the susceptibility. One finds at the point $j=j_\text{i}$
\begin{equation}
\begin{split}
  \chi^{(T, \mu)}_{mn} = & \frac{1}{Z} \int_0^{\beta} d\tau \, \text{Tr} \left\{ \exp(-(\beta-\tau) (H_\text{i}-\mu N)) \phi_m \exp(-\tau (H_\text{i}-\mu N)) \phi_n \right\} \\ & - \beta \Phi_m \Phi_n + \text{Tr} \left\{ \rho_\text{i} \frac{\partial \phi_n(j)}{\partial j^m} \right\} \\
  = & \beta \int_0^1 d\lambda \text{Tr} \{ \phi_m \, \rho_\text{i}^{1-\lambda} \, \phi_n \rho_\text{i}^{\lambda}  \} - \beta \Phi_m \Phi_n + \text{Tr} \left\{ \rho_\text{i} \phi_{mn} \right\} \\
  = & \beta\langle \phi_m \,;\,  \phi_n \rangle - \beta \Phi_m \Phi_n + \text{Tr} \left\{ \rho_\text{i} \phi_{mn} \right\}.
\end{split}\label{eq:susceptibilitiesFixedTmu}
\end{equation}
The last term arises here from taking another derivative of the source-dependent operator in eq.\ \eqref{eq:SourceDependentOperator}. In the last line we have introduced the notation 
\begin{equation}
  \langle \phi_m \,;\,  \phi_n \rangle = \int_0^1 d\lambda \text{Tr} \{ \phi_m \, \rho_\text{i}^{1-\lambda} \, \phi_n \rho_\text{i}^{\lambda}  \},
\label{eq:BKMTwoPointStatic}
\end{equation}
for what is known as Bogoliubov-Kobu-Mori (BKM) inner product \cite{PetzToth1993} %\textcolor{red}{\bf [Petz, D., Toth, G. The Bogoliubov inner product in quantum statistics, Lett Math Phys 27, 205–216 (1993), https://doi.org/10.1007/BF00739578, more references needed]} 
of the operators $\phi_m$ and $\phi_n$ with respect to the state $\rho_\text{i}$. It can be seen as one possible quantum generalization of a classical correlation function, albeit this generalization is not unique, see section \ref{sec:QuantumCorrelationFunctions} for a more detailed discussion. The basic point here is that the non-commuting nature of operator products allows to generalize classical expressions in different ways. One has the symmetry $\langle \phi_m \,;\,  \phi_n \rangle = \langle \phi_n \,;\,  \phi_m \rangle$.

We also introduce the notation
\begin{equation}
  \langle  \phi_m ;  \phi_n \rangle_c = \langle  \phi_m ; \phi_n \rangle - \Phi_m \Phi_n ,
\end{equation}
for the connected part of the BKM quantum correlation function. With this, the isothermal susceptibility becomes
\begin{equation}
  \chi^{(T, \mu)}_{mn} = \beta \langle \phi_m \,;\, \phi_n \rangle_c + \langle \phi_{mn} \rangle.
  \label{eq:IsothermalSusceptConnectedCorr}
\end{equation}

\subsection{Higher-order Bogoliubov-Kobu-Mori quantum correlation functions}

At this point we may pause for a moment and define a generalization of the Bogoliubov-Kubo-Mori correlation function beyond quadratic order,
\begin{equation}
  \langle  \phi_{n_1} ;  \phi_{n_2} ; \ldots ;  \phi_{n_N} \rangle = \int_0^1 ds_2 \cdots ds_{N} \text{Tr} \left\{ \phi_{n_1}(t_\text{i}) \rho_\text{i} \, \mathscr{T} \left\{ \phi_{n_2}^\text{H}(t_\text{i} - is_2\beta) \cdots \phi_{n_N}^\text{H}(t_\text{i} - i s_N\beta) \right\} \right\},
\label{eq:quantumCorrelationKuboMoriOrderN}
\end{equation}
where $\mathscr{T}\{ \ldots \}$ indicates an imaginary-time ordering where the operator with largest value of $s\beta$ in its time argument is ordered to the left and so on, and we are employing Heisenberg-type operators
\begin{equation}
  \phi_{n}^\text{H}(t_\text{i} - i s \beta) = \rho_\text{i}^{-s} \phi_{n} \rho_\text{i}^s.
\end{equation}
This kind of correlation $\langle  \phi_{n_1} ;  \phi_{n_2} ; \ldots ;  \phi_{n_N} \rangle$ is symmetric with respect to any interchange of indices, because of the time ordering. We have used here that the cyclic property of the trace always allows to have $\phi_{n_1}$ as first operator, and the imaginary time coordinates on the Matsubara circle can be chosen such that $\phi_{n_1}$ has time argument $t_\text{i}$.

Consider as an example the three-point quantum correlation  function
\begin{equation}
\begin{split}
  \langle  \phi_{m} ;  \phi_{n} ; \phi_{k} \rangle = & \int_0^1 ds_2 ds_3 \text{Tr} \left\{ \phi_m(t_\text{i}) \rho_\text{i} \, \mathscr{T} \left\{ \phi_{n}^\text{H}(t_\text{i} - i s_2\beta) \phi_{k}^\text{H}(t_\text{i} - i s_3\beta)  \right\} \right\} \\
  = & \int_0^1 ds_2 ds_3 \text{Tr} \left\{ \phi_m \, \mathscr{T} \left\{ \rho_\text{i}^{1-s_2} \phi_{n} \rho_\text{i}^{s_2 - s_3} \phi_{k} \rho_\text{i}^{s_3}  \right\} \right\}.  
\end{split}
\label{eq:quantumCorrelationOrder3}
\end{equation}
The time ordering makes sure that the exponents of $\rho_\text{i}$ are all positive. One can also write this as
\begin{equation}
  \langle  \phi_{m} ;  \phi_{n} ; \phi_{k} \rangle = \int_0^1 d\lambda_1 d\lambda_2 d\lambda_3 \, \delta(\lambda_1+\lambda_2+\lambda_3 -1) \, \text{Tr}\left\{ \phi_m \rho_\text{i}^{\lambda_1} \phi_n \rho_\text{i}^{\lambda_2} \phi_k \rho_\text{i}^{\lambda_3} + \phi_m \rho_\text{i}^{\lambda_1} \phi_k \rho_\text{i}^{\lambda_2} \phi_n \rho_\text{i}^{\lambda_3} \right\}.
\end{equation}
This is indeed fully symmetric with respect to any permutation of $m$, $n$ and $k$. The generalization to higher order is now clear,
\begin{equation}
\begin{split}
  \langle  \phi_{n_1} ;  \ldots ; \phi_{n_N} \rangle = & \int_0^1 d\lambda_1 \ldots d\lambda_N \, \delta(\lambda_1 + \ldots + \lambda_N - 1)\, \text{Tr}{\Big \{}  \phi_{n_1} \rho_\text{i}^{\lambda_1} \phi_{n_2} \cdots \phi_{n_N} \rho_\text{i}^{\lambda_N} \\
  & +  \text{permutations of }(n_2, \ldots, n_N) {\Big \}}.
\end{split}
\end{equation}
Again full permutation symmetry is manifest because of the cyclic property of the trace.

The second order susceptibility at fixed temperature and chemical potential defined in \eqref{eq:defEqQuadSuscept} becomes
\begin{equation}
\begin{split}
  \chi^{(T, \mu)}_{mnk} = \beta^2 \langle \phi_m \,;\, \phi_n \,;\, \phi_k \rangle_c + \beta \langle \phi_{mn} \,;\, \phi_k \rangle_c + \beta \langle \phi_{nk} \,;\, \phi_m \rangle_c + \beta \langle \phi_{km} \,;\, \phi_n \rangle_c  + \langle \phi_{mnk} \rangle.
\end{split}
\end{equation}
Besides the connected third-order BKM correlation function it features also lower order connected correlation functions from higher derivatives of the Hamiltonian in eq.\ \eqref{eq:ExpansionHamiltonianResponseOperators}. 

\subsection{Static response in conditions of fixed entropy and particle number}
So far we have calculated the static response functions at fixed temperature and chemical potential, i.\ e.\ with a coupling to a heat bath. However, it is also interesting to consider the static response in an isolated situation, i.\ e.\ without the coupling to an external heat bath. In an isolated situation the response must be induced by unitary dynamics that keeps the entropy $S$ fixed, instead of the temperature $T$. We also assume that a particle number or charge $N$ is fixed, instead of the corresponding chemical potential. Depending on the physical situation there might be further conservation laws that need to be obeyed, and which have to be treated similar to $N$ here. The generalization to an arbitrary finite number of conserved quantities is straight forward.\footnote{This includes the case where there is no conserved charge besides entropy $S$. A special situation arises when the number of conservation laws is as large as the number of degrees of freedom so that the system becomes integrable. This prevents thermalization in the standard sense and goes beyond the scope of the discussion here.}

For simplicity we assume in the following that there is just a single conserved quantity $N$, in addition to entropy $S$. In this case one expects the static susceptibilities $\chi^{(S,N)}$ at fixed entropy $S$ and fixed $N$ to be equal to the \textit{isolated} or Kubo susceptibilities \cite{KuboTodaHashitsume1991}.

To develop the formalism, it is best to perform a Legendre transform, from $\ln Z(\beta, \alpha, j_\text{i})$ to
\begin{equation}
  S(E_\text{i}, N, j_\text{i}) = \ln Z(\beta, \alpha, j_\text{i}) + \beta E_\text{i} - \alpha N,
\end{equation}
or from $\Omega(T, \mu, j_\text{i})$ to
\begin{equation}
  E_\text{i}(S, N, j_\text{i}) = \Omega(T, \mu, j_\text{i}) + TS + \mu N,
\end{equation}
so that the differentials are
\begin{equation}
\begin{split}
  d S = & \beta dE_\text{i} - \alpha dN - \beta \sum_m \Phi_m^\text{i} dj^m_\text{i},\\
  d E_\text{i} = & T dS + \mu dN - \sum_m \Phi_m^\text{i} dj^m_\text{i}.
\end{split}
\end{equation}
Response functions at fixed entropy $S$ and particle number $N$ follow now as
\begin{equation}
\begin{split}
  \chi^{(S,N)}_{mn} = & \left(\frac{\partial \Phi_m^\text{i}}{\partial j^n} \right)_{S,N}  %= T \frac{\partial^2}{\partial j^n \partial j^m} S(E_\text{i}, N, j) {\big |}_{j=j_\text{i}} 
  = - \frac{\partial^2}{\partial j^n \partial j^m} E_\text{i}(S, N, j) {\big |}_{j=j_\text{i}}, \\
  \chi^{(S,N)}_{mnk} = & \left(\frac{\partial^2 \Phi_m^\text{i}}{\partial j^n\partial j^k} \right)_{S,N}  
  = - \frac{\partial^3}{\partial j^n \partial j^m\partial j^k} E_\text{i}(S, N, j) {\big |}_{j=j_\text{i}},
\end{split}
  \label{eq:defStaticSusceptibility}
\end{equation}
and so on at higher orders. 

Using the properties of a Legendre transform leads to relations of the kind
\begin{equation}
\begin{split}
  \chi^{(S,N)}_{mn}  = & \chi^{(T, \mu)}_{mn} + \frac{\partial S}{\partial j^m} \frac{\partial T}{\partial j^n} + \frac{\partial N}{\partial j^m} \frac{\partial \mu}{\partial j^n} = \chi^{(T, \mu)}_{mn} - \frac{\partial \Phi_m}{\partial T} \frac{\partial \Phi_n}{\partial S} - \frac{\partial \Phi_m}{\partial \mu} \frac{\partial \Phi_n}{\partial N} \\
  = & \chi^{(T, \mu)}_{mn} - \begin{pmatrix} \frac{\partial \Phi_m}{\partial T} & \frac{\partial \Phi_m}{\partial \mu} \end{pmatrix} \begin{pmatrix} \frac{\partial T}{\partial S} && \frac{\partial\mu}{\partial S} \\ \frac{\partial T}{\partial N} && \frac{\partial \mu}{\partial N}  \end{pmatrix} \begin{pmatrix} \frac{\partial \Phi_n}{\partial T} \\ \frac{\partial \Phi_n}{\partial \mu} \end{pmatrix},
\end{split}\label{eq:RelationChiSNChiTmu}
\end{equation}
that relate both kinds of susceptibilities. The expressions on the right hand side can be evaluated with the density matrix and partition function of the grand-canonical ensemble. The Jacobi matrix appearing in the second term is best evaluated through its inverse,
\begin{equation}
  \begin{pmatrix}
    \frac{\partial S}{\partial T} && \frac{\partial N}{\partial T} \\ \frac{\partial S}{\partial \mu} && \frac{\partial N}{\partial\mu} 
  \end{pmatrix} = \begin{pmatrix}
    - \frac{\partial^2 \Omega}{\partial T^2} && - \frac{\partial^2 \Omega}{\partial T \partial \mu} \\ - \frac{\partial^2 \Omega}{\partial T \partial\mu} && - \frac{\partial^2 \Omega}{\partial \mu^2}
  \end{pmatrix}.
\end{equation}
For the next order susceptibility one finds similarly
\begin{equation}
\begin{split}
  \chi^{(S,N)}_{mnk} = & \chi^{(T,\mu)}_{mnk} - \left[ \begin{pmatrix} \frac{\partial \chi_{mk}^{(T,\mu)}}{\partial T} & \frac{\partial \chi_{mk}^{(T,\mu)}}{\partial \mu} \end{pmatrix} \begin{pmatrix} \frac{\partial T}{\partial S} && \frac{\partial\mu}{\partial S} \\ \frac{\partial T}{\partial N} && \frac{\partial \mu}{\partial N}  \end{pmatrix} \begin{pmatrix} \frac{\partial \Phi_n}{\partial T} \\ \frac{\partial \Phi_n}{\partial \mu} \end{pmatrix} + \text{2 perm.} \right] \\
  & + \left[\begin{pmatrix} \frac{\partial \Phi_m}{\partial T} & \frac{\partial \Phi_m}{\partial \mu} \end{pmatrix} \begin{pmatrix} \frac{\partial T}{\partial S} && \frac{\partial\mu}{\partial S} \\ \frac{\partial T}{\partial N} && \frac{\partial \mu}{\partial N}  \end{pmatrix} 
  \begin{pmatrix}
    \frac{\partial^2 \Phi_k}{\partial T^2} && \frac{\partial^2 \Phi_k}{\partial T \partial \mu} \\ \frac{\partial^2 \Phi_k}{\partial T \partial\mu} && \frac{\partial^2 \Phi_k}{\partial\mu^2}   
  \end{pmatrix}
  \begin{pmatrix} \frac{\partial T}{\partial S} && \frac{\partial\mu}{\partial S} \\ \frac{\partial T}{\partial N} && \frac{\partial \mu}{\partial N}  \end{pmatrix}
  \begin{pmatrix} \frac{\partial \Phi_n}{\partial T} \\ \frac{\partial \Phi_n}{\partial \mu} \end{pmatrix} + \text{2 perm.}\right] \\
  & + \left[\begin{pmatrix} \frac{\partial \Phi_m}{\partial T} & \frac{\partial \Phi_m}{\partial \mu} \end{pmatrix} \begin{pmatrix} \frac{\partial T}{\partial S} && \frac{\partial\mu}{\partial S} \\ \frac{\partial T}{\partial N} && \frac{\partial \mu}{\partial N}  \end{pmatrix} 
  \begin{pmatrix}
    \frac{\partial^3 \Omega}{\partial T^3} && \frac{\partial^3 \Omega}{\partial T^2 \partial \mu} \\ \frac{\partial^3 \Omega}{\partial T^2 \partial\mu} && \frac{\partial^3 \Omega}{\partial T\partial\mu^2}   
  \end{pmatrix}
  \begin{pmatrix} \frac{\partial T}{\partial S} && \frac{\partial\mu}{\partial S} \\ \frac{\partial T}{\partial N} && \frac{\partial \mu}{\partial N}  \end{pmatrix}
  \begin{pmatrix} \frac{\partial \Phi_n}{\partial T} \\ \frac{\partial \Phi_n}{\partial \mu} \end{pmatrix} \right] \times \begin{pmatrix} \frac{\partial T}{\partial S} & \frac{\partial \mu}{\partial S} \end{pmatrix} \begin{pmatrix} \frac{\partial \Phi_k}{\partial T} \\ \frac{\partial \Phi_k}{\partial \mu} \end{pmatrix} \\
  & + \left[\begin{pmatrix} \frac{\partial \Phi_m}{\partial T} & \frac{\partial \Phi_m}{\partial \mu} \end{pmatrix} \begin{pmatrix} \frac{\partial T}{\partial S} && \frac{\partial\mu}{\partial S} \\ \frac{\partial T}{\partial N} && \frac{\partial \mu}{\partial N}  \end{pmatrix} 
  \begin{pmatrix}
    \frac{\partial^3 \Omega}{\partial T^2\partial\mu} && \frac{\partial^3 \Omega}{\partial T \partial \mu^2} \\ \frac{\partial^3 \Omega}{\partial T \partial\mu^2} && \frac{\partial^3 \Omega}{\partial\mu^3}   
  \end{pmatrix}
  \begin{pmatrix} \frac{\partial T}{\partial S} && \frac{\partial\mu}{\partial S} \\ \frac{\partial T}{\partial N} && \frac{\partial \mu}{\partial N}  \end{pmatrix}
  \begin{pmatrix} \frac{\partial \Phi_n}{\partial T} \\ \frac{\partial \Phi_n}{\partial \mu} \end{pmatrix} \right] \times \begin{pmatrix} \frac{\partial T}{\partial N} & \frac{\partial \mu}{\partial N} \end{pmatrix} \begin{pmatrix} \frac{\partial \Phi_k}{\partial T} \\ \frac{\partial \Phi_k}{\partial \mu} \end{pmatrix}.
\end{split}\label{eq:RelationChiSNChiTmu2ndOrder}
\end{equation}
This scheme can be extended to higher orders, although the algebraic complexity grows quickly.

\subsubsection{Additional operator expressions for response at fixed entropy and particle number}
To evaluate eqs.\ \eqref{eq:RelationChiSNChiTmu} and \eqref{eq:RelationChiSNChiTmu2ndOrder} we need additional quantities, like
\begin{equation}
\begin{split}
  \frac{\partial \Phi_m}{\partial T} = & \frac{1}{T^2} \left[ \text{Tr}\{ \rho_\text{i} (H_\text{i}-\mu N) \phi_m\} - \text{Tr}\{ \rho_\text{i} (H_\text{i}- \mu N ) \} \text{Tr}\{ \rho_\text{i} \phi_m \}  \right] = \frac{1}{T^2} \langle (H_\text{i} - \mu N) \phi_m \rangle_c, \\
  \frac{\partial \Phi_m}{\partial \mu} = & \frac{1}{T} \left[ \text{Tr}\{ \rho_\text{i} N \phi_m\} - \text{Tr}\{ \rho_\text{i} N \} \text{Tr}\{ \rho_\text{i} \phi_m \} \right] = \frac{1}{T} \langle N \phi_m \rangle_c.
\end{split}
\end{equation}
On the right hand side we have connected correlation functions of the standard kind, because $H_\text{i}-\mu N$ as well as $N$ commute with $\rho_\text{i}$.

Similarly, second derivatives of expectation values with respect to $T$ and $\mu$ are found to be
\begin{equation}
\begin{split}
  \frac{\partial^2 \Phi_m}{\partial T^2} = & \frac{1}{T^4} \left[ \langle (H_\text{i}-\mu N)^2 \phi_m \rangle - 2 \langle (H_\text{i}-\mu N) \rangle \langle (H-\mu N) \phi_m \rangle \right. \\
  & \quad\quad \left.- \langle (H_\text{i}-\mu N)^2\rangle \Phi_m + 2 \langle H_\text{i}-\mu N \rangle^2 \Phi_m \right],\\
  \frac{\partial^2 \Phi_m}{\partial T \partial \mu} = & \frac{1}{T^3} \left[ \langle (H-\mu N) N \phi_m \rangle - \langle (H-\mu N) \phi_m \rangle \langle N \rangle - \langle N \phi_m \rangle \langle H-\mu N\rangle \right. \\
  & \quad\quad\left.- \langle (H-\mu N) N \rangle \Phi_m + 2 \langle H-\mu N \rangle \langle N \rangle \Phi_m \right] - \frac{1}{T^2} \left[ \langle N \phi_m \rangle - \langle N \rangle \Phi_m \right], \\
  \frac{\partial^2 \Phi_m}{\partial \mu^2} = & \frac{1}{T^2} \left[ \langle N^2 \phi_m \rangle - 2 \langle N \rangle \langle N \phi_m \rangle - \langle N^2 \rangle \Phi_m + 2 \langle N\rangle^2 \Phi_m \right].
\end{split}
\end{equation}
It remains to evaluate the derivatives of the susceptibilities in \eqref{eq:susceptibilitiesFixedTmu} with respect to $T$ and $\mu$, which can be done with the help of
\begin{equation}
\begin{split}
  \frac{\partial}{\partial T}\left( \frac{1}{T} \langle \phi_m ; \phi_n \rangle \right) = & - \frac{1}{T^2} \langle \phi_m ; \phi_n \rangle - \frac{1}{T^3} \langle H_\text{i}-\mu N \rangle \langle \phi_m ; \phi_n \rangle, \\
  & + \frac{1}{T^3} \int_0^1 d\lambda \, \text{Tr} \left\{ \phi_n \rho_\text{i}^{(1-\lambda)} \phi_m \rho_\text{i}^\lambda (H_\text{i}-\mu N) \lambda + \phi_m \rho_\text{i}^{(1-\lambda)} \phi_n \rho_\text{i}^\lambda (H_\text{i}-\mu N) \lambda \right\} \\
  \frac{\partial}{\partial \mu}\left( \frac{1}{T} \langle \phi_m ; \phi_n \rangle \right) = & - \frac{1}{T^2} \langle N \rangle \langle \phi_m ; \phi_n \rangle, \\
  & + \frac{1}{T^2} \int_0^1 d\lambda \, \text{Tr} \left\{ \phi_n \rho_\text{i}^{(1-\lambda)} \phi_m \rho_\text{i}^\lambda N \lambda + \phi_m \rho_\text{i}^{(1-\lambda)} \phi_n \rho_\text{i}^\lambda N \lambda \right\}, \\
  \frac{\partial}{\partial T} \text{Tr}\{ \rho_\text{i} \phi_{mn} \} = & \frac{1}{T^2} \left\langle (H_\text{i}-\mu N) \phi_{mn} \right\rangle  - \frac{1}{T^2} \left\langle (H_\text{i}-\mu N) \right\rangle \left\langle \phi_{mn} \right\rangle,\\
  \frac{\partial}{\partial \mu} \text{Tr}\{ \rho_\text{i} \phi_{mn} \} = & \frac{1}{T} \left\langle N \phi_{mn} \right\rangle  - \frac{1}{T} \left\langle N \right\rangle \left\langle \phi_{mn} \right\rangle.
\end{split}
\end{equation}
This closes this discussion and we have now all ingredients to determine static response functions at fixed entropy and particle number, at least up to third order.

\section{Real time evolution}
We now discuss the time evolution of density matrices in an isolated situation where the evolution is unitary. The discussion prepares the derivation of expressions for dynamical response functions in a subsequent step. 

Assuming an isolated system is a restriction, of course. While really isolated systems are rare, one can sometimes take further degrees of freedom explicitly into account to make this assumption more realistic.

On the other side, dropping the isolation assumption and considering an open quantum system leads to a lot of freedom, because everything could depend on the details of the coupling to the environment. Also for this reason we concentrate on situations where isolation, possibly after an extension to further degrees of freedom, is a good assumption.

\subsection{Evolution of density matrix}
Starting from the time $t=t_\text{i}$ we allow time-dependent functions $j^n(t)$. This makes the Hamiltonian time-dependent and as a consequence the density matrix as well. The latter evolves according to the von-Neumann equation,
\begin{equation}
  i \frac{d}{d t} \rho(t) = H(t) \rho(t) - \rho(t) H(t) = [ H(t), \rho(t)].
  \label{eq:TimeEvRho}
\end{equation}
The formal solution is $\rho(t) = U(t, t_\text{i})[j] \, \rho_\text{i} \, U(t, t_\text{i})[j]^\dagger$, with a unitary evolution operator $U(t, t_\text{i})[j]$ so that
\begin{equation}
  i \frac{d}{dt} U(t, t_\text{i})[j] = H(j(t)) \; U(t, t_\text{i})[j].
\label{eq:EOMU}
\end{equation}
The evolution operator should be seen as a functional of $j(t)$, where causality implies that only $j(t^\prime)$ with $t^\prime < t$ can influence $U(t, t_\text{i})[j]$. The matrix elements can be determined with the functional integral formalism,
\begin{equation}
  \langle \chi_\text{f} | U(t_\text{f}, t_\text{i})[j] | \chi_\text{i} \rangle = \int_{\chi_\text{f}, \chi_\text{i}} D\chi \exp\left(i S[\chi, j]\right),
\end{equation}
where the dynamical fields are kept fixed at the initial time $t_\text{i}$ and final time $t_\text{f}$. We note that the conjugate operator has the matrix elements
\begin{equation}
  \langle \chi_\text{f} | (U(t_\text{f}, t_\text{i})[j])^\dagger | \chi_\text{i} \rangle = \langle \chi_\text{i} | U(t_\text{f}, t_\text{i})[j] | \chi_\text{f} \rangle^* = \int_{\chi_\text{i}, \chi_\text{f}} D\chi \exp\left(-iS^*[\chi, j]\right).
\end{equation}
One should see this as a time evolution from the final time $t_\text{f}$ backwards to the initial time $t_\text{i}$, with the evolution depending on the time-dependent source $j(t)$. Accordingly we occasionally use the notation
\begin{equation}
  \langle \chi_\text{f} | (U(t_\text{f}, t_\text{i})[j])^\dagger | \chi_\text{i} \rangle = \langle \chi_\text{i} | U(t_\text{i}, t_\text{f})[j] | \chi_\text{f} \rangle,
\end{equation}
where the reversed order of time arguments implements the complex conjugation.

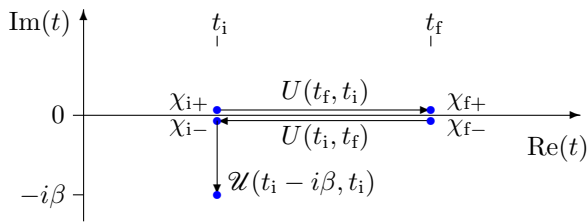
\begin{figure}
\begin{center}
\begin{picture}(200,90)(-100,-45)
% axis
\put(-94,0){\vector(1,0){190}} % x-Achse
\put(-90,-40){\vector(0,1){80}} % y-Achse
% axis labels
\put(76,-15){{Re$(t)$}}
\put(-118,32){{Im$(t)$}}
\put(-102,-3){$0$}
% Label on y-axis
\put(-94,-30){\line(1,0){4}} % short line
\put(-114,-33){$-i\beta$} % label
% label on x-axis
\put(-40,26){\line(0,1){4}} % short line
\put(40,26){\line(0,1){4}} % short line
\put(-42,32){$t_\text{i}$} % label
\put(38,32){$t_\text{f}$} % label
% forward evolution
\put(-40,2){\vector(1,0){80}}
\put(-40,2){\color{blue}\circle*{3}}
\put(40,2){\color{blue}\circle*{3}}
% backward evolution
\put(40,-2){\vector(-1,0){80}}
\put(-40,-2){\color{blue}\circle*{3}}
\put(40,-2){\color{blue}\circle*{3}}
% downward evolution
\put(-40,-2){\vector(0,-1){28}}
\put(-40,-30){\color{blue}\circle*{3}}
% labels inside
\put(-58,4){$\chi_{\text{i}+}$} % label
\put(46,4){$\chi_{\text{f}+}$} % label
\put(-58,-6){$\chi_{\text{i}-}$} % label
\put(46,-6){$\chi_{\text{f}-}$} % label
\put(-16,6){$U(t_\text{f}, t_\text{i})$} % label
\put(-16,-11){$U(t_\text{i}, t_\text{f})$} % label
\put(-36,-28){$\mathscr{U}(t_\text{i}-i\beta, t_\text{i})$} % label
\end{picture}
\captionof{figure}{Illustration of evolution path in the complex time plane to represent the density matrix.}\label{fig:1}
\end{center}
\end{figure}

With this we can write the density matrix elements at time $t_\text{f}$,
\begin{equation}
\begin{split}
  \langle \chi_{\text{f}+} | \rho_\text{f} | \chi_{\text{f}-} \rangle = & \frac{1}{Z(\beta, j_\text{i})} \int D \chi_{\text{i}+} D \chi_{\text{i}-} \langle \chi_{\text{f}+} | U(t_\text{f}, t_\text{i})[j] | \chi_{\text{i}+} \rangle \\
  & \times \langle \chi_\text{i+} | \mathscr{U}(t_\text{i}-i \beta, t_\text{i}, j_\text{i}) | \chi_{\text{i}-} \rangle \langle \chi_{\text{i}-} | U (t_\text{i}, t_\text{f})[j] | \chi_{\text{f}-} \rangle.
\end{split}
\label{eq:FunctIntegralEvolvedDM}
\end{equation}
We note that the evolution operators $U$ depend here in a functional way on $j(t)$, while the operator $\mathscr{U}$ for the initial state is simply a function of $j_\text{i}$. We illustrate the complex time contour in Fig.\ \ref{fig:1}.

\subsubsection{Some remarks}
\begin{enumerate}
  \item For an isolated quantum system with unitary dynamics the von-Neumann entropy is conserved, $S(t) = - \text{Tr}\{ \rho(t) \ln \rho(t) \} = - \text{Tr}\{ \rho_\text{i} \ln \rho_\text{i} \} = S_\text{i}$, also at $t>t_\text{i}$ where the Hamiltonian is time-dependent.
  \item Energy is in the region $t_\text{i}<t<t_\text{f}$ not conserved. Through the time-dependent external sources $j^n(t)$ one can transmit energy to the system, or occasionally extract energy. One would believe the former to be much more likely than the latter. Both the precise amount of the inserted energy, as well as the sign, are subject to statistical fluctuations.
  \item At times $t>t_\text{i}$ the density matrix $\rho(t)$ is not constant, because thermal equilibrium is lost. For very large times $t\gg t_\text{f}$ one expects many macroscopic interacting quantum systems to thermalize again, in the sense that macroscopic or coarse-grained observables are well approximated by an appropriate equilibrium ensemble.
  \item Assume that a density matrix $\rho(t)$ that can be written in the form $\rho(t) = \sum_n p_n \rho_n(t)$, where $\rho_n(t)$ has the charge or particle number eigenvalue $n$, i.\ e.\ $N\rho_n = \rho_n N = n \rho_n$. This is usually the case when particle number is conserved, $[H(t), N]=0$. In this case one can replace $H(t)$ in \eqref{eq:TimeEvRho} and \eqref{eq:EOMU} by $H(t)-\mu N$, without any change in $\rho(t)$. In a quantum field theory this change in the Hamiltonian amounts to a gauge transformation (see also section \ref{sec:ComplexTimeEvolution}).
\end{enumerate}

\subsection{Causal response}
We are now interested in the expectation value of an observable $\phi_m$ for a time $t > t_\text{i}$,
\begin{equation}
  \Phi_m(t) = \text{Tr} \{ \rho(t) \phi_m(j(t)) \}.
\end{equation}
The question is: How does $\Phi_m(t)$ depend on the form of the function $j^n(t)$? Or, with other words, what is the response of the expectation value $\Phi_m(t)$ to a perturbation in the Hamiltonian or action proportional to $j^n(t^\prime)$ at some time $t^\prime$? It is here important that both the density matrix $\rho(t)$ and the operator $\phi_m(j(t))$ depend on time as a consequence of the time-dependent source $j$.

For reasons of causality, $\Phi_m(t)$ can depend on $j^n(t^\prime)$ only at times $t^\prime \leq t$, but not on perturbations at times $t^\prime > t$.

\subsection{Volterra series}\label{sec:VolterraSeries}
The general idea of response theory is to write a functional Taylor series for the deviation from the initial equilibrium value,
\begin{equation}
\begin{split}
 \Phi_m(t) - \Phi_m^\text{i} = &  \sum_n \int_{t_\text{i}}^t d t^\prime \; \Delta^R_{mn}(t-t^\prime)   \left(j^n(t^\prime) - j^n_\text{i} \right) \\
 & + \frac{1}{2} \sum_{n,k} \int_{t_\text{i}}^t dt^\prime dt^{\prime\prime} \; \Delta^R_{mnk}(t-t^\prime, t-t^{\prime\prime})   \left( j^{n}(t^\prime) - j^n_\text{i} \right) \left( j^{k}(t^{\prime\prime}) - j^k_\text{i} \right) + \ldots
\end{split}\label{eq:Volterra}
\end{equation}
This series is named after Vito Volterra %(1860 -- 1940) 
\cite{Volterra1959,Schetzen1980}. For reasons of causality, the integrals over $t^\prime$ and $t^{\prime\prime}$ are restricted to values $t^\prime, t^{\prime\prime} \leq t$, and they are restricted to $t^\prime, t^{\prime\prime} > t_\text{i}$ because the source is only there deviating from its initial value.

One can assume that the lower orders of this series describe the response $\Phi_m(t) - \Phi_m^\text{i}$ reasonably well when  $\rho_\text{i}$ is a stable equilibrium state, and if the perturbation amplitudes $j^n(t)-j^n_\text{i}$ are small enough.

The function $\Delta^R_{mn}(t-t^\prime)$ quantifies linear response, similarly $\Delta^R_{mnk}(t-t^\prime, t-t^{\prime\prime})$ quadratic response, and so on. For quadratic response one can assume the symmetry $\Delta^R_{mnk}(t-t^\prime, t-t^{\prime\prime}) = \Delta^R_{mkn}(t-t^{\prime\prime}, t-t^{\prime})$, known as intrinsic permutation symmetry, and similar at higher orders.

One can consider the expectation value $\Phi_m(t)$ as a functional of the source, $\Phi_m(t)[j]$. The response functions are then functional derivatives at $j^n(t) = j^n_\text{i}$. The linear response function is
\begin{equation}
  \Delta^R_{mn}(t-t^\prime) = \frac{\delta}{\delta j^n(t^\prime)} \Phi_m(t)[j] {\big |}_{j=j_\text{i}},
\label{eq:linearResponseFunctionalDerivative}
\end{equation}
the quadratic response function
\begin{equation}
  \Delta^R_{mnk}(t-t^\prime, t-t^{\prime\prime}) = \frac{\delta^2}{\delta j^n(t^\prime)\delta j^k(t^{\prime\prime})} \Phi_m(t)[j] {\big |}_{j=j_\text{i}},
\label{eq:quadraticResponseFunctionalDerivative}
\end{equation}
and so on.

\subsubsection{Translational symmetry at the expansion point}
At the expansion point $j(t)=j_\text{i}$ our system is in global thermal equilibrium. This implies for the response functions a translational symmetry so that $\Delta^R_{mn}(t-t^\prime)$ does not depend separately on $t$ and $t^\prime$ but ony on the difference $t-t^\prime$, and similarly $\Delta^R_{mnk}(t-t^\prime, t-t^{\prime\prime})$ only on $t-t^\prime$ and $t-t^{\prime\prime}$, and so on. Our notation is already taking this into account.

\subsection{Instantaneous and delayed response}
One is often interested in a description of a physical system on a typical time scale $T$. The response functions might have a contribution for time differences much shorter than this scale, $t-t^\prime \ll T$, which can be approximated as instantaneous, and a contribution on the scale of $t-t^\prime \approx T$, or larger. This motivates the decomposition
\begin{equation}
  \Delta^R_{mn}(t-t^\prime) = \delta(t-t^\prime) \Delta^\infty_{mn}  + \Delta^{\mathscr{R}}_{mn}(t-t^\prime),
  \label{eq:DecompositionLinearRetardedInstDelayed}
\end{equation}
where $\Delta^\infty_{mn}$ parametrizes the instantaneous response, and $\Delta^{\mathscr{R}}_{mn}(t-t^\prime)$ the part of the response that is delayed by at least a short time of order $T$. 

For the quadratic response function we can introduce a similar decomposition, 
\begin{equation}
\begin{split}
  \Delta^R_{mnk}(t-t^\prime, t-t^{\prime\prime}) = & \delta(t-t^\prime) \delta(t-t^{\prime\prime}) \Delta^{\infty}_{mnk} + \delta(t-t^\prime) \Delta^{\infty\mathscr{R}}_{mnk}(t-t^{\prime\prime}) \\
  & + \delta(t-t^{\prime\prime}) \Delta^{\infty\mathscr{R}}_{mkn}(t-t^{\prime}) + \Delta^{\mathscr{R}}_{mnk}(t-t^\prime, t-t^{\prime\prime}),
\end{split}\label{eq:DecompositionQuadraticRetardedInstDelayed}
\end{equation}
where we have the symmetry $\Delta^{\infty}_{mnk} = \Delta^{\infty}_{mkn}$ and $\Delta^{\mathscr{R}}_{mnk}(t-t^\prime, t-t^{\prime\prime}) = \Delta^{\mathscr{R}}_{mkn}(t-t^{\prime\prime}, t-t^{\prime})$. Similar at higher order.

The Volterra series becomes with this
\begin{equation}
\begin{split}
  \Phi_m(t) - \Phi_m^\text{i} = & \sum_n (j^n(t)-j^n_\text{i}) \Delta^\infty_{mn} + \sum_n \int_{t_\text{i}}^t d t^\prime \; \Delta^{\mathscr{R}}_{mn}(t-t^\prime)   \left(j^n(t^\prime) - j^n_\text{i} \right) \\
  & + \frac{1}{2} \sum_{n,k}  \left( j^{n}(t) - j^n_\text{i} \right) \left( j^{k}(t) - j^k_\text{i} \right) \Delta^\infty_{mnk} \\
  &  + \sum_{n,k} \left( j^{n}(t) - j^n_\text{i} \right)  \int_{t_\text{i}}^t dt^\prime \;  \Delta^{\infty\mathscr{R}}_{mnk}(t-t^\prime) \left( j^{k}(t^{\prime}) - j^k_\text{i} \right) \\
 & + \frac{1}{2} \sum_{n,k} \int_{t_\text{i}}^t dt^\prime dt^{\prime\prime} \; \Delta^{\mathscr{R}}_{mnk}(t-t^\prime, t-t^{\prime\prime})   \left( j^{n}(t^\prime) - j^n_\text{i} \right) \left( j^{k}(t^{\prime\prime}) - j^k_\text{i} \right) + \ldots .
\end{split}\label{eq:ResponseInstDelayed}
\end{equation}
On the right hand side one has now instantaneous and delayed contributions. One expects that the kernels for the delayed contributions vanish for larger arguments of time differences, so that the time integrals are dominated by the regions close to the time $t$.

\subsection{Relaxation functions}
It is useful to consider a source $j(t)$ that deviate from its initial value $j_\text{i}$ through a linear superpositions of step functions, or more specifically of functions that grow very slowly until they go to zero abruptly at some time $t_A$, where $A$ is an index in a convenient set,
\begin{equation}
  j^n(t) - j^n_\text{i} = \sum_{A}\Delta j^n_A \, \theta(t_A-t) \, \exp( (t-t_A) \epsilon). 
\label{eq:perturbationRelaxationFunctionDef}
\end{equation}
As usual, $\epsilon$ is a very small positive number, and we will eventually take the limit $\epsilon\to 0$. In that case the Volterra series becomes
\begin{equation}
\begin{split}
 \Phi_m(t) - \Phi_m^\text{i} = &  \sum_A \sum_n \int_{-\infty}^{t_A} d t^\prime \; \Delta^R_{mn}(t-t^\prime) e^{(t^\prime-t_A)\epsilon}  \Delta j^n_A \\
 & + \frac{1}{2} \sum_{A,B} \sum_{n,k} \int_{-\infty}^{t_A} dt^\prime \int_{-\infty}^{t_B} dt^{\prime\prime} \; \Delta^R_{mnk}(t-t^\prime, t-t^{\prime\prime}) e^{(t^\prime-t_A+t^{\prime\prime}-t_B)\epsilon}  \Delta j^{n}_A \Delta j^{k}_B \\
 & + \ldots \\
 = & \sum_A \sum_n \Psi_{mn}(t-t_A) \Delta j^n_A + \frac{1}{2} \sum_{A,B}\sum_{n, k} \Psi_{mnk}(t-t_A, t-t_B) \Delta j^{n}_A \Delta j^{k}_B + \ldots.
\end{split}\label{eq:VolterraRelaxation}
\end{equation}
In the last equation we employed the linear relaxation function (we drop a factor that goes to unity for $\epsilon \to 0$ in the second step)
\begin{equation}
  \Psi_{mn}(t-t_A) = \int_{-\infty}^{t_A} d t^\prime \; \Delta^R_{mn}(t-t^\prime) e^{(t^\prime-t_A)\epsilon}=  \int_{t-t_A}^\infty ds\;  \Delta^R_{mn}(s) e^{-s\epsilon},
\end{equation}
as well as the quadratic relaxation function,
\begin{equation}
\begin{split}
  \Psi_{mnk}(t-t_A, t-t_B) = & \int_{-\infty}^{t_A} dt^\prime \int_{-\infty}^{t_B} dt^{\prime\prime} \; \Delta^R_{mnk}(t-t^\prime, t-t^{\prime\prime}) e^{(t^\prime - t_A + t^{\prime\prime}-t_B)\epsilon} \\
  = & \int_{t-t_A}^\infty ds_1 \int_{t-t_B}^\infty ds_2 \; \Delta^R_{mnk}(s_1, s_2) e^{-(s_1+s_2)\epsilon},
\end{split}
\end{equation}
and similar at higher order.

With the decomposition in eq.\ \eqref{eq:DecompositionLinearRetardedInstDelayed} one finds
\begin{equation}
  \Psi_{mn}(t-t_A) = \theta(t_A-t) \Delta^\infty_{mn} + \int_{t-t_A}^\infty ds \; \Delta^{\mathscr{R}}_{mn}(s) e^{-s\epsilon}.
\end{equation}
The instantaneous part of the response manifests as a step in the relaxation function. In the vicinity of $t-t_A=0$ one has 
\begin{equation}
\begin{split}
  \Psi_{mn}(0_-) = & \Delta^\infty_{mn} + \int_0^\infty   ds \; \Delta^{\mathscr{R}}_{mn}(s) e^{-s\epsilon}, \\
  \Psi_{mn}(0_+) = & \int_0^\infty   ds \; \Delta^{\mathscr{R}}_{mn}(s) e^{-s\epsilon}.
\end{split}\label{eq:PsimnBoundary}
\end{equation}
For negative argument $t-t_A<0$ the relaxation function is constant, $\Psi_{mn}(t-t_A) = \Psi_{mn}(0_-)$. This is a consequence of causality.

For the quadratic response function, the decomposition \eqref{eq:DecompositionQuadraticRetardedInstDelayed} yields
\begin{equation}
\begin{split}
  \Psi_{mnk}(t-t_A, t-t_B) = & \theta(t_A-t) \theta(t_B-t) \Delta^\infty_{mnk} \\
  & + \theta(t_A-t) \int_{t-t_B}^\infty ds \; \Delta^{\infty \mathscr{R}}_{mnk}(s) e^{-s\epsilon} + \theta(t_B-t) \int_{t-t_A}^\infty ds \; \Delta^{\infty \mathscr{R}}_{mkn}(s) e^{-s\epsilon} \\
  & + \int_{t-t_A}^\infty ds_1 \int_{t-t_B}^\infty ds_2 \; \Delta^{\mathscr{R}}_{mnk}(s_1, s_2) e^{-(s_1+s_2)\epsilon}.
\end{split}
\end{equation}
Again there are step functions corresponding to the instantaneous response terms and the function $\Psi_{mnk}(t-t_A, t-t_B)$ is independent of its argument $t-t_A$ or $t-t_B$ when the latter is negative.

Note that the linear response functions follows from the linear relaxation function through
\begin{equation}
  \Delta^R_{mn}(t-t^\prime) %= - \frac{d}{dt} \Psi_{mn}(t-t^\prime) 
  = \frac{d}{dt^\prime} \Psi_{mn}(t-t^\prime),
\label{eq:RelaxationFunctionResponseFunctionRelationDerivative}
\end{equation}
and at quadratic order one has
\begin{equation}
  \Delta^R_{mnk}(t-t^\prime, t-t^{\prime\prime}) = \frac{d^2}{dt^\prime dt^{\prime\prime}} \Psi_{mnk}(t-t^\prime, t-t^{\prime\prime}),
\label{eq:QuadraticRelaxationFunctionResponseFunctionRelationDerivative}
\end{equation}
and similar at higher order. Together these relations show that the knowledge of the relaxation functions is equivalent to the knowledge of the response functions.

For times $t$ much larger than $t_A$ one expects $\Psi_{mn}(t-t_A) \to 0$, and similarly for higher order relaxation functions. The physical reason is that in this limit the perturbation in the system brought in through the perturbation \eqref{eq:perturbationRelaxationFunctionDef} is washed out again, through relaxation back to a global equilibrium state. This expectation should hold for macroscopic observables $\phi_n$ on interacting quantum systems in the case where the perturbation has not changed any of the  conserved quantum numbers that define global equilibrium states.

More concretely, we have assumed here that the evolution dynamics is unitary, so that entropy $S$ remains constant, and that the particle number $N$ remains unchanged. A question mark must be put on energy conservation because a Hamiltonian that depends on time through the external perturbation does not conserve energy in general. As we will discuss below, the expectation value of the work done on the system during a period $t_\text{i} < t < t_\text{f}$, where the Hamiltonian is time-dependent through external sources, is $\langle H^\text{H}(t_\text{f}) - H^\text{H}(t_\text{i}) \rangle$,  with Hamiltonians $H^\text{H}(t)$ in the interaction picture. This expectation value vanishes indeed for the perturbation \eqref{eq:perturbationRelaxationFunctionDef} so that the expectation value of energy at the time where all perturbations have ceased is the same as at very early times. Fluctuations around this expectation value can arise, though.

\subsection{Static response}
For the special case where $j^n(t)-j_\text{i}$ is very slowly turned on from early time on and switched off at time $t=\epsilon$, $j^n(t)-j^n_\text{i} = (j^n-j^n_\text{i})e^{\epsilon t}\theta(\epsilon-t)$, one obtains at $t=0$
\begin{equation}
\begin{split}
  \Phi_m(0) - \Phi^\text{i}_m = \sum_n  \Psi_{mn}(0_-)(j^n-j^n_\text{i})+ \frac{1}{2} \sum_{nk} \Psi_{mnk}(0_-,0_-) (j^n-j^n_\text{i}) (j^k-j^k_\text{i})  + \ldots.
\end{split}
\end{equation}
The linear relaxation function at vanishing argument is a static response function or susceptibility
\begin{equation}
  \Psi_{mn}(0_-) = \int_{0_-}^\infty dt \,  e^{- t\epsilon}\Delta^R_{mn}(t) %= \int_{-\infty}^\infty dt \Delta^R_{mn}(t) e^{-\epsilon t} 
  = \Delta^R_{mn}(\omega){\big |}_{\omega = i \epsilon} = \chi^{(S,N)}_{mn},
\label{eq:InitialLinearRelaxationFunctionSuscept}
\end{equation}
with
\begin{equation}
  \chi^{(S,N)}_{mn} = \Delta^\infty_{mn} + \int_{-\infty}^t dt^\prime \Delta^{\mathscr{R}}_{mn}(t-t^\prime) e^{-(t-t^\prime)\epsilon} = \Delta^\infty_{mn} + \Delta^\mathscr{R}_{mn}(\omega){\big |}_{\omega = i \epsilon}.
\end{equation}
This must be the susceptibility at constant entropy $S$ and particle number $N$ because these quantities cannot be changed by the slow unitary time evolution from $t-t^\prime = \infty$ to $t-t^\prime=0$. In general, the static susceptibilities defined via such a protocol are known as \textit{isolated} static susceptibilities. Under the assumption that there are no conserved quantities besides $S$ and $N$ they agree with $\chi^{(S,N)}_{mn}$ \cite{KuboTodaHashitsume1991}.

Similarly one has
\begin{equation}
  \Psi_{mnk}(0_-, 0_-) = \Delta^R_{mnk}(\omega_1, \omega_2){\big |}_{\omega_1=\omega_2=i \epsilon} = \chi^{(S,N)}_{mnk},
\end{equation}
with
\begin{equation}
\begin{split}
  \chi^{(S,N)}_{mnk} = & \Delta^\infty_{mnk} + \int_{-\infty}^t dt^\prime \left[ \Delta^{\infty\mathscr{R}}_{mnk}(t-t^\prime) + \Delta^{\infty\mathscr{R}}_{mkn}(t-t^\prime) \right] e^{(t-t^\prime)\epsilon} \\
  & + \int_{-\infty}^t dt^\prime dt^{\prime\prime} \Delta^{\mathscr{R}}_{mnk}(t-t^\prime, t-t^{\prime\prime}) e^{-(2t-t^\prime-t^{\prime\prime})\epsilon},
\end{split}
\end{equation}
and so on at higher order. Interestingly this shows that $\Psi_{mn}(0_-)$ and $\Psi_{mnk}(0_-,0_-)$ are fully symmetric in the indices. 

Qualitatively one expects relaxation functions like $\Psi_{mn}(t-t_A)$ to start from equilibrium expressions like $\chi^{(S,N)}_{mn}$ at $t=t_A$, and to decay to zero for large separation $t-t_A$.

\subsection{Volterra series in terms of relaxation functions}
One can insert the relations \eqref{eq:RelaxationFunctionResponseFunctionRelationDerivative} and \eqref{eq:QuadraticRelaxationFunctionResponseFunctionRelationDerivative} in eq.\ \eqref{eq:Volterra} and perform partial integrations. This leads to
\begin{equation}
\begin{split}
  \Phi_m(t) - \Phi_m^\text{i} = & \sum_n \Psi_{mn}(0_-) \, (j^n(t)-j^n_\text{i}) - \sum_n \int_{t_\text{i}}^t dt^\prime \, \Psi_{mn}(t-t^\prime) \frac{d}{dt^\prime} j^n(t^\prime) \\
  & + \frac{1}{2} \sum_{n,k} \Psi_{mnk}(0_-,0_-) (j^n(t)-j^n_\text{i}) (j^k(t)-j^k_\text{i}) \\
  & - \frac{1}{2} \sum_{n,k} \int_{t_\text{i}}^t dt^\prime \Psi_{mnk}(t-t^\prime, 0_-) \left[ \frac{d}{dt^\prime} j^n(t^\prime) \right] (j^k(t) - j^k_\text{i}) \\
  & - \frac{1}{2} \sum_{n,k} \int_{t_\text{i}}^t dt^{\prime\prime} \Psi_{mnk}(0_-,t-t^{\prime\prime}) (j^n(t)-j^n_\text{i}) \left[ \frac{d}{dt^{\prime\prime}} j^k(t^{\prime\prime}) \right] \\
  & + \frac{1}{2} \sum_{n,k} \int_{t_\text{i}}^t dt^\prime dt^{\prime\prime} \Psi_{mnk}(t-t^\prime, t-t^{\prime\prime}) \left[ \frac{d}{dt^\prime} j^n(t^\prime) \right] \left[ \frac{d}{dt^{\prime\prime}} j^k(t^{\prime\prime}) \right]\\
  & + \ldots
\end{split}
\end{equation}
Here one recognizes instantaneous terms, as well as memory terms going with the derivative of the source. A reorganization leads to
\begin{equation}
\begin{split}
  \Phi_m(t) - \Phi_m^\text{i} = & \sum_n \chi_{mn}^{(S,N)} \, (j^n(t)-j^n_\text{i}) + \frac{1}{2} \sum_{n,k} \chi^{(S,N)}_{mnk} (j^n(t)-j^n_\text{i}) (j^k(t)-j^k_\text{i}) + \ldots \\
  & - \sum_n \int_{t_\text{i}}^t dt^\prime \left[ \Psi_{mn}(t-t^\prime) + \sum_k \Psi_{mnk}(t-t^\prime, 0_-)  (j^k(t)-j^k_\text{i}) + \ldots \right] \frac{d}{dt^\prime} j^n(t^\prime) \\
  & + \frac{1}{2} \sum_{n,k} \int_{t_\text{i}}^t dt^\prime dt^{\prime\prime} \left[ \Psi_{mnk}(t-t^\prime, t-t^{\prime\prime}) + \ldots \right] \left[ \frac{d}{dt^\prime} j^n(t^\prime) \right] \left[ \frac{d}{dt^{\prime\prime}} j^k(t^{\prime\prime}) \right] + \ldots
\end{split}
\end{equation}
This shows the separation into instantaneous terms at time $t$ and memory terms even more clearly. The instantaneous terms are here governed by the static susceptibilities $\Psi_{mn}(0_-) = \chi_{mn}^{(S,N)}$ and $\Psi_{mnk}(0_-,0_-) = \chi^{(S,N)}_{mnk}$ and so on. Corrections to this static response part are here proportional to time derivatives of $j^n(t)$, as one would assume intuitively.

\subsection{Volterra series frequency space}
One can rewrite the Volterra series in eq.\ \eqref{eq:Volterra} in the form
\begin{equation}
\begin{split}
  \Phi_m(t) - \Phi_m^\text{i} = & \sum_n \int_{-\infty}^\infty d t^\prime \; \Delta^R_{mn}(t-t^\prime)   \left( j^n(t^\prime) - j^n_\text{i} \right) \\
  & + \frac{1}{2} \sum_{n,k} \int_{-\infty}^\infty dt^\prime dt^{\prime\prime} \; \Delta^R_{mnk}(t-t^\prime, t-t^{\prime\prime})   \left( j^{n}(t^\prime) - j^n_\text{i} \right) \left( j^{k}(t^{\prime\prime}) - j^k_\text{i} \right) + \ldots
\end{split}\label{eq:AmtTime}
\end{equation}
We have extended the time integrals to $(-\infty, \infty)$, using that $\Delta^R_{mn}(t-t^\prime)=0$ for $t-t^\prime<0$, and similarly $\Delta^R_{mnk}(t-t^\prime, t-t^{\prime\prime})=0$ for $t-t^\prime<0$ or $t-t^{\prime\prime}<0$, and so on, as implied by causality, and $j^n(t^\prime) = j^n_\text{i}$ when $t^\prime<t_\text{i}$. 

In this form one can conveniently go to Fourier space. To that end we now express the field expectation value and the source field through Fourier transforms,
\begin{equation}
\begin{split}
  \Phi_m(t) - \Phi_m^\text{i} = & \int \frac{d\omega}{2\pi} e^{-i\omega t} \Phi_m(\omega), \\
  j^n(t^\prime) - j^n_\text{i} = & \int \frac{d\omega}{2\pi} e^{-i\omega t^\prime} j^n(\omega),
\end{split}\label{eq:LinearResponseFourierRep}
\end{equation}
Similarly, we write the linear response function as
\begin{equation}
  \Delta^R_{mn}(t-t^\prime) = \int \frac{d\omega}{2\pi} e^{-i\omega(t-t^\prime)} \Delta^R_{mn}(\omega),
\end{equation}
the quadratic response function as
\begin{equation}
  \Delta^R_{mnk}(t-t^\prime, t-t^{\prime\prime}) = \int \frac{d\omega^\prime d\omega^{\prime\prime}}{(2\pi)^2} e^{-i\omega^\prime (t-t^\prime)-i \omega^{\prime\prime}(t-t^{\prime\prime})} \Delta^R_{mnk}(\omega^\prime, \omega^{\prime\prime}).
\label{eq:retardedQuadraticFourier}
\end{equation}
and so on at higher order.

The Volterra series \eqref{eq:AmtTime} becomes in Fourier space
\begin{equation}
\begin{split}
  \Phi_m(\omega) = & \Delta^R_{mn}(\omega) j^n(\omega) \\
  & + \frac{1}{2} \sum_{n,k} \int_{-\infty}^\infty \frac{d\omega^\prime d\omega^{\prime\prime}}{(2\pi)^2} 2\pi \delta(\omega - \omega^\prime - \omega^{\prime\prime}) \Delta^R_{mnk}(\omega^\prime, \omega^{\prime\prime}) j^n(\omega^\prime) j^k(\omega^{\prime\prime}) + \ldots
\end{split}\label{eq:VolterraSeriesFrequencySpace}
\end{equation}
We note in particular that the the convolution governing the linear response in \eqref{eq:AmtTime} becomes a product in frequency space \eqref{eq:VolterraSeriesFrequencySpace}. A periodic perturbation implies a linear response at the same frequency. The function $\Delta^R_{mn}(\omega)$ is sometimes called \textit{dynamic susceptibility} or \textit{complex admittance}. %For the special case where the response function $\Delta^R_{mn}(t-t^\prime)$ is real, one has $\Delta^R_{mn}(-\omega)^* = \Delta^R_{mn}(\omega)$.

The decomposition \eqref{eq:ResponseInstDelayed} reads in Fourier space
\begin{equation}
  \Delta^R_{mn}(\omega) = \Delta^\infty_{mn} + \Delta^{\mathscr{R}}_{mn}(\omega).
\label{eq:DeltaROmegaDecomposition}
\end{equation}
The instantaneous part is independent of frequency. The remaining part $\Delta^{\mathscr{R}}_{mn}(\omega)$ decays for large frequency $\omega$, so that $\Delta^\infty_{mn}$ corresponds to the high-frequency limit of $\Delta^R_{mn}(\omega)$.

For the quadratic response the frequencies $\omega^\prime$ and $\omega^{\prime\prime}$ need to sum up to $\omega$ to have a non-vanishing contribution. Similar to \eqref{eq:DeltaROmegaDecomposition}, the decomposition \eqref{eq:DecompositionQuadraticRetardedInstDelayed} becomes in Fourier space
\begin{equation}
\begin{split}
  \Delta^R_{mnk}(\omega^\prime, \omega^{\prime\prime}) = &  \Delta^{\infty}_{mnk} + \Delta^{\infty\mathscr{R}}_{mnk}(\omega^{\prime\prime}) + \Delta^{\infty\mathscr{R}}_{mkn}(\omega^{\prime}) + \Delta^{\mathscr{R}}_{mnk}(\omega^\prime, \omega^{\prime\prime}),
\end{split}
\end{equation}
where $\Delta^{\infty\mathscr{R}}_{mnk}(\omega^{\prime\prime})$ decays for $\omega^{\prime\prime} \to \infty$ and $\Delta^{\mathscr{R}}_{mnk}(\omega^\prime, \omega^{\prime\prime})$ for $\omega^\prime \to \infty$ or $\omega^{\prime\prime} \to \infty$. Let us also note the intrinsic permutation symmetries in the Fourier representation $\Delta^R_{mnk}(\omega^\prime, \omega^{\prime\prime}) = \Delta^R_{mkn}(\omega^{\prime\prime}, \omega^{\prime})$, $\Delta^{\infty}_{mnk} = \Delta^{\infty}_{mkn}$ and $\Delta^{\mathscr{R}}_{mnk}(\omega^\prime, \omega^{\prime\prime}) = \Delta^{\mathscr{R}}_{mkn}(\omega^{\prime\prime}, \omega^\prime)$. 

\section{Causality and analyticity properties}\label{sec:CausalityAndAnalyticityProperties}

\subsection{From causality to analyticity and back}
In Fourier space, the causality condition on $\Delta^R_{mn}(t-t^\prime)$ becomes an analyticity property. Indeed, the integral
\begin{equation}
  \Delta^R_{mn}(\omega) = \int_0^\infty dt e^{i \omega t} \Delta_{mn}^R(t),
\label{eq:DeltamnOmegaDef}
\end{equation}
is for $\omega$ with $\text{Im}(\omega)>0$ convergent for large times $t$, if $\Delta_{mn}^R(t)$ does not grow exponentially or faster. On the other side, if $\Delta^R_{mn}(\omega)$ is analytic in the upper half of the complex $\omega$ plane, and tends to zero on the real $\omega$-axis for $\omega\to \pm \infty$, the residue theorem implies that $\Delta^R_{mn}(t-t^\prime)$ as defined through \eqref{eq:LinearResponseFourierRep} vanishes for $t-t^\prime <0$. A limiting case is a constant part $\Delta^\infty_{mn}$ in $\Delta^R_{mn}(\omega)$ which yields $\delta(t-t^\prime) \Delta^\infty_{mn}$ in the time domain. To exclude this limiting case, one may subtract this part and consider $\Delta^{\mathscr{R}}_{mn}(\omega)$ instead. One can assume that $\Delta^{\mathscr{R}}_{mn}(\omega)$ vanishes for large real $\omega$. 

For the case where $\Delta^R_{mn}(t-t^\prime)\in \mathbbm{R}$ is real one has $\Delta^R_{mn}(\omega) = \Delta^R_{mn}(-\omega^*)^*$. This holde for $\omega$ on the real axis or in the upper half of the complex plan where eq.\ \eqref{eq:DeltamnOmegaDef} can be used as a definition.

Similarly, for the quadratic response function one could use a Fourier space representation of the form \eqref{eq:retardedQuadraticFourier}. Causality in the time domain implies that $\Delta^R_{mnk}(\omega^\prime, \omega^{\prime\prime})$ is analytic in the upper half of the complex $\omega^\prime$ and $\omega^{\prime\prime}$ planes, and \textit{vice versa}. This can be extended to higher orders.

\subsection{Kramers-Kronig relations}
The analytic properties of $\Delta^{\mathscr{R}}_{mn}(\omega)$ can be used to obtain the relation
\begin{equation}
  \Delta^{\mathscr{R}}_{mn}(\omega) = \frac{i}{\pi} \mathscr{P} \int_{-\infty}^\infty d\omega^\prime\,  \frac{\Delta^{\mathscr{R}}_{mn}(\omega^\prime)}{\omega - \omega^\prime},
\end{equation}
where $\mathscr{P}$ denotes the Cauchy principal value. This can be shown with a closed integral contour in the upper half of the complex $\omega^\prime$-plane that goes around the singularity at $\omega^\prime = \omega$ in a small half circle.

Dividing now into real and imaginary parts,
\begin{equation}
  \Delta^{\mathscr{R}}_{mn}(\omega) = \text{Re}\, \Delta^{\mathscr{R}}_{mn}(\omega) + i \, \text{Im} \, \Delta^{\mathscr{R}}_{mn}(\omega),
\end{equation}
yields the Kramers-Kronig relations
\begin{equation}
\begin{split}
  \text{Re}\, \Delta^{\mathscr{R}}_{mn}(\omega) = & - \frac{1}{\pi} \mathscr{P} \int_{-\infty}^\infty d\omega^\prime\,  \frac{\text{Im}\, \Delta^{\mathscr{R}}_{mn}(\omega^\prime)}{\omega - \omega^\prime},\\
  \text{Im}\, \Delta^{\mathscr{R}}_{mn}(\omega) = & + \frac{1}{\pi} \mathscr{P} \int_{-\infty}^\infty d\omega^\prime\,  \frac{\text{Re}\, \Delta^{\mathscr{R}}_{mn}(\omega^\prime)}{\omega - \omega^\prime}.
\end{split}\label{eq:KramersKronigRelations}
\end{equation}
This shows that the real and imaginary parts of $\Delta^{\mathscr{R}}_{mn}(\omega)$ are not independent and in fact the imaginary part is the \textit{Hilbert transform} of the real part.

This holds similarly for $\Delta^{\infty\mathscr{R}}_{mnk}(\omega)$, the Fourier transform of $\Delta^{\infty\mathscr{R}}_{mnk}(t-t^\prime)$ as used in eq.\ \eqref{eq:DecompositionQuadraticRetardedInstDelayed} and corresponding higher-order terms.

\subsection{Properties of the Hilbert transform}
The Hilbert transform is a convolution,
\begin{equation}
  \tilde g(\omega) = \mathcal{H}[\tilde f](\omega) = \mathscr{P}\int_{-\infty}^\infty \frac{d\omega^\prime}{2\pi} \tilde K(\omega-\omega^\prime) \tilde f(\omega^\prime),
\end{equation}
with the kernel
\begin{equation}
  \tilde K(\omega - \omega^\prime) = \frac{2}{\omega - \omega^\prime}.
\end{equation}
After a Fourier transform,
\begin{equation}
\begin{split}
  f(t) = \int \frac{d\omega}{2\pi} e^{-i\omega t} \tilde f(\omega), \\
  g(t) = \int \frac{d\omega}{2\pi} e^{-i\omega t} \tilde g(\omega),
\end{split}
\end{equation}
the convolution becomes a product,
\begin{equation}
  g(t) = K(t) f(t),
\end{equation}
with
\begin{equation}
  K(t) = \int \frac{d\omega}{2\pi} e^{-i \omega t} \tilde K(\omega) = \int \frac{d\omega}{2\pi}  e^{-i\omega t}\frac{2}{\omega} = - i\, \text{sign}(t).
\label{eq:KernelHilbertTransform}
\end{equation}
This shows that the Hilbert transform induces a kind of analytic structure and squares to minus one. In the present context eq.\ \eqref{eq:KernelHilbertTransform} is an interesting relation because it allows to translate multiplication by step functions to the frequency domain. 

\subsection{Spectral relations from Hilbert transform}
\label{sec:SpectralRelationsFormHilberTransform}
As an application, consider two functions $x(t)$ and $y(t)$ related in the time domain through
\begin{equation}
  x(t) = i \theta(t) y(t) = \frac{i}{2} y(t) + \frac{i}{2} \text{sign}(t) y(t).
\label{eq:GtFtRelation}
\end{equation}
This implies, due to the above relations, in the frequency domain
\begin{equation}
  \tilde x(\omega) = \frac{i}{2} \tilde y(\omega) - \frac{1}{2} \mathcal{H}[\tilde y](\omega) = \frac{i}{2} \tilde y(\omega) - \mathscr{P}\int_{-\infty}^\infty \frac{d\omega^\prime}{2\pi} \frac{\tilde y(\omega^\prime)}{\omega - \omega^\prime}.
\end{equation}
With the Sokhotski-Plemelj relation
\begin{equation}
  \frac{1}{u \pm i \varepsilon} = \mp i \pi \delta(u) + \mathscr{P} \frac{1}{u},
  \label{SokhotskiPlemeljRelation}
\end{equation}
this can also be written as
\begin{equation}
  \tilde x(\omega) = - \int_{-\infty}^\infty \frac{d\omega^\prime}{2\pi}\frac{\tilde y(\omega^\prime)}{\omega+i\varepsilon - \omega^\prime}.
\label{eq:GFSpectralRelation}
\end{equation}

One may define a function $G(z)$ with complex argument $z$ through the integral
\begin{equation}
G(z) = - \int_{-\infty}^\infty \frac{d\omega^\prime}{2\pi}\frac{\rho(\omega^\prime)}{z - \omega^\prime},
\label{eq:GzSpectralRep}
\end{equation}
This is known as a spectral representation, where $\rho(\omega) = \tilde y(\omega)$ plays here the role of a spectral density. Note that in this version $rho(\omega)$ is in general complex; we will show below how it can be made real. Because the integral over $\omega^\prime$ goes along the real line, $G(z)$ is analytic in the complex plane, except for a branch cut along the real axis, with the discontinuity
\begin{equation}
  \text{Disc} \, G(z) = G(z + i \epsilon) - G(z - i \epsilon) = i \rho(\omega),
  \label{eq:DiscInG}
\end{equation}
and possibly poles that can arise from terms proportional to a Dirac delta distribution in $\tilde y(\omega^\prime)$.

Evaluating $G(z)$ just above the real axis yields $\tilde x(\omega)$,
\begin{equation}
  \tilde x(\omega) = G(\omega+i \epsilon).
\label{eq:xwInTermsOfG}
\end{equation}
The spectral representation \eqref{eq:GzSpectralRep} now ensures that $\tilde x(\omega)$ is analytic in the upper half of the complex plane, corresponding to the causality condition.

This also shows how the Fourier transform of $y(t)$ fully determines the Fourier transform of $x(t)$, as expected. On the other side, $\tilde y(\omega)$ cannot be fixed by $\tilde x(\omega)$ without further information. This is clear because one could arbitrarily change $y(t)$ at $t<0$ without changing $x(t)$.  This also shows that the spectral density $\rho(\omega)$ representing $\tilde x(\omega)$ through eqs.\ \eqref{eq:xwInTermsOfG} and \eqref{eq:GzSpectralRep} is not unique. Instead of taking $\rho(\omega)$ to be the Fourier transform of $y(t)$ we could take it to be the Fourier transform of any function that agrees with $y(t)$ for $t\leq 0$, but differs at $t<0$. In the frequency domain that means that one could add any function to $\rho(\omega)$ that is analytic in the lower half of the complex plane without changing the resulting $\tilde x(\omega)$. One can use this freedom to make $\rho(\omega)$ real. Indeed, the choice
\begin{equation}
  \rho(\omega) = 2\, \text{Im}\, \tilde x(\omega),
\end{equation}
fulfills all requirements. This can be seen using \eqref{SokhotskiPlemeljRelation} and the first relation in \eqref{eq:KramersKronigRelations} for $\tilde x(\omega)$. The discontinuity in eq.\ \eqref{eq:DiscInG} is then purely imaginary.

\subsection{Example electric circuit}
As an example we consider a simple electric cricuit with a resistance $R$ and a capacity $C$ in series. We are interested in the charge $Q(t)$ in the capacity as a response to an applied voltage $U(t)$. The Hamiltonian contains a term $\Delta H(t) = U(t) Q$ so that this problem fits indeed into the scheme. In thermal equilibrium one has vanishing expectation value, $\langle Q \rangle = \text{Tr}\{ \rho_\text{i} Q \} = 0$.  

We have $U=RI + Q/C$ and with $I=dQ/dt$ we obtain the differential equation
\begin{equation}
  U(t) =  R \frac{d}{dt} Q(t) + \frac{Q(t)}{C}.
\label{eq:DGLStromkeis}
\end{equation}
In Fourier space this becomes
\begin{equation}
  Q(\omega) = \Delta^R(\omega) U(\omega),
\end{equation}
with the response function
\begin{equation}
  \Delta^R(\omega) = \frac{1}{- i \omega R + \frac{1}{C}}.
  \label{eq:StromkreisAntwortFrequenzraum}
\end{equation}
As a function of frequency $\omega$ this is indeed analytic in the upper half of the complex plane, with a single pole at $\omega = -i /(RC)$. Fourier transformation to the time domain gives
\begin{equation}
  Q(t) = \int_{-\infty}^\infty dt^\prime \Delta^R(t-t^\prime) U(t^\prime),
\end{equation}
with the response function
\begin{equation}
  \Delta^R(t-t^\prime) = \int \frac{d\omega}{2\pi} \frac{e^{-i\omega(t-t^\prime)}}{- i \omega R + \frac{1}{C}} = \theta(t-t^\prime) \frac{1}{R} \exp\left( - \frac{t-t^\prime}{RC} \right).
  \label{eq:AntwortfunktionStromkreis}
\end{equation}
Indeed it is causal. One can also obtain \eqref{eq:AntwortfunktionStromkreis} directly as a retarded Green's function to the linear differential equation \eqref{eq:DGLStromkeis}.

\subsection{Example damped harmonic oscillator}
The equation of motion for a damped harmonic oscillator with an accelerating force per unit mass $f(t)$ is
\begin{equation}
  \ddot x(t) + 2 \zeta \omega_0 \dot x(t) + \omega_0^2 x(t) = f(t).
\end{equation}
With the Fourier space representations defined by
\begin{equation}
  x(t) = \int \frac{d\omega}{2\pi} e^{-i\omega t} \tilde x(\omega), \quad\quad\quad
  f(t) = \int \frac{d\omega}{2\pi} e^{-i\omega t} \tilde f(\omega),
\end{equation}
this becomes
\begin{equation}
  \left[-\omega^2 - 2 i \zeta \omega_0 \omega + \omega_0^2 \right] \tilde x(\omega) = \tilde f(\omega).
\end{equation}
Accordingly one has
\begin{equation}
  x(t) = \int_{-\infty}^\infty dt^\prime \Delta^R(t-t^\prime) f(t^\prime),
\end{equation}
with the retarded response function
\begin{equation}
  \Delta^R(t-t^\prime) = \int \frac{d\omega}{2\pi} \frac{e^{-i\omega(t-t^\prime)}}{-\omega^2 - 2i \zeta \omega_0 \omega + \omega_0^2} = \theta(t-t^\prime) \exp(-\zeta \omega_0 (t-t^\prime)) \frac{\sin\left(\sqrt{1-\zeta^2}\omega_0 (t-t^\prime)\right)}{\sqrt{1-\zeta^2} \omega_0}.
\end{equation}
The frequency poles were at $\omega = [\pm \sqrt{1-\zeta^2} -i \zeta ] \omega_0$, indeed below the real axis. One recognizes for $\zeta<1$ an oscillating part, as well as an exponential damping.

For example, one may start with an oscillator at rest, and then accelerate it with a pulse $f(t) = \delta(t)$. This implies that $\dot x(t)$ jumps from zero to unity at $t=0$. The solution at $t>0$ is
\begin{equation}
  x(t) =  \exp(-\zeta \omega_0 (t-t^\prime))\frac{\sin\left(\sqrt{1-\zeta^2}\omega_0 t\right)}{\sqrt{1-\zeta^2} \omega_0},
\end{equation}
which indeed satisfies the correct boundary conditions $x(0)= 0$ and $\dot x(0) = 1$.

\section{Time-dependent perturbation theory}
\label{sec:TimedependentPerturbationTheory}
We develop now a time-dependent perturbation theory in order to find expressions for response functions within the microscopic quantum theory. We start with a version of the interaction picture, as a mix of the Schrödinger and Heisenberg pictures of quantum dynamics.

\subsection{Interaction picture}
We start with the unitary evolution operator $U(t, t_\text{i})$, defined through the evolution equation
\begin{equation}
  i \frac{d}{dt} U(t, t_\text{i}) = \left[ H(j(t))-\mu N \right] U(t, t_\text{i}).
\end{equation}
The unusual term $-\mu N$ on the right hand side leads to additional $N$-dependent phases in $U(t, t_\text{i})$ that do not influence observables as long as we are concerned with quantum states that are block-diagonal with respect to the particle number $N$ and as long as there is no interference between sectors of different $N$, see also section \ref{sec:ComplexTimeEvolution}.

Because $H(j(t))$ depends on the sources $j^n(t)$, this is also the case for the evolution operator $U(t, t_\text{i})$. We want to expand it around the constant configuration $j^n_\text{i}$ to find concrete expressions for response functions.

It is convenient to introduce the operator $\tilde U(t, t_\text{i}) = \exp(i (t-t_\text{i}) [H_\text{i}-\mu N]) U(t, t_\text{i})$, and we find for its time derivative,
\begin{equation}
\begin{split}
  i \frac{d}{dt} \tilde U(t, t_\text{i}) = & \exp\left(i (t-t_\text{i}) [H_\text{i}-\mu N]\right) \left[- H_\text{i} + \mu N + i \frac{d}{dt}\right] U(t, t_\text{i}) \\
  = &  \exp\left(i (t-t_\text{i}) [H_\text{i}-\mu N]\right) [H(t)-H_\text{i}] \exp(-i (t-t_\text{i}) [H_\text{i}-\mu N]) \tilde U(t, t_\text{i})\\
  = & [H^\text{H}(t)-H_\text{i}] \tilde U(t, t_\text{i}).
\end{split}\label{eq:timeEvolutionUTilde}
\end{equation}
In the last step we are using the time-dependent Hamiltonian in the Heisenberg picture (or better, interaction picture) with respect to $H_\text{i}-\mu N$,
\begin{equation}
  H^\text{H}(t) = \exp\left(i (t-t_\text{i}) [H_\text{i}-\mu N]\right) H(t) \exp\left(-i (t-t_\text{i}) [H_\text{i}-\mu N]\right).
\end{equation}
Note that the time-independent piece of the Hamiltonian $H_\text{i}$ is the same in the Schrödinger and the interaction picture. The evolution equation \eqref{eq:timeEvolutionUTilde} has a formal solution as a time-ordered exponential
\begin{equation}
\tilde U(t, t_\text{i}) = \mathscr{T}\exp\left(-i\int_{t_\text{i}}^t dt^\prime \{H^\text{H}(t^\prime)-H_\text{i}\}\right),
\end{equation}
and the full time evolution is thus
\begin{equation}
  U(t, t_\text{i}) = \exp\left(-i (t-t_\text{i}) [H_\text{i}-\mu N] \right) \mathscr{T}\left\{\exp\left(-i\int_{t_\text{i}}^t dt^\prime \{H^\text{H}(t^\prime)-H_\text{i}\}\right) \right\}.
\label{eq:EvOpInteractionPicture}
\end{equation}
The second piece contains the contributions from deviations $j^n(t) - j^n_\text{i}$ of the sources from their initial value. Similarly one has
\begin{equation}
  U(t_\text{i}, t) = U(t,t_\text{i})^\dagger = \bar{\mathscr{T}} \left\{\exp\left(i\int_{t_\text{i}}^t dt^\prime \{H^\text{H}(t^\prime)-H_\text{i}\}\right) \right\} \exp\left(i (t-t_\text{i}) [H_\text{i} - \mu N] \right),
\label{eq:EvOpInteractionPictureHC}
\end{equation}
where the first exponential is now anti-time-ordered.

The density matrix evolves according to
\begin{equation}
  \rho(t) = U(t, t_\text{i}) \rho_\text{i} U(t_\text{i}, t),
\end{equation}
and for the expectation value we find
\begin{equation}
\begin{split}
  & \Phi_m(t) = \text{Tr}\{ \phi_m(j(t)) \rho(t) \} \\
  & = \text{Tr} \left\{ \phi_m^\text{H}(t, j(t)) \mathscr{T}\left\{\exp\left(-i\int_{t_\text{i}}^t dt^\prime \{H^\text{H}(t^\prime)-H_\text{i}\}\right) \right\} \rho_\text{i} % \right. \\
  % & \times \left.  
  \bar{\mathscr{T}} \left\{\exp\left(i\int_{t_\text{i}}^t dt^\prime \{H^\text{H}(t^\prime)-H_\text{i}\}\right) \right\} \right\},
\end{split}\label{eq:ExpValueInteractionPicture}
\end{equation}
with the modified Heisenberg, or better interaction picture, operator
\begin{equation}
\begin{split}
  \phi_m^\text{H}(t, j(t)) = & \exp\left(i (t-t_\text{i}) [H_\text{i} - \mu N] \right) \phi_m(j(t)) \exp\left(-i (t-t_\text{i}) [H_\text{i} - \mu N] \right) \\
  = & \phi_m^\text{H}(t) + \sum_n \phi_{mn}^\text{H}(t) (j^n(t)-j^n_\text{i}) + \ldots .
\end{split}
\end{equation}
Note that this operator has an explicit time dependence as an operator in the Heisenberg or interaction picture, and it also depends on the source $j(t)$. In the second line we have used the expansion \eqref{eq:ExpansionHamiltonianResponseOperators} and defined Heisenberg or interaction picture operators 
\begin{equation}
  \phi_m^\text{H}(t) =  \exp\left(i (t-t_\text{i}) [H_\text{i} - \mu N] \right) \phi_m \exp\left(-i (t-t_\text{i}) [H_\text{i} - \mu N] \right),
\label{eq:defHeisenbergOperator}
\end{equation}
with $\phi_m=\phi_m(j_\text{i})$, and similar for $\phi_{mn}^\text{H}(t)$ etc.

\subsection{Operator expressions for linear response function}
By expanding eq.\ \eqref{eq:ExpValueInteractionPicture} in the interaction term one can obtain operator expressions for response functions. 

For the linear response function one finds two contributions 
\begin{equation}
\begin{split}
  \Delta^R_{mn}(t-t^\prime) = & \delta(t-t^\prime) \text{Tr}\left\{ \rho_\text{i} \phi_{mn} \right\} + \frac{\theta(t-t^\prime)}{i} \text{Tr}\left\{ \phi_m^\text{H}(t) \left[\rho_\text{i}, \phi_n^\text{H}(t^\prime)\right] \right\} \\
  = & \delta(t-t^\prime) \text{Tr}\left\{ \rho_\text{i} \phi_{mn} \right\} + \frac{\theta(t-t^\prime)}{i} \text{Tr}\left\{ \rho_\text{i}  \left[ \phi_n^\text{H}(t^\prime), \phi_m^\text{H}(t)\right] \right\}. 
\end{split}
\end{equation}
Interestingly, this contains an instantaneous piece,
\begin{equation}
  \Delta^{\infty}_{mn} = \text{Tr}\{ \rho_\text{i} \phi_{mn} \},
\label{eq:InstResponseExpValue}
\end{equation}
and a delayed piece, cf.\ eq.\ \eqref{eq:DecompositionLinearRetardedInstDelayed},
\begin{equation}
  \Delta^\mathscr{R}_{mn}(t-t^\prime) = \frac{\theta(t-t^\prime)}{i} \text{Tr}\left\{ \phi_m^\text{H}(t) \left[\rho_\text{i}, \phi_n^\text{H}(t^\prime)\right] \right\}.
\label{eq:delayedTwoPointResponse}
\end{equation}
One should, however, keep in mind that the decomposition of the response function into instantaneous and delayed parts is a question of the time scales one is interested in.

\subsection{Response function from time-dependent Bogoliubov-Kubo-Mori correlation functions}

To rewrite the delayed part of the response function one can employ the relation \cite{Kubo:1957mj}
\begin{equation}
\begin{split}
  \frac{1}{i} \left[ \rho_\text{i}, \phi_n^\text{H}(t^\prime) \right] = & i \rho_\text{i} \left( \phi_n^\text{H}(t^\prime - i \beta) - \phi_n^\text{H}(t^\prime) \right) =  \frac{d}{dt^\prime} \int_0^{\beta} d\tau \rho_\text{i} \phi_n^\text{H}(t^\prime - i \tau) \\
  = & \beta \frac{d}{dt^\prime} \int_0^1 d\lambda \, \rho_\text{i} \phi^\text{H}_n(t^\prime - i \lambda \beta).
\end{split}
\end{equation}
For the delayed part of the linear response function this yields
\begin{equation}
\begin{split}
  \Delta^{\mathscr{R}}_{mn}(t-t^\prime) = & \theta(t-t^\prime) \frac{d}{dt^\prime} \beta \int_0^1 d\lambda \, \text{Tr}\{ \phi_m^\text{H}(t) \rho_\text{i} \phi_n^\text{H}(t^\prime - i \lambda \beta) \} =  \theta(t-t^\prime) \beta \frac{d}{dt^\prime}\Delta^B_{mn}(t - t^\prime).
\end{split}\label{eq:LinearDelayedResponseTimeDerivative}
\end{equation}
In the last equation we are using a generalization of the BKM correlation function \eqref{eq:BKMTwoPointStatic} to time-dependent operators in the interaction picture,
\begin{equation}
\begin{split}
  \Delta^B_{mn}(t - t^\prime) = & \left\langle \phi_m^\text{H}(t) \,; \phi_n^H(t^\prime) \right\rangle = 
  \int_0^1 d\lambda \text{Tr}\left\{ \phi_m^\text{H}(t) \rho_\text{i} \phi_n^H(t^\prime-i\lambda\beta) \right\} \\
  = & \int_0^1 d\lambda \text{Tr}\left\{ \rho_\text{i}^{\lambda}\phi_m^\text{H}(t) \rho_\text{i}^{1-\lambda} \phi_n^H(t^\prime) \right\}.
\end{split}\label{eq:DeltaBDef}
\end{equation}
We note the symmetry $\Delta^B_{mn}(t - t^\prime) = \Delta^B_{nm}(t^\prime - t)$ following from the definition.

\subsection{Relaxation functions in terms of BKM correlation functions}

Interestingly, comparison to \eqref{eq:PsimnBoundary} shows now that the relaxation function is
\begin{equation}
  \Psi_{mn}(t-t^\prime) = \theta(t^\prime - t) \Delta^\infty_{mn} +  \theta(t-t^\prime) \left[ \beta \left\langle \phi_m^\text{H}(t) \,; \, \phi_n^\text{H}(t^\prime)\right\rangle_c - \beta \left\langle \phi_m \,; \, \phi_n \right\rangle_c \right] + c_{mn}.
\end{equation}
We have used here that for $t=t^\prime$ the correlation function $\left\langle \phi_m^\text{H}(t) \,; \, \phi_n^\text{H}(t^\prime)\right\rangle$ equals the equilibrium correlation function $ \left\langle \phi_m \,; \, \phi_n \right\rangle$ and that the disconnected parts of both correlation functions cancel independent of $t-t^\prime$, being simply $\Phi_m \Phi_n$. The first term on the right hand side is only present at $t-t^\prime<0$ and induces a step at $t=t^\prime$ proportional to $\Delta^\infty_{mn}$, while the second term is only present for $t-t^\prime>0$ and vanishes at $t=t^\prime$. The last term $c_{mn}$ is an integration constant that still needs to be determined. 

We can find the integration constant $c_{mn}$ is two different ways. First, using the boundary condition \eqref{eq:InitialLinearRelaxationFunctionSuscept} gives $c_{mn} = \chi_{mn}^{(S,N)} - \Delta^\infty_{mn}$. Using also eq.\ \eqref{eq:IsothermalSusceptConnectedCorr}, as well as \eqref{eq:InstResponseExpValue} yields
\begin{equation}
  \Psi_{mn}(t-t^\prime) = \chi_{mn}^{(S,N)} + \theta(t-t^\prime) \left[ \beta \left\langle \phi_m^\text{H}(t) \,; \, \phi_n^\text{H}(t^\prime)\right\rangle_c - \chi^{(T, \mu)}_{mn}  \right].
  \label{eq:PsimnIntegratedOne}
\end{equation}

Alternatively, one may use that $\Psi_{mn}(t-t^\prime) \to 0$ for $t-t^\prime \to \infty$ because all perturbations are assumed to have relaxed then. We define the long time interval limit of the connected BKM correlation function 
\begin{equation}
\begin{split}
  L_{mn} = & \lim_{\Delta t \to\infty} \beta \left\langle \phi_m^\text{H}(\Delta t) \,; \, \phi_n^\text{H}(0)\right\rangle_c \\
  = & \lim_{\epsilon\to 0} \epsilon \int_0^\infty d\Delta t \, e^{-\epsilon \Delta t} \beta \left\langle \phi_m^\text{H}(\Delta t) \,; \, \phi_n^\text{H}(0)\right\rangle_c.
\end{split}
\end{equation}
The second relation holds when the limit exists, because the divergent part $\sim 1/\epsilon$ comes only from the upper boundary of the integral. With this one finds $c_{mn} = \beta \langle \phi_m \,;\, \phi_n \rangle - L_{mn}$ and thus
\begin{equation}
\begin{split}
  \Psi_{mn}(t-t^\prime) = & \theta(t-t^\prime) \left[ \beta \left\langle \phi_m^\text{H}(t) \,; \, \phi_n^\text{H}(t^\prime)\right\rangle_c - \beta \langle \phi_m \,;\, \phi_n \rangle_c - \langle \phi_{mn} \rangle \right] \\
  &  +  \beta \langle \phi_m \, ; \, \phi_n \rangle_c + \langle \phi_{mn} \rangle - L_{mn}.
\end{split}\label{eq:RelaxationFunctionInTermsOfBKM}
\end{equation}
It is remarkable that eq.\ \eqref{eq:RelaxationFunctionInTermsOfBKM} expresses the relaxation function fully in terms of BKM correlation functions and its limits.

Using \eqref{eq:IsothermalSusceptConnectedCorr} the expression in \eqref{eq:RelaxationFunctionInTermsOfBKM} is of the form \eqref{eq:PsimnIntegratedOne} precisely if the additional relation
\begin{equation}
  \chi^{(T,\mu)}_{mn} = \chi^{(S,N)}_{mn} + L_{mn},
\end{equation}
holds. Using now also \eqref{eq:RelationChiSNChiTmu} yields
\begin{equation}
\begin{split}
  L_{mn} = & \frac{\partial \Phi_m}{\partial T} \frac{\partial \Phi_n}{\partial S} + \frac{\partial \Phi_m}{\partial \mu} \frac{\partial \Phi_n}{\partial N} = \begin{pmatrix} \frac{\partial \Phi_m}{\partial T} , & \frac{\partial \Phi_m}{\partial \mu} \end{pmatrix} \begin{pmatrix} \frac{\partial T}{\partial S} && \frac{\partial\mu}{\partial S} \\ \frac{\partial T}{\partial N} && \frac{\partial \mu}{\partial N}  \end{pmatrix} \begin{pmatrix} \frac{\partial \Phi_n}{\partial T} \\ \frac{\partial \Phi_n}{\partial \mu} \end{pmatrix} \\
  = & \begin{pmatrix} \beta^2 \langle (H_\text{i}-\mu N) \phi_m \rangle_c , & \beta \langle N \phi_m \rangle_c \end{pmatrix} \begin{pmatrix} \frac{\partial T}{\partial S} && \frac{\partial\mu}{\partial S} \\ \frac{\partial T}{\partial N} && \frac{\partial \mu}{\partial N}  \end{pmatrix} \begin{pmatrix} \beta^2 \langle (H_\text{i} - \mu N) \phi_n \rangle_c \\ \beta \langle N \phi_n \rangle_c \end{pmatrix},
\end{split}
\end{equation}
which is in agreement with the Suzuki equality \cite{SUZUKI1971277}, see also \cite{DHAR2021110618}.

\subsection{Operator expressions for quadratic response function}
For the quadratic response one finds terms in the form of the decomposition in eq.\ \eqref{eq:DecompositionQuadraticRetardedInstDelayed}, with the fully instantaneous term
\begin{equation}
  \Delta^{\infty}_{mnk} = \text{Tr}\{ \rho_\text{i} \phi_{mnk}\},
\label{eq:quadraticResponseInstPartOperator}
\end{equation}
the mixed instantaneous delayed term
\begin{equation}
  \Delta^{\infty\mathscr{R}}_{mnk}(t-t^{\prime\prime}) = \frac{\theta(t-t^{\prime\prime})}{i} \text{Tr}\left\{ \phi_{mn}^\text{H}(t) [\rho_\text{i}, \phi^\text{H}_k(t^{\prime\prime})] \right\},
  \label{eq:mixedInstDelayedQuadraticResponse}
\end{equation}
and the fully delayed response function
\begin{equation}
\begin{split}
  \Delta^\mathscr{R}_{mnk}(t-t^\prime, t-t^{\prime\prime}) = & \frac{\theta(t-t^\prime)\theta(t^\prime-t^{\prime\prime})}{i^2} \text{Tr}\left\{ \phi_m^\text{H}(t) \left[ \left[\rho_\text{i}, \phi_n^\text{H}(t^\prime)\right], \phi_k^\text{H}(t^{\prime\prime}) \right] \right\} \\
  + &  \frac{\theta(t-t^{\prime\prime})\theta(t^{\prime\prime}-t^{\prime})}{i^2} \text{Tr}\left\{ \phi_m^\text{H}(t) \left[ \left[\rho_\text{i}, \phi_k^\text{H}(t^{\prime\prime})\right], \phi_n^\text{H}(t^{\prime}) \right] \right\}.
\end{split}\label{eq:DelayedPartResponseFunctionCommutators}
\end{equation}
This can be easily extended to higher orders.

\subsection{Quadratic response function as second time derivative}
Let us first note that for the mixed instantaneous delayed quadratic response function \eqref{eq:mixedInstDelayedQuadraticResponse} one can use essentially eq.\ \eqref{eq:LinearDelayedResponseTimeDerivative}, so that
\begin{equation}
  \Delta^{\infty\mathscr{R}}_{mnk}(t-t^{\prime\prime}) = \theta(t-t^{\prime\prime}) \beta \frac{d}{dt^{\prime\prime}} \left\langle \phi^\text{H}_{mn}(t) \,;\, \phi^\text{H}_k(t^{\prime\prime}) \right\rangle.
\end{equation}
In contrast, for the fully delayed response at quadratic order one can use a generalization of Kubos relation,
\begin{equation}
\begin{split}
  & \frac{1}{i^2} \left[\left[ \rho_\text{i}, \phi_n^\text{H}(t^\prime) \right] , \phi_k^\text{H}(t^{\prime\prime}) \right] = \beta \frac{d}{dt^\prime} \int_0^1 d\lambda_1 \frac{1}{i}\left[ \rho_\text{i} \phi_n^\text{H}(t^\prime - i \lambda_1 \beta), \phi_k^\text{H}(t^{\prime\prime}) \right] \\
  & = \beta i \frac{d}{dt^\prime} \int_0^1 d\lambda_1 \rho_\text{i} \left( \phi_k^\text{H}(t^{\prime\prime}-i\beta) \phi_n^\text{H}(t^\prime - i \lambda_1 \beta) - \phi_n^\text{H}(t^\prime-i\lambda_1 \beta) \phi_k^\text{H}(t^{\prime\prime}) \right) \\
  & = \beta i \frac{d}{dt^\prime} \int_0^1 d\lambda_1 \, \rho_\text{i} \mathscr{T}\left\{ \phi^\text{H}_n(t^\prime - i \lambda_1 \beta) \phi_k^\text{H}(t^{\prime\prime}-i\beta) - \phi_n^\text{H}(t^\prime - i \lambda_1 \beta) \phi_k^\text{H}(t^{\prime\prime}) \right\} \\
  & = \beta i \frac{d}{dt^\prime} \int_0^1 d\lambda_1 \int_0^1 d\lambda_2 \frac{d}{d\lambda_2} \rho_\text{i} \mathscr{T}\left\{ \phi_n^\text{H}(t^\prime-i\lambda_1 \beta) \phi_k^\text{H}(t^{\prime\prime}-i\lambda_2\beta) \right\}.
\end{split}
\end{equation}
Here $\mathscr{T}\{ \ldots \}$ stands for ordering with respect to the imaginary part of time so that operators with largest imaginary part are moved to the left. Replacing the derivative with respect to $\lambda_2$ by a derivative with respect to $t^{\prime\prime}$ yields
\begin{equation}
\begin{split}
  & \frac{1}{i^2} \left[\left[ \rho_\text{i}, \phi_n^\text{H}(t^\prime) \right] , \phi_k^\text{H}(t^{\prime\prime}) \right]  = \beta^2 \frac{d^2}{dt^\prime dt^{\prime\prime}}  \int_0^1 d\lambda_1 \int_0^1 d\lambda_2 \, \rho_\text{i} \mathscr{T}\left\{ \phi_n^\text{H}(t^\prime-i\lambda_1 \beta) \phi_k^\text{H}(t^{\prime\prime}-i\lambda_2\beta) \right\}.
\end{split}
\end{equation}
Interestingly, the integral is a BKM three-point function, see eq.\ \eqref{eq:quantumCorrelationOrder3}. 
On this basis we find for the quadratic response function an expression as time derivative of a BKM three-point function of Heisenberg operators,
\begin{equation}
\begin{split}
  \Delta^\mathscr{R}_{mnk}(t-t^\prime, t-t^{\prime\prime}) = \beta^2 \left[ \theta(t-t^\prime) \theta(t^\prime - t^{\prime\prime}) + \theta(t-t^{\prime\prime}) \theta(t^{\prime\prime}-t^\prime) \right] \frac{d^2}{dt^\prime dt^{\prime\prime}}\left\langle \phi_m^\text{H}(t) \,;\, \phi_n^\text{H}(t^\prime) \,;\, \phi_k^\text{H}(t^{\prime\prime}) \right\rangle.
\end{split}
\end{equation}
This principle can now be extended to higher order response functions, and they are given by time derivatives of higher order BKM correlation functions.

What we have established here is a close relation between response functions and symmetrically defined BKM quantum correlation functions. The former are essentially time derivatives of the latter. This is an extended quantum version of the fluctuation-dissipation relation at linear and non-linear order, see also section \ref{sec:QuantumCorrelationFunctions}.

\section{Spectral representations}

\subsection{Spectral representation for linear response function}
\label{sec:SpectralRepresentationForLinearResponseFunction}

We define the \textit{spectral two-point correlation function} at $j=j_\text{i}$ as
\begin{equation}
  \Delta^\rho_{mn}(t-t^\prime) = - \text{Tr} \left\{ \phi_m^\text{H}(t) \left[\rho_\text{i} , \phi^\text{H}_n(t^\prime) \right] \right\} = \text{Tr} \left\{ \rho_\text{i} \left[ \phi_m^\text{H}(t), \phi_n^\text{H}(t^\prime) \right] \right\}= \int \frac{d\omega}{2\pi} e^{-i\omega(t-t^\prime)} \Delta^\rho_{mn}(\omega).
\label{eq:SpectralTwoPointFunctionDef}
\end{equation}
Note that one has 
\begin{equation}
\Delta^\rho_{mn}(t-t^\prime) = - \Delta^\rho_{nm}(t^\prime-t)  
\label{eq:antisymmetrySpectralFunction}
\end{equation}
and 
\begin{equation}
\Delta^\rho_{mn}(\omega) = - \Delta^\rho_{nm}(-\omega),
\end{equation}
following directly from the definition.

For hermitian operators $\phi^\text{H}_m(t)$ and $\phi^\text{H}_n(t^\prime)$ the function defined by \eqref{eq:SpectralTwoPointFunctionDef} is purely imaginary in the time domain,
\begin{equation}
  \Delta^\rho_{mn}(t-t^\prime) = - \Delta^\rho_{mn}(t-t^\prime)^*.
\end{equation}
In the frequency domain this implies
\begin{equation}
  \Delta^\rho_{mn}(\omega) = - \Delta^\rho_{mn}(-\omega)^*.
\label{eq:SpectralFourierComplexConjugation}
\end{equation}

In parallel to eq.\ \eqref{eq:GtFtRelation} one can write  the delayed part of the response function \eqref{eq:delayedTwoPointResponse}
\begin{equation}
  \Delta^\mathscr{R}_{mn}(t-t^\prime) = i \theta(t-t^\prime) \Delta^\rho_{mn}(t-t^\prime),
\label{eq:retardedCorrelationInTermsOfSpectralFunction}
\end{equation}
and in parallel to eq.\ \eqref{eq:GFSpectralRelation} one finds
\begin{equation}
  \Delta^\mathscr{R}_{mn}(\omega) = G_{mn}(\omega + i \varepsilon) ,
\end{equation}
where the \textit{complex-argument two-point correlation function} has the spectral representation
\begin{equation}
  G_{mn}(z) = - \int_{-\infty}^\infty \frac{d\omega}{2\pi} \frac{\Delta^\rho_{mn}(\omega)}{z - \omega}.
\end{equation}
Because the integral goes along the real $\omega$-axis, the spectral representation shows that $G_{mn}(z)$ has all poles and branch cuts along the real axis. As a consequence $\Delta^\mathscr{R}_{mn}(\omega)$ is analytic in the upper half of the complex frequency plane as expected.

For situations with time-reflection invariance we will show in eq.\ \eqref{eq:spectralTwoPointTimeReflection} that $\Delta^\rho_{mn}(\omega)$ is either real or imaginary.

Also time-ordered Feynman or anti-time-ordered Dyson correlation functions can be obtained from the spectral representation. Finally, the Matsubara propagator is obtained by evaluating $G_{mn}(z)$ with imaginary discrete frequency argument.

\subsection{Spectral representation for quadratic response}\label{sec:SpectralRepQuadraticResponse}

We make the tentative definition of a second order spectral density function (see e.\ g.\ \cite{Rostami2021, Bradlyn2024SpectralDensity}),
\begin{equation}
\begin{split}
  \Delta^\rho_{mnk}(t-t^\prime, t^\prime - t^{\prime\prime}) = & \text{Tr} \left\{ 
  \phi_m^\text{H}(t) \left[ \left[\rho_\text{i}, \phi_n^\text{H}(t^\prime)\right], \phi_k^\text{H}(t^{\prime\prime}) \right]  
  %\rho_\text{i} \left[ \left[ \phi^\text{H}_m(t), \phi_n^\text{H}(t^\prime) \right], \phi_k^\text{H}(t^{\prime\prime}) \right] 
  \right\} \\ = & \int \frac{d\omega_1}{2\pi} \frac{d\omega_2}{2\pi} e^{- i \omega_1 (t-t^{\prime}) - i \omega_2 (t^\prime-t^{\prime\prime})} \Delta^\rho_{mnk}(\omega_1, \omega_2).
\end{split}
\label{eq:GmnkDef}
\end{equation}
Note that we are using $t^\prime-t^{\prime\prime}$ instead of $t-t^{\prime\prime}$ as the second argument. Also the Fourier representation has been chosen slightly different than in eq.\ \eqref{eq:retardedQuadraticFourier}.

Directly from the definition one finds the alternative form
\begin{equation}
  \Delta^\rho_{mnk}(t - t^\prime, t^\prime - t^{\prime\prime}) = \text{Tr}\left\{ \rho_\text{i} \left[ \left[ \phi_m^\text{H}(t), \phi_k^\text{H}(t^{\prime\prime}) \right], \phi_n^\text{H}(t^{\prime}) \right] \right\},
\end{equation}
and the permutation symmetry
\begin{equation}
  \Delta^\rho_{mnk}(t - t^\prime, t^\prime - t^{\prime\prime}) = - \Delta^\rho_{knm}(t^{\prime\prime} - t^\prime, t^\prime-t).
\end{equation}

For hermitian operators the function defined by eq.\ \eqref{eq:GmnkDef} is purely real in the time domain, $\Delta^\rho_{mnk}(t - t^\prime, t^\prime - t^{\prime\prime}) = \Delta^\rho_{mnk}(t - t^\prime, t^\prime - t^{\prime\prime})^*$. In the frequency domain this implies
\begin{equation}
  \Delta^\rho_{mnk}(\omega_1, \omega_2) = \Delta^\rho_{mnk}(-\omega_1, -\omega_2)^*.
\label{eq:ThreePointSpectralComplexConjugate}
\end{equation}
In a time-reflection symmetric situation we will show in eq.\ \eqref{eq:TimeReflectionThreePointSpectral} that $\Delta^{\rho}_{mnk}(\omega_1, \omega_2)$ is either real or imaginary.

The delayed part of the retarded quadratic response function in eq.\ \eqref{eq:DelayedPartResponseFunctionCommutators} follows in terms of this spectral function as
\begin{equation}
\begin{split}
  \Delta^\mathscr{R}_{mnk}(t-t^\prime, t-t^{\prime\prime}) = & i^2 \theta(t-t^\prime) \theta(t^\prime - t^{\prime\prime})\Delta^\rho_{mnk}(t - t^\prime, t^\prime - t^{\prime\prime}) \\
  & + i^2 \theta(t-t^{\prime\prime}) \theta(t^{\prime\prime} - t^{\prime}) \Delta^\rho_{mkn}(t - t^{\prime\prime}, t^{\prime\prime} - t^{\prime}),
\end{split}
\end{equation}
where we observe that the second line is obtained from the first line after the permutation of $(n,t^\prime)$ with $(k, t^{\prime\prime})$. 

This can be brought to a spectral representation in Fourier space using again a variant of the Hilbert transform trick in section \ref{sec:SpectralRelationsFormHilberTransform},
\begin{equation}
  \Delta^\mathscr{R}_{mnk}(\omega, \omega^{\prime}) = G_{mnk}(\omega + \omega^\prime+i\varepsilon, \omega^\prime + i \varepsilon) + G_{mkn}(\omega + \omega^\prime+ i\varepsilon, \omega + i \varepsilon),
\end{equation}
with the complex-argument second order Greens function defined by the spectral representation
\begin{equation}
  G_{mnk}(z_1, z_2) = (-1)^2\int_{-\infty}^\infty \frac{d\omega_1}{2\pi} \frac{d\omega_2}{2\pi} \frac{\Delta_{mnk}^\rho(\omega_1, \omega_2)}{[z_1 - \omega_1][z_2 - \omega_2]}.
\end{equation}
Indeed one finds that the so-defined $\Delta^\mathscr{R}_{mnk}(\omega, \omega^{\prime})$ is analytic in the upper halves of the complex $\omega$- and $\omega^\prime$-planes as needed for causality.

In a similar way one can go on and find higher-order response functions and the corresponding spectral representations, see e.\ g.\ \cite{Rostami2021}.

Using the spectral representation one can also easily obtain imaginary-time correlation functions by evaluating them on the discrete imaginary Matsubara frequencies. % $\omega = i 2\pi T n$ etc.

\section{More on measurements}
\label{sec:MoreOnMeasurements}

So far we have considered unitary dynamics until the time $t$, where we assumed that an observable $\phi_m$ is measured. This observable was influenced by the change in external parameters, either in a static or in a time-dependent situation. In the present section we generalize this in two directions. First, we allow for more general measurement at the final time, so that also equal-time correlation functions or full counting statistics becomes observable. Second, we develop a scheme to describe additional, non-destructive measurements at intermediate times. 

\subsection{A weak measurement scheme}
\label{sec:AWeakMeasurementsScheme}
Allowing measurements at intermediate times brings a couple of challenges, because in a quantum theory they necessarily modify the evolution, and it is not easy to tell in general how this happens. 
An exception would be a complete projective measurement, but this is usually not feasible for quantum systems with many degrees of freedom. Instead, we concentrate here on a \textit{weak measurement} (see e. g. \cite{Jordan:2024zbn}) at an intermediate time, where a subsequent evolution is possible in a well defined sense. Because we are here mainly interested in conceptual questions, our measurement scheme will be simple and certainly idealized, but the discussion will help to clarify what can be measured, at least in principle.

In any case, to allow measurements we need to a coupling of our system to an external device, or measurement apparatus. We will consider a measurement scheme of the \textit{selective tunneling} type \cite{Jordan:2024zbn}. Here, an auxiliary system used for the measurement is in a metastable state, such that a decay towards a more stable situation via tunneling is induced through an operator we wish to measure. For our purpose it is convenient to assume that the tunneling rate $\Gamma$ is proportional to the combination of energy and particle number of our system, described by the modified Hamiltonian, $H(j_\text{m})-\mu N$, i.\ e.\ 
\begin{equation}
  \Gamma = g \left[ H(j_\text{m{}})-\mu N \right].
\end{equation}
The measurement protocol will consist in switching the coupling $g$ on, waiting for some amount of time $\Delta t$, and observing whether a decay has happened, or not. If no decay has happened the coupling is switched off again, and our system evolves further in isolation, i. e. with unitary dynamics. 

If a decay has happened the experiment is stopped, and subsequently started with a freshly prepared sample. (This is the destructive part of the scheme.) This protocol is repeated many times to determine the decay rate. During the short period where the coupling is switched on we will assume that the external source parameters $j^n(t)=j^n_\text{m}$ are constant, but we allow them to be changed from run to run. In this way also the dependence of the decay probability on these parameters can be studied, at least in principle.

The above measurement can be seen as a generalized measurement with two possible outcomes, `decay' or `no decay', and it is formally described in terms of two corresponding Kraus operators (see e.\ g.\ ref.\ \cite{Jordan:2024zbn}) $K$ and $\bar K$, such that $K^\dagger K = \exp(-\zeta (H(j_\text{m})-\mu N))$ and $\bar K^\dagger \bar K = \mathbbm{1} - \exp(-\zeta (H(j_\text{m})-\mu N))$, where $\zeta$ is proportional to the measurement period $\Delta t$ and to the coupling between system and measurement device. Note that $K^\dagger K$ and $\bar K^\dagger \bar K$ are positive operators, albeit not projectors. Just after the measurement with outcome `no decay' the density matrix is, 
\begin{equation}
\rho(t+\Delta t) = \frac{K \rho(t) K^\dagger}{\text{Tr}\{ K \rho(t) K^\dagger \}} = \frac{K \rho(t) K^\dagger}{\text{Tr}\{ \rho(t) \exp(-\zeta (H(j_\text{m}))-\mu N) \}},
\end{equation}
when the density matrix was $\rho(t)$ just before the measurement.

In principle the Kraus operators are not unique, but we assume here for definiteness that $K = \exp(-\zeta (H(j_\text{m})-\mu N)/2)$. In that case $K$ can be represented mathematically with the same methods as the initial thermal density matrix, i.\ e. evolution in negative imaginary time direction, from $t=t_\text{m}$ to $t=t_\text{m}-i\zeta  /2$,
\begin{equation}
  K = \mathscr{U}(t_\text{m}-i \zeta/2, t_\text{m}, j_\text{m}).
\end{equation}
This operator depends also on the source parameters $j_\text{m}$.

As stated above, we assume that the measurement can not only be done for a fixed set of parameters $\zeta$ and $j_\text{m}$, but that these can be varied in a small range, so that derivatives of observables with respect to these parameters become observable, as well. This leads us to the idea of a \textit{generating observable}.

\subsection{Generating observable}
In a construction that resembles the one for a thermal density matrix, we may define a kind of \textit{generating observable} as an hermitian operator 
\begin{equation}
 \sigma(\zeta, \mu, j_\text{f}) = \exp(-\zeta (H(j_\text{f})-\mu N)).
\label{eq:generatingObservable}
\end{equation}
This could be seen as $K^\dagger K$ of a weak measurement, and by taking derivatives with respect to $j^n_\text{f}$ one can generate other interesting operators. For example, 
\begin{equation}
  \phi_m(j_\text{f}) = - \frac{1}{\zeta} \frac{\partial }{\partial j^m_\text{f}} \sigma(\zeta, \mu, j_\text{f}) {\Big |}_{\zeta = 0}, \quad\quad\quad \phi_{mn}(j_\text{f}) = - \frac{1}{\zeta} \frac{\partial^2}{\partial j^m_\text{f}\partial j^n_\text{f}} \sigma(\zeta, \mu, j_\text{f}){\Big |}_{\zeta = 0},
\end{equation}
and so on. Similarly, derivatives with respect to $\zeta$ and $\mu$ yield Hamiltonian and particle number operators.

One can also obtain composite operators from higher order derivatives of \eqref{eq:generatingObservable}, for example
\begin{equation}
\begin{split}
  \frac{1}{2}\phi_m(j_\text{f}) \phi_n(j_\text{f}) + \frac{1}{2} \phi_n(j_\text{f}) \phi_m(j_\text{f}) = & \left[\frac{1}{\zeta^2} \frac{\partial^2 \sigma(\zeta, \mu, j_\text{f})}{\partial j^m_\text{f} \partial j^n_\text{f}} + \frac{1}{\zeta} \phi_{mn}(j_\text{f}) \right]_{\zeta = 0} \\ & - \left( \sigma_{mn}(j_\text{f}) \left[ H(j_\text{f})-\mu N \right] + \left[ H(j_\text{f}) - \mu N \right] \sigma_{mn}(j_\text{f}) \right).
\end{split}\label{eq:equalTimeTwoPointOperator}
\end{equation}
The expectation value of this would be an equal-time symmetric correlation function. Expressions like \eqref{eq:equalTimeTwoPointOperator} can be obtained from an expansion of the right hand side of eq.\ \eqref{eq:generatingObservable}.

The generating observable has the matrix elements
\begin{equation}
  \langle \chi_{\text{f}-} | \sigma(\zeta, \mu, j_\text{f}) | \chi_{\text{f}+} \rangle = \langle \chi_{\text{f}-} | \mathscr{U}(t_\text{f}-i \zeta, t_\text{f}, \mu, j_\text{f}) | \chi_{\text{f}+} \rangle = \int_{\chi_{\text{f}-}, \chi_{\text{f}+}} D \chi \exp(-S_\text{E}[\chi]).
\label{eq:matrixElementsSuperObservable}
\end{equation}
The Euclidean action in the last equality depends also on the chemical potential $\mu$ and the source $j_\text{f}$. 

\subsection{Measurement partition function}

We may now consider a partition function that combines the thermal equilibrium initial state with time evolution operators, and the generating observable,
\begin{equation}
\begin{split}
  Z_\text{M}(\beta, \zeta, \mu)[j] = & \text{Tr} \{ \exp(-\beta (H(j_\text{i})-\mu N)) \, U(t_\text{f}, t_\text{i})[j]^\dagger \, \exp(-\zeta (H(j_\text{f})-\mu N)) \, U(t_\text{f}, t_\text{i})[j]\}.
\end{split}
\label{eq:defFinalStateGeneratingFunctional}
\end{equation}
We refer to this as \textit{measurement partition function}. It depends on the initial temperature through $\beta$, the parameter $\zeta$, the initial chemical potential $\mu$, and the entire set of source functions $j(t)$ for $t_\text{i}<t<t_\text{f}$, including the initial values $j_\text{i}$ and final values $j_\text{f}$.

One can also read this as
\begin{equation}
  Z_\text{M}(\beta, \zeta, \mu)[j] = Z(\beta, \mu, j_\text{i}) \, \text{Tr}\{ \rho_\text{f} \exp(-\zeta H(j_\text{f})) \}.
\end{equation}
Here $\rho_\text{f} = \rho(t_\text{f})$ is the time-dependent density matrix evolved until $t=t_\text{f}$ and $H(j_\text{f})$ is the Hamiltonian with a final set of source parameters. 

In the functional integral formalism, the measurement partition function corresponds to a closed time evolution contour that starts at $t=t_\text{i}$, evolves forward to to $t=t_\text{f}$, from there downwards to $t=t_\text{f}-i \zeta$, from there backwards to $t=t_\text{i}-i\zeta$ and then down to $t=t_\text{i}-i\beta-i\zeta$, a point that is to be identified with $t=t_\text{i}$,
\begin{equation}
\begin{split}
  Z_\text{M}(\beta, \zeta, \mu)[j] = & \int D \chi_{\text{i}+} D \chi_{\text{i}-} D \chi_{\text{f}+} D \chi_{\text{f}-} \langle \chi_{\text{f}-} | \mathscr{U}(t_\text{f}-i\zeta, t_\text{f}, \mu, j_\text{f}) | \chi_{\text{f}+} \rangle  \langle \chi_{\text{f}+} | U(t_\text{f}, t_\text{i})[j] | \chi_{\text{i}+} \rangle \\
  & \times \langle \chi_\text{i+} | \mathscr{U}(t_\text{i}- i (\beta+\zeta), t_\text{i}-i\zeta, \mu, j_\text{i}) | 
  \chi_{\text{i}-} \rangle  \langle \chi_{\text{i}-} | U(t_\text{i}-i\zeta, t_\text{f}-i\zeta)[j] | \chi_{\text{f}-} \rangle.
\end{split}\label{eq:FunctIntFinalStateGeneratingFunctional}
\end{equation}
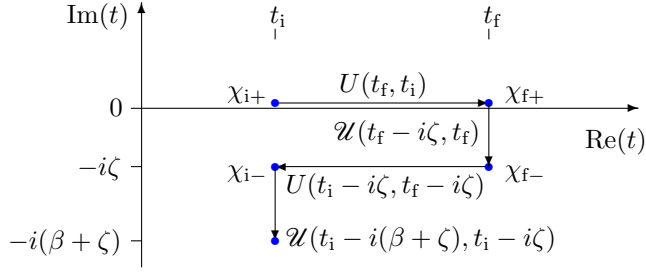
\begin{figure}
\begin{center}
\begin{picture}(200,110)(-100,-65)
% axis
\put(-94,0){\vector(1,0){190}} % x-Achse
\put(-90,-60){\vector(0,1){100}} % y-Achse
% axis labels
\put(76,-15){{Re$(t)$}}
\put(-118,32){{Im$(t)$}}
\put(-102,-3){$0$}
% Label on y-axis
\put(-94,-50){\line(1,0){4}} % short line
\put(-140,-53){$-i(\beta+\zeta)$} % label
\put(-94,-22){\line(1,0){4}} % short line
\put(-114,-25){$-i\zeta$} % label
% label on x-axis
\put(-40,26){\line(0,1){4}} % short line
\put(40,26){\line(0,1){4}} % short line
\put(-42,32){$t_\text{i}$} % label
\put(38,32){$t_\text{f}$} % label
% forward evolution
\put(-40,2){\vector(1,0){80}}
\put(-40,2){\color{blue}\circle*{3}}
\put(40,2){\color{blue}\circle*{3}}
% backward evolution
\put(40,-22){\vector(-1,0){80}}
\put(-40,-22){\color{blue}\circle*{3}}
\put(40,-22){\color{blue}\circle*{3}}
% downward evolution
\put(-40,-22){\vector(0,-1){28}}
\put(-40,-50){\color{blue}\circle*{3}}
\put(40,2){\vector(0,-1){24}}
% labels inside
\put(-58,4){$\chi_{\text{i}+}$} % label
\put(46,4){$\chi_{\text{f}+}$} % label
\put(-58,-26){$\chi_{\text{i}-}$} % label
\put(46,-26){$\chi_{\text{f}-}$} % label
\put(-16,6){$U(t_\text{f}, t_\text{i})$} % label
\put(-36,-32){$U(t_\text{i}-i\zeta, t_\text{f}-i\zeta)$} % label
\put(-36,-52){$\mathscr{U}(t_\text{i}-i(\beta+\zeta), t_\text{i}-i\zeta)$} % label
\put(-18,-12){$\mathscr{U}(t_\text{f}-i\zeta, t_\text{f})$} % label
\end{picture}
\captionof{figure}{Illustration of evolution path in the complex time plane to calculate the measurement partition function in eq.\ \eqref{eq:FunctIntFinalStateGeneratingFunctional}.}\label{fig:2}
\end{center}
\end{figure}
We illustrate the contour in Fig.\ \ref{fig:2}.

Variants of this contour could also be used for situations with additional measurements at intermediate times, following the scheme described above in subsection \ref{sec:AWeakMeasurementsScheme}.

\subsubsection{Limits of measurement partition function}
From the definition in eq.\ \eqref{eq:defFinalStateGeneratingFunctional} one finds that the measurement partition function at $\zeta=0$ is simply the static partition function,
\begin{equation}
  Z_\text{M}(\beta, 0, \mu)[j] = Z(\beta, \mu, j_\text{i}) = \exp(-\beta \Omega(T, \mu, j_\text{i})) = \text{Tr}\{ \exp(-\beta (H(j_\text{i})-\mu N)) \}.
\end{equation}
On the other side, one can also set $\beta = 0$ and finds a partition function for the final state Hamiltonian, with inverse temperature $\zeta$,
\begin{equation}
  Z_\text{M}(0, \zeta, \mu)[j] = Z(\zeta, \mu, j_\text{f}) = \exp(-\zeta \Omega(1/\zeta, \mu, j_\text{f})) = \text{Tr}\{ \exp(-\zeta (H(j_\text{f})-\mu N)) \}.
\end{equation}
%These relations generalize to imaginary-time-dependent sources.

\subsubsection{Functional derivatives of the measurement partition function}

The measurement partition function \eqref{eq:FunctIntFinalStateGeneratingFunctional} allows to determine all kinds of dynamical observables. Here we compile a few.

One has here the first derivatives
\begin{equation}
\begin{split}
  E_\text{i} - \mu N = & \text{Tr}\{ \rho_\text{i} (H_\text{i} -\mu N) \} = - \frac{\partial}{\partial \beta} \ln Z_\text{M}(\beta, \zeta, \mu)[j]{\big |}_{\zeta=0},\\
  E_\text{f} - \mu N = & \text{Tr}\{ \rho_\text{f} (H_\text{f} -\mu N) \} = -\frac{\partial}{\partial \zeta} \ln Z_\text{M}(\beta, \zeta, \mu)[j]{\big |}_{\zeta=0},
\end{split}\label{eq:energyExptInitialFinal}
\end{equation}
for any choice of $j^n(t)$.  Interestingly, the combination allows to obtain an expression for the expectation value of the change in energy, $E_\text{f} - E_\text{i}$. The relations in \eqref{eq:energyExptInitialFinal} can be extended to higher order cumulants with higher order derivatives. The measurement partition function $Z_\text{M}(\beta, \zeta, \mu)[j]$ contains the full information about energy statistics in the initial and final state.

Similarly, other expectation values at initial time follow as,
\begin{equation}
  \Phi_m^\text{i} = \Phi_m(t_\text{i}) = \frac{1}{\beta} \frac{\partial}{\partial j^m_\text{i}} \ln Z_\text{M}(\beta, \zeta, \mu)[j]{\big |}_{\zeta=0}, %= \frac{1}{\beta} \int_0^\beta d\tau \frac{\delta}{\delta j^m(t_\text{i}-i \tau)} \ln Z(\beta, \zeta)[j]{\big |}_{\zeta=0} .
\end{equation}
and those at final time similarly as
\begin{equation}
  \Phi_m^\text{f}[j] = \Phi_m(t_\text{f}) = \frac{1}{\zeta} \frac{\partial}{\partial j^m_\text{f}} \ln Z_\text{M}(\beta, \zeta, \mu)[j]{\big |}_{\zeta=0}. % = \frac{1}{\zeta} \int_0^\zeta d\tau \frac{\delta}{\delta j^m(t_\text{f}-i \tau)} \ln Z(\beta, \zeta)[j]{\big |}_{\zeta=0}
\end{equation}
Higher order moments or cumulants at initial or final time can be obtained along these lines, as  well.

Functional derivatives of expectation values with respect to $j(t)$ at intermediate times $t_\text{i} \leq t \leq t_\text{f}$ around $j=j_\text{i}$ lead to response functions, see eqs.\ \eqref{eq:linearResponseFunctionalDerivative} and \eqref{eq:quadraticResponseFunctionalDerivative}. This implies for the linear response function
\begin{equation}
  \Delta^R_{mn}(t_\text{f}-t^\prime) =  \frac{1}{\zeta} \frac{\partial}{\partial j^m_\text{f}} \frac{\delta}{\delta j^n(t^\prime)} \ln Z_\text{M}(\beta, \zeta, \mu)[j] {\big |}_{j=j_\text{i}, \zeta=0},
\label{eq:linearResponseFunctionalDerivativeDynPartFunct}
\end{equation}
and for the quadratic response function
\begin{equation}
  \Delta^R_{mnk}(t_\text{f}-t^\prime, t_\text{f}-t^{\prime\prime}) = \frac{1}{\zeta} \frac{\partial}{\partial j^m_\text{f}}\frac{\delta^2}{\delta j^n(t^\prime)\delta j^k(t^{\prime\prime})} \ln Z_\text{M}(\beta, \zeta, \mu)[j] {\big |}_{j=j_\text{i}, \zeta=0},
\label{eq:quadraticResponseFunctionalDerivativeDynPartFunct}
\end{equation}
and similar for higher orders.

By combining derivatives at the final time as in eq.\ \eqref{eq:equalTimeTwoPointOperator}, now acting on $Z_\text{M}(\beta, \zeta, \mu)[j]$, with functional derivatives with respect to $j(t)$ one can also calculate the response of composite operators to changes in the external source at intermediate times $t_\text{i} < t < t_\text{f}$. 

\subsubsection{Bogoliubov-Kubo-Mori statistical two-point correlation function}
Another version of a statistical correlation function is obtained by derivative with respect to an initial and a final source parameter,
\begin{equation}
  \Delta^{B}_{mn}(t_\text{f}, t_\text{i})[j] = \frac{1}{\beta\zeta} \frac{\partial}{\partial j^m_\text{f}} \frac{\partial}{\partial j^n_\text{i}} \ln Z_\text{M}(\beta, \zeta, \mu)[j]{\big |}_{\zeta=0}. 
\end{equation}
In terms of operators this is again a Bogoliubov-Kubo-Mori correlation function,
\begin{equation}
\begin{split}
  \Delta^{B}_{mn}(t_\text{f}, t_\text{i})[j] = & \int_0^1 d\lambda \text{Tr}\left\{ \rho_\text{i}^{1-\lambda} \phi_n(j_\text{i}) \rho_\text{i}^{\lambda}  U(t_\text{f}, t_\text{i})[j]^\dagger \phi_m(j_\text{f}) U(t_\text{f}, t_\text{i})[j]  \right\} \\
  = & \left\langle \phi_n(j_\text{i}) \, ; \, U(t_\text{f}, t_\text{i})[j]^\dagger \phi_m(j_\text{f}) U(t_\text{f}, t_\text{i})[j] \right\rangle.
\end{split}
\end{equation}
When evaluated at $j(t)=j_\text{i}$ this is a special case of Bogoliubov-Kubo-Mori correlation function of Heisenberg- or interaction-picture operators,
\begin{equation}
\begin{split}
  \Delta^{B}_{mn}(t- t^\prime) = & \int_0^1 d\lambda \text{Tr}\left\{ \rho_\text{i}^{1-\lambda} \phi_m^\text{H}(t) \rho_\text{i}^\lambda \phi_n^H(t^\prime) \right\} = \int \frac{d\omega}{2\pi} e^{-i\omega(t-t^\prime)} \Delta_{mn}^{B}(\omega) \\
  = & \int_0^1 d\lambda \text{Tr}\left\{ \rho_\text{i} \phi_m^\text{H}(t-i\lambda\beta) \phi_n^H(t^\prime) \right\} = \langle \phi_m^\text{H}(t) \,;\, \phi_n^H(t^\prime) \rangle.
\end{split}\label{eq:DefBogoliubovCorrelationFunction}
\end{equation}
Now there is a symmetry $\Delta^{B}_{mn}(t- t^\prime) = \Delta^{B}_{nm}(t^\prime-t)$. This can also be extended to higher order correlation functions.

\section{Statistics of work}
In the quantum theory, work done on the system through the external driving between the times $t_\text{i}$ and $t_\text{f}$ can only be quantified with two measurements of energy: one at the initial time $t_\text{i}$ represented by the operator $H(t_\text{i}) = H^\text{H}(t_\text{i})$ and another at the finial time $t_\text{f}$ represented by the Heisenberg operator $H^\text{H}(t_\text{f})$. The difference between the two measurement outcomes quantifies the work $W$ \cite{Talkner2007, Esposito2009, Campisi2011}. 

\subsection{Partition function of work}
A characteristic function for the statistics of this so-defined two-time observable is\footnote{See ref.\ \cite{PhysRevLett.100.230404} for a similar construction.}
\begin{equation}
  Z_\text{W}(\xi)[j] = \langle\!\langle \exp(-\xi W) \rangle\!\rangle  = \text{Tr} \left\{ \rho_\text{i} \exp(-\xi H^\text{H}(t_\text{f}))  \exp(\xi H^\text{H}(t_\text{i}))   \right\}.
\end{equation}
We use the notation $\langle\!\langle \ldots \rangle\!\rangle$ for the expectation value associated to the two-time measurement protocol. There is a functional dependence on the source $j(t)$, very much as for the measurement partition function \eqref{eq:defFinalStateGeneratingFunctional}. Because the particle number $N$ is assumed to be conserved, such that the corresponding operator commutes with the Hamiltonian at all times, we can replace in these expressions $H^\text{H}(t)$ by $H^\text{H}(t)-\mu N$. Such a replacement is convenient if one prefers to work with the grand-canonical ensemble. This yields
\begin{equation}
\begin{split}
  Z_\text{W}(\xi)[j] % = & \langle\!\langle \exp(-\xi W) \rangle\!\rangle  = \text{Tr} \left\{ \rho_\text{i} \exp(-\xi H^\text{H}(t_\text{f}))  \exp(\xi H^\text{H}(t_\text{i}))   \right\} \\
  = &  \frac{1}{Z(\beta, \mu, j_\text{i})}\text{Tr} \left\{ \exp(-\beta (H(j_\text{i})-\mu N)) U(t_\text{i}, t_\text{f})[j] \right. \\
  & \left. \times \exp(-\xi (H(j_\text{f})-\mu N)) U(t_\text{f}, t_\text{i})[j] \exp(\xi (H(j_\text{i})-\mu N)) \right\}.
\end{split}\label{eq:WorkStatisticsCharacteristicFunction}
\end{equation}
We observe a relation to the measurement partition function \eqref{eq:defFinalStateGeneratingFunctional} and initial state partition function \eqref{eq:thermalPartitionFunctionHamiltonianAndAction},
\begin{equation}
  Z_\text{W}(\xi)[j] = \frac{Z_\text{M}(\beta - \xi, \xi, \mu)[j]}{Z(\beta, \mu, j_\text{i})}.
  \label{eq:ZWTTPartFunctRel}
\end{equation}
From this characteristic function one can obtain for example the expectation value of work
\begin{equation}
  \langle\!\langle W \rangle\!\rangle = \left\langle H^\text{H}(t_\text{f}) - H^\text{H}(t_\text{i}) \right\rangle = - \frac{\partial}{\partial \xi} Z_\text{W}(\xi)[j] {\big |}_{\xi=0},
\end{equation}
and similarly for higher order moments (see below). Equation \eqref{eq:ZWTTPartFunctRel} shows that the measurement partition function contains the information about work statistics.

One can also write eq.\ \eqref{eq:WorkStatisticsCharacteristicFunction} as a time-ordered expectation value,
\begin{equation}
  Z_\text{W}(\xi)[j] = \left\langle \mathscr{T} \{ \exp(-\xi H^\text{H}(t_\text{f})) \exp(\xi H^\text{H}(t_\text{i})) \} \right\rangle = \left\langle \mathscr{T} \{ \exp(-\xi [H^\text{H}(t_\text{f}) - H^\text{H}(t_\text{i}) ] \} \right\rangle.
\end{equation}
In the second step we have used that the two exponentials could be combined under the time ordering. By taking derivatives one finds time-ordered moments, for example
\begin{equation}
  \frac{\partial^2}{\partial \xi^2} Z_\text{W}(\xi)[j]{\big |}_{\xi=0} = \langle\!\langle H^\text{H}(t_\text{f})^2 - 2  H^\text{H}(t_\text{f}) H^\text{H}(t_\text{i}) +  H^\text{H}(t_\text{i})^2 \rangle\!\rangle = \langle\!\langle \mathscr{T}\left\{ \left(H^\text{H}(t_\text{f}) - H^\text{H}(t_\text{i}) \right)^2 \right\} \rangle\!\rangle.
\end{equation}

Now one can use in the exponential
\begin{equation}
  H^\text{H}(t_\text{f}) - H^\text{H}(t_\text{i}) = \int_{t_\text{i}}^{t_\text{f}} dt \frac{d}{dt} H^\text{H}(t) =  \int_{t_\text{i}}^{t_\text{f}} dt \frac{\partial H^\text{H}(t)}{\partial j^n(t)}\frac{d}{dt} j^n(t) = -  \sum_n \int_{t_\text{i}}^{t_\text{f}} dt  \phi^\text{H}_n(t, j(t)) \frac{d}{dt}  j^n(t). 
\end{equation}
This implies in particular an expression for the expectation value of dissipated energy
\begin{equation}
  \langle\!\langle W \rangle\!\rangle  = - \sum_m \int_{t_\text{i}}^{t_\text{f}} dt \; \Phi_m(t) \frac{d}{dt} j^m(t),
\end{equation}
with $\Phi_m(t) = \langle \phi^\text{H}_m(t, j(t)) \rangle$ the expectation value of the operator coupling to the Hamiltonian. For the latter one could insert the Volterra series \eqref{eq:Volterra} to obtain an expression depending on response functions and sources, only. %Note, however, that due to memory effects this does not lead to an expression that is local in time without further approximations.

\subsection{Dissipated energy in linear response approximation}
For the expectation value of the dissipated energy one finds with the Volterra series \eqref{eq:Volterra}
\begin{equation}
\begin{split}
  & \langle\!\langle W \rangle\!\rangle = - \sum_m \Phi_m^\text{i} [j^m(t_\text{f}) - j^m(t_\text{i}) ]  - \sum_{m,n} \int_{t_\text{i}}^{t_\text{f}} dt \int_{t_\text{i}}^t dt^\prime \Delta^R_{mn}(t-t^\prime) \frac{d}{dt} [j^m(t) - j^m_\text{i}] [j^n(t^\prime) - j^n_\text{i}] \\
  & - \frac{1}{2}\sum_{m,n,k} \int_{t_\text{i}}^{t_\text{f}} dt \int_{t_\text{i}}^t dt^\prime dt^{\prime\prime} \Delta^R_{mnk}(t-t^\prime, t-t^{\prime\prime}) \frac{d}{dt} [j^m(t) - j^m_\text{i}] [j^n(t^\prime)-j^n_\text{i}] [j^k(t^{\prime\prime}) - j^k_\text{i}] - \ldots
\end{split}
\end{equation}
The ellipses are for terms of quartic and higher order in $j^n(t)-j^n_\text{i}$. 

With the Fourier space representations \eqref{eq:LinearResponseFourierRep}, and under the assumption that $j^n(t)-j^n_\text{i}$ is non-vanishing only within a finite time interval, this becomes
\begin{equation}
\begin{split}
  \langle\!\langle W \rangle\!\rangle = & \sum_{m,n} \int \frac{d\omega}{2\pi} (-i \omega) \Delta_{mn}^R(\omega) j^m(-\omega) j^n(\omega) \\
  & + \frac{1}{2} \sum_{m,n,k} \int \frac{d\omega}{2\pi} \frac{d\omega^\prime}{2\pi} (-i\omega-i\omega^\prime) \Delta^R_{mnk}(\omega, \omega^\prime) j^m(-\omega-\omega^\prime) j^n(\omega) j^k(\omega^\prime) + \ldots
\end{split}\label{eq:WorkProductionExpValueFourierSpace}
\end{equation}
We will show below that the right hand side must be non-negative, and this poses constraints on the response functions.

\subsection{Quantum version of Jarzynski equation}
The characteristic function for work \eqref{eq:WorkStatisticsCharacteristicFunction} evaluated for $\zeta=\beta$ gives the expectation value of $\exp(-\beta W)$,
\begin{equation}
  Z_\text{W}(\beta)[j] = \langle\!\langle \exp(-\beta W) \rangle\!\rangle = \frac{Z(\beta, \mu, j_\text{f})}{Z(\beta, \mu, j_\text{i})} = \exp(-\beta \Omega(T, \mu, j_\text{f}) + \beta \Omega(T, \mu, j_\text{i})).
\end{equation}
On the right hand side we have written the equilibrium partition functions in terms of the corresponding grand canonical potentials $\Omega = \langle H \rangle - TS - \mu \langle N\rangle$. The above relation is a quantum version of the Jarzynski equation \cite{Jarzynski1997} (see also \cite{Kurchan2000, Tasaki2000}) adapted to our initial state in the form of a grand-canonical equilibrium state.

Because $\exp(-x)$ is convex, one can infer with Jensens inequality
\begin{equation}
  \exp(-\beta \langle\!\langle W \rangle\!\rangle) \leq \langle\!\langle \exp(-\beta W) \rangle\!\rangle = \exp(-\beta \Omega(T, \mu, j_\text{f}) + \beta \Omega(T, \mu, j_\text{i})),
\end{equation}
or, for the expectation value of the dissipated energy,
\begin{equation}
  \langle\!\langle W \rangle\!\rangle \geq \Omega(T, \mu, j_\text{f}) - \Omega(T, \mu, j_\text{i}).
\end{equation}
In other words, the expectation value of the dissipated work is larger than the difference in grand potential of the two equilibrium states associated to the final and initial Hamiltonian at temperature $T$ and chemical potential $\mu$. It is important to remember here that the final state after the external driving is \textit{not} an equilibrium state.

For the special case $j_\text{f}=j_\text{i}$ one finds $\langle\!\langle W \rangle\!\rangle \geq 0$.

\section{Time reversal and consequences of micro-reversibility}\label{sec:TimeReversal}
For response theory, the study of time reversal is particularly fruitful, as first demonstrated by Onsager.

\subsection{Time reversal}
For most microscopic physics theories one can define a time reversal operation, or inversion of the direction of motion, that leaves them invariant. This symmetry is in general not realized macroscopically because of the second law of thermodynamics, except in thermal equilibrium. For thermal equilibrium, time reversal symmetry is closely connected to detailed balance between the two directions of a process or reaction.

Recall that in quantum mechanics time reversal $t\to - t$ is realized with an antilinear operator $\Theta$ so that $\Theta i \Theta^{-1} = -i$. For antilinear operators $\Theta$ the adjoint $\Theta^\dagger$ is defined such that $\langle \Psi | \Theta^\dagger \Phi \rangle = \langle \Theta \Psi | \Phi \rangle^*$. 

The operator $\Theta$ for time reversal is also antiunitary, which means
\begin{equation}
  \langle \Theta \Psi | \Theta \Phi \rangle = \langle \Psi | \Phi \rangle^* = \langle \Phi | \Psi \rangle.
\end{equation}
This implies $\Theta^{-1} = \Theta^\dagger$. 

The tine reversal transformation in quantum mechanics has the property that position eigenstates $\delta(x-x_0)$ remain unchanged, while momentum eigenstates $\sim\exp(i p x)$ change according to $p\to -p$. In general, one can decompose observables like scalars, vectors, tensors, and fields into even and odd components with respect to time reversal. The behavior of spinors under time reversal is more involved and will not be discussed here.

For the Hamiltonian we assume
\begin{equation}
  \Theta H(j^1(t), \ldots j^N(t)) \Theta^\dagger = H(\varepsilon_1 j^1(t), \ldots, \varepsilon_N j^N(t) = H(\varepsilon j(t)),
\label{eq:TimereflectionParityHamiltonian}
\end{equation}
where $\varepsilon_n$ is the intrinsic time parity of the source component $j^n(t)$. From $\partial H(j(t)) / \partial j^n(t) = - \phi_n(j(t))$ and \eqref{eq:TimereflectionParityHamiltonian} one finds
\begin{equation}
  \Theta \phi_n(j(t)) \Theta^\dagger = \varepsilon_n \phi_n(\varepsilon j(t)).
\end{equation}
Because $H$ must be even with respect to time reflections, $\phi_n$ has the same intrinsic time parity $\varepsilon_n$ as $j^n(t)$. This can be extended to operators that are obtained from higher derivatives of the Hamiltonian according to eq.\ \eqref{eq:ExpansionHamiltonianResponseOperators}. One finds
\begin{equation}
  \Theta \phi_{mn}(j(t)) \Theta^\dagger = \varepsilon_m \varepsilon_n \phi_{mn}(\varepsilon j(t)),
\end{equation}
and so on. Along these lines one can also show for the initial thermal density matrix $\Theta \rho(j_\text{i}) \Theta^\dagger = \rho(\varepsilon j_\text{i})$. 

Examples for even quantities are positions, accelerations, forces, scalar potentials, electric fields, charge density, energy density or the space-space components of the energy-momentum tensor. Examples for odd quantities are velocities, momenta, angular momenta, the electromagnetic vector potential, the magnetic field, magnetization, current densities, the Poynting vector and the time-space components of the energy-momentum tensor.

From the equation of motion \eqref{eq:EOMU} one obtains
\begin{equation}
  i \frac{d}{d(-t)} \Theta U(t, t_\text{i})[j] \Theta^\dagger = H(\varepsilon j(t)) \Theta U(t, t_\text{i})[j] \Theta^\dagger.
\end{equation}
This means that one has
\begin{equation}
  \Theta U(t_\text{f}, t_\text{i})[j] \Theta^\dagger = U(t_\text{i}, t_\text{f})[j_\mathsf{T}],
\label{eq:TimeReverseU}
\end{equation}
which is an evolution operator in backward time direction, and with the time-reversed source configuration,
\begin{equation}
  j_\mathsf{T}^n(t)  = \varepsilon_n j^n(-t).
\end{equation}

For the matrix element of an operator $A$ one has
\begin{equation}
\langle j | A | k \rangle = \langle A^\dagger j | k \rangle = \langle \Theta k | \Theta A^\dagger j \rangle = \langle \Theta k | \Theta A^\dagger \Theta^\dagger | \Theta j\rangle,
\end{equation}
and for $\Theta A^\dagger \Theta^\dagger = \varepsilon_A A^\dagger$ one finds 
\begin{equation}
  \langle j | A | k \rangle = \varepsilon \langle \Theta k | A^\dagger | \Theta j\rangle.
\end{equation}
For the trace one finds
\begin{equation}
  \text{Tr}\{ A \} = \text{Tr}\{ \Theta A \Theta^\dagger \}^* = \text{Tr}\{ \Theta A^\dagger \Theta^\dagger \} = \varepsilon_A \text{Tr} \{ A^\dagger \}.
\label{eq:TimeParityTrace}
\end{equation}
For a thermal expectation value this implies a relation
\begin{equation}
  \text{Tr}\{ \rho(j_\text{i}) A \} = \text{Tr} \{ \Theta A^\dagger \Theta^\dagger \Theta \rho(j_\text{i}) \Theta^\dagger \} = \text{Tr}\{ \rho(\varepsilon j_\text{i}) \Theta A^\dagger \Theta^\dagger \}.
\label{eq:TimeReflectionThermalExpVal}
\end{equation}

For Heisenberg or interaction-picture operators, at $j=j_\text{i}$,
\begin{equation}
  \phi_n^\text{H}(t) = \exp\left(i (t-t_\text{i}) H(j_\text{i})\right)  \phi_n  \exp\left(-i (t-t_\text{i}) H(j_\text{i})\right),
\end{equation}
one finds with $\Theta \exp(i(t-t_\text{i})H(j_\text{i})) \Theta^\dagger = \exp(-i(t-t_\text{i})H(\varepsilon j_\text{i}))$ that
\begin{equation}
  \Theta \phi_n^\text{H}(t) \Theta^\dagger = \varepsilon_n \phi_n^\text{H}(2t_\text{i}-t) {\big |}_{\varepsilon j_\text{i}}.
\label{eq:ThetanTimeReflection}
\end{equation}
Time reflections invert the direction of time, $t\to 2t_\text{i} - t$ (the translation part $\sim 2 t_\text{i}$ is a matter of convention), lead to an intrinsic time reflection parity factor $\varepsilon_n$, and also invert external fields with odd intrinsic time parity, $j_\text{i} \to \varepsilon j_\text{i}$.

\subsection{Time reflections and reality constraints on spectral  functions}
Using the definition \eqref{eq:SpectralTwoPointFunctionDef}, together with eq.\ \eqref{eq:TimeReflectionThermalExpVal} and eq.\ \eqref{eq:ThetanTimeReflection} yields
\begin{equation}
\begin{split}
  \Delta^\rho_{mn}(t-t^\prime) = & \text{Tr}\{ \rho(\varepsilon j_\text{i}) \left[ \Theta \phi^\text{H}_m(t) \Theta^\dagger, \Theta \phi_n^\text{H}(t^\prime) \Theta^\dagger \right]^\dagger \} = \text{Tr}\{ \rho(\varepsilon j_\text{i}) \left[ \phi_n^\text{H}(2t_\text{i} -t^\prime), \phi_m^\text{H}(2t_\text{i}-t) \right] \} \\
  = & \varepsilon_m \varepsilon_n \Delta^\rho_{nm}(t-t^\prime){\big |}_{\varepsilon j_\text{i}} = - \varepsilon_m \varepsilon_n \Delta^\rho_{mn}(t^\prime - t){\big |}_{\varepsilon j_\text{i}},
\end{split}
\label{eq:TimreReflectionSpectralFunction}
\end{equation}
or, in Fourier space,
\begin{equation}
  \Delta^\rho_{mn}(\omega) = - \varepsilon_m \varepsilon_n \Delta^\rho_{mn}(-\omega) {\big |}_{\varepsilon j_\text{i}}.
\end{equation}
Combined with eq.\ \eqref{eq:SpectralFourierComplexConjugation}, one finds thus
\begin{equation}
  \Delta^\rho_{mn}(\omega) = \varepsilon_m \varepsilon_n \Delta^\rho_{mn}(\omega)^* {\big |}_{\varepsilon j_\text{i}}. 
\label{eq:spectralTwoPointTimeReflection}
\end{equation}
For a time-reversal symmetric situation where $\varepsilon j_\text{i} = j_\text{i}$ this shows that $\Delta^\rho_{mn}(\omega)$ is real when $\varepsilon_m \varepsilon_n=1$, and imaginary when $\varepsilon_m \varepsilon_n=-1$. The former case is much more important, because a time reversal even (odd) quantity reacts to a time reversal even (odd) perturbation.

For the three-point spectral function defined in \eqref{eq:GmnkDef} one obtains in a similar way
\begin{equation}
  \Delta^\rho_{mnk}(t - t^\prime, t^\prime - t^{\prime\prime}) = \varepsilon_m \varepsilon_n \varepsilon_k   \Delta^\rho_{mnk}(-t+t^\prime, -t^\prime + t^{\prime\prime}){\big |}_{\varepsilon j_\text{i}},
\label{eq:TimeReflectionThreePointSpectralTimeDomain}
\end{equation}
which reads in Fourier space
\begin{equation}
  \Delta^\rho_{mnk}(\omega_1, \omega_2) = \varepsilon_m \varepsilon_n \varepsilon_k   \Delta^\rho_{mnk}(- \omega_1, -\omega_2){\big |}_{\varepsilon j_\text{i}} = \varepsilon_m \varepsilon_n \varepsilon_k   \Delta^\rho_{mnk}(\omega_1, \omega_2)^*{\big |}_{\varepsilon j_\text{i}}.
\label{eq:TimeReflectionThreePointSpectral}
\end{equation}
In the last equation we have used \eqref{eq:ThreePointSpectralComplexConjugate}. In particular, when $\varepsilon_m \varepsilon_n \varepsilon_k=1$ the three-point spectral function is real in Fourier space. 

\subsection{Time reflections and response functions}
From eqs.\ \eqref{eq:retardedCorrelationInTermsOfSpectralFunction}, \eqref{eq:antisymmetrySpectralFunction} and \eqref{eq:TimreReflectionSpectralFunction} we obtain for the delayed part of the linear response function
\begin{equation}
\begin{split}
  \Delta^\mathscr{R}_{mn}(t-t^\prime) & = - i \theta(t-t^\prime) \varepsilon_m \varepsilon_n \Delta^\rho_{mn}(t^\prime-t){\big |}_{\varepsilon j_\text{i}} = i \theta(t-t^\prime) \varepsilon_m \varepsilon_n \Delta^\rho_{nm}(t-t^\prime){\big |}_{\varepsilon j_\text{i}} \\
  & = \varepsilon_m \varepsilon_n \Delta^\mathscr{R}_{nm}(t-t^\prime) {\big |}_{\varepsilon j_\text{i}}.
\end{split}
\end{equation}
In a similar way one finds for the instantaneous part with eq.\ \eqref{eq:InstResponseExpValue}
\begin{equation}
  \Delta^\infty_{mn} = \text{Tr}\{ \Theta \rho(j_\text{i}) \Theta^\dagger \Theta \phi_{mn}(j_\text{i}) \Theta^\dagger \}^* = \varepsilon_m \varepsilon_n \Delta^\infty_{mn} {\big |}_{\varepsilon j_\text{i}}.
\end{equation}
Taken together, with the decomposition \eqref{eq:DecompositionLinearRetardedInstDelayed}, this implies the \textit{Onsager-Casimir reciprocal relation} for the linear response function,
\begin{equation}
  \Delta^R_{mn}(t-t^\prime) = \varepsilon_m \varepsilon_n \Delta^R_{nm}(t-t^\prime) {\big |}_{\varepsilon j_\text{i}}.
\end{equation}
A simple generalization of this relation to higher order response functions is not available.

\subsection{Measurement partition function and time reversal}
We now derive an identity for the measurement partition function \eqref{eq:defFinalStateGeneratingFunctional} based on a consideration of time reversal. For that purpose it is useful to have the reflection point or origin of time $t=0$ between $t_\text{i}$ and $t_\text{f}$, so we choose coordinates here such that $t_\text{f} = - t_\text{i}$. 

We have basically two possibilities to use the measurement partition function \eqref{eq:defFinalStateGeneratingFunctional}. The first is to consider $\exp(-\beta (H(j_\text{i})-\mu N))/Z(\beta,\mu, j_\text{i})$ as an initial density matrix that is evolved forward with the time evolution operator corresponding to the source $j^n(t)$ to $t_\text{f}$, where it is probed by the operator $\exp(-\zeta (H(j_\text{f})-\mu N))$. Alternatively, one may consider $\exp(-\zeta (H(j_\text{f})-\mu N))/Z(\zeta, \mu, j_\text{f})$ as specifying an ``initial'' thermal state  with inverse temperature $\zeta$ at time $t_\text{f}$, that is propagated back to $t_\text{i}$ where it is probed by $\exp(-\beta (H(j_\text{i})-\mu N))$.

There is one crucial difference between the two pictures, and that relates to time dependence of $j^n(t)$: The backward time evolution is following the time-reversed protocol. But this can be changed with replacing $j^n(t) \to j^n_\mathsf{T}(t)$. There should accordingly be an identity
\begin{equation}
  Z_\text{M}(\beta, \zeta, \mu)[j] = Z_\text{M}(\zeta, \beta, \mu)[j_\mathsf{T}].
\label{eq:TimeReversalPropertyDynamicalPartitionFunction}
\end{equation}
This can indeed be proven starting from eq.\ \eqref{eq:defFinalStateGeneratingFunctional}, which we rewrite using eqs.\ \eqref{eq:ThetanTimeReflection} and \eqref{eq:TimeReverseU},
\begin{equation}
\begin{split}
  Z_\text{M}(\beta, \zeta, \mu)[j] = & \text{Tr}\left\{ \Theta U(t_\text{f}, t_\text{i})[j] \Theta^\dagger \Theta \exp( -\beta (H(j_\text{i})-\mu N)) \Theta^\dagger \right. \\
  & \left. \quad \times \Theta U(t_\text{i}, t_\text{f})[j] \Theta^\dagger \Theta \exp(- \zeta (H(j_\text{f})-\mu N)) \Theta^\dagger \right\}^* \\
  = & \text{Tr} \left\{ U(t_\text{i}, t_\text{f})[j_\mathsf{T}] \exp(-\beta (H(\varepsilon j_\text{i})-\mu N)) \right. \\
  & \left. \quad \times U(t_\text{f}, t_\text{i})[j_\mathsf{T}] \exp(-\zeta (H(\varepsilon j_\text{f})-\mu N)) \right\}^*.
\end{split}
\end{equation}
Through the cyclic property of the trace, and $Z_\text{M}(\beta,\zeta,\mu)[j]\in \mathbbm{R}$, this indeed implies eq.\ \eqref{eq:TimeReversalPropertyDynamicalPartitionFunction}.

One may take \eqref{eq:TimeReversalPropertyDynamicalPartitionFunction} as a starting point for deriving different consequences of microscopic time reversal symmetry.

\subsection{Statistics of work and Crooks fluctuation theorem}
By inserting eq.\ \eqref{eq:TimeReversalPropertyDynamicalPartitionFunction} into eq.\ \eqref{eq:ZWTTPartFunctRel} one finds the relation
\begin{equation}
  Z_\text{W}(\xi)[j] = \frac{Z(\beta, \mu, j_\text{f})}{Z(\beta, \mu, j_\text{i})} Z_\text{W}(\beta - \xi)[j_\mathsf{T}],
\end{equation}
between the characteristic function $Z_\text{W}(\xi)[j]$ for work in the process described by $j(t)$ and the characteristic function $ Z_\text{W}(\xi)[j_\mathsf{T}]$ of the time-reversed process $j_\mathsf{T}(t)$. This in turn implies a quantum version of Crooks fluctuation theorem \cite{Crooks1999},
\begin{equation}
  \frac{p(W)[j]}{p(-W)[j_\mathsf{T}]} = \exp(\beta W - \beta \Omega(T, \mu, j_\text{f})+ \beta \Omega(T, \mu, j_\text{i})),
\end{equation}
for the ratio of probability distributions $p(W)[j]$ for work $W$ with driving $j(t)$ or $j_\mathsf{T}(t)$. In other words, the dissipation of work is exponentially more likely than the extraction of work in the time reversed protocol.

\section{Quantum correlation functions and fluctuation-dissipation relations}
\label{sec:QuantumCorrelationFunctions}
Besides response functions, also correlation functions are of interest to characterize quantum states and quantum dynamics. Due to the non-commutative character of quantum theory, there is a large class of correlation functions, or generalized covariances. We have already seen one kind, the Bogoliubov-Kubo-Mori quantum correlation functions, playing an interesting role in section \ref{sec:TimedependentPerturbationTheory}. The following discussion is partly based on refs.\ \cite{petz2002,10.1109/TIT.2008.2008142, petz2010quantum, bengtsson2017}. Subsequently we will relate them to response functions through (generalized) fluctuation-dissipation relations, following here partly ref.\ \cite{shitara2016}.

We concentrate in this section on somewhat formal, mathematical relations, without a detailed discussion of the relevance of the different correlation functions to specific experimental measurement schemes.

\subsection{Quantum two-point correlation functions}

In classical statistics, a two-point correlation function or covariance matrix can be seen as the expectation value of a composite observable,
\begin{equation}
  \mathcal{C}_p(a,b) = \sum_j p_j a_j b_j.
\end{equation}
Here $j$ labels the possible states, which are realized with probability $p_j$, and $a_j$ and $b_j$ are the values taken by the observables $a$ and $b$ in the state $j$. The set of states can also be continuous, and the sum in the above expression becomes an integral. The domain of $\mathcal{C}_p(a,b)$ follows from the domain of the observables $a_j$ and $b_j$, given that $p_j \geq 0$. For example, when $a_j, b_j \in \mathbbm{R}$ one has also $\mathcal{C}_p(a,b)\in \mathbbm{R}$. 

In quantum theory the probability distribution gets replaced by a density operator $\rho$ and the observables are represented by self-adjoint operators $A$ and $B$. A generalized covariance matrix can be seen as a bilinear map from the space of self-adjoint operators to the domain of the product of observables, typically $\mathbbm{R}$, which also depends on the density matrix $\rho$. We denote this as 
\begin{equation}
\mathcal{C}_\rho(A,B).  
\end{equation}
One example would be $\text{Tr}\{ \rho A B\}$, but when $A$ and $B$ do not commute with $\rho$, and possibly not among themselves, there are many more possibilities. Our goal will now be to find sensible physical criteria for the construction of generalized covariance matrices. 

\subsubsection{Hilbert-Schmidt inner product}

As a preparatory step, recall that the space of Hilbert-Schmidt operators (not necessarily self-adjoint) has an inner product,
\begin{equation}
  (A, B) = \text{Tr}\{ A^\dagger B \},
\label{eq:HSInnerProduct}
\end{equation}
which makes it a Hilbert space. This allows to consider linear maps between such operators using basically the same concepts as for operators on Hilbert spaces themselves. 

\subsubsection{Two first quantum correlation functions}

To have something concrete in mind, let us first consider two examples for quantum correlation functions,
\begin{equation}
\begin{split}
    \mathcal{C}_\rho^R(A, B)= & \text{Tr}\{ A B \rho \} = \text{Tr}\{ A R_\rho(B) \},\\
    \mathcal{C}_\rho^L(A, B)= & \text{Tr}\{ A \rho B \} = \text{Tr}\{ A L_\rho(B) \}.
\end{split}\label{eq:CRAndCLDef}
\end{equation}
We note that both choices are ``linear'' in the density matrix $\rho$ in the sense that they scale proportional to a linear rescaling of the latter, $\mathcal{C}_{\lambda \rho}(A,B) = \lambda \mathcal{C}_\rho(A,B)$.

\subsubsection{Two superoperators involing the density matrix}
In the following we need the definition of several \textit{superoperators} involving the density matrix. We start from multiplication with $\rho$ from the right or from the left, already used in eq.\ \eqref{eq:CRAndCLDef}, 
\begin{equation}
  R_\rho(A) = A \rho, \quad\quad\quad
    L_\rho(A) = \rho A. 
\label{eq:defSuperOperatorsLR}
\end{equation}
Here $\rho$ is the density matrix and we observe that both $R_\rho$ and $L_\rho$ scale linearly with the normalization of $\rho$, in the sense that for a real coefficient $\lambda > 0$ one has $R_{\lambda\rho}(A) = \lambda R_\rho(A)$ and $L_{\lambda\rho}(A) = \lambda L_\rho(A)$. We also note that the two operators commute, $[R_\rho, L_\rho] = %[R_\rho, L_\rho^{-1}] = [R_\rho^{-1}, L_\rho] = [R_\rho^{-1}, L_\rho^{-1}] = 
0$, and that they are both self-adjoint with respect to the Hilbert-Schmidt inner product \eqref{eq:HSInnerProduct}, $R_\rho^\dagger = R_\rho$ and $L_\rho^\dagger = L_\rho$.

\subsubsection{Functions of superoperators}
In a basis where $\rho$ is diagonal, $\rho = \sum_j p_j |j\rangle \langle j|$, the action of the superoperator is simply
\begin{equation}
  \langle j | R_\rho(A) | k \rangle = \langle j | A | k \rangle p_k, \quad\quad\quad  
  \langle j | L_\rho(A) | k \rangle = \langle j | A | k \rangle p_j.
\end{equation}
In this basis, the matrix components of $A$ are eigenstates of the superoperators $R_\rho$ and $L_\rho$. We may then also define functions of these superoperators, which are again superoperators, for example
\begin{equation}
  \langle j | f(R_\rho)(A) | k \rangle = \langle j | A | k \rangle f(p_k), \quad\quad\quad  
  \langle j | f(L_\rho)(A) | k \rangle = \langle j | A | k \rangle f(p_j).
\end{equation}
For this, the function $f(z)$ must be defined on the domain of the eigenvalues $p_k$, here the real and non-negative numbers.

For example, when the $p_j$ are all positive, such that $\rho$ can be inverted, one can also invert the two superoperators with $f(z)=1/z$. In an arbitrary basis these inverses act like
\begin{equation}
  R_\rho^{-1}(A) = R_{\rho^{-1}}(A) = A \rho^{-1}, \quad\quad\quad L_\rho^{-1}(A) = L_{\rho^{-1}}(A) = \rho^{-1} A.
\end{equation}

In the following we are particularly interested in superoperators defined by a function $f(z)$ on the domain of real and non-negative variables, $z\geq 0$, 
\begin{equation}
  K_\rho^f = f\big(L_\rho R^{-1}_\rho\big) R_\rho.
\label{eq:defKfRho}
\end{equation}
Like $R_\rho$ and $L_\rho$, these have the property $K^f_{\lambda\rho} = \lambda K^f_\rho$. In the basis where $\rho$ is diagonal,
\begin{equation}
  \langle j | K_\rho^f(A) | k \rangle = \langle j | f\left( L_\rho R_\rho^{-1} \right) R_\rho | k \rangle = \langle j | A | k \rangle f(p_j/p_k) p_k.
\label{eq:explicitRepKf}
\end{equation}

A special situation arises for operators $A$ that commute with $\rho$. They can be diagonalized together with $\rho$, and we find the matrix elements
\begin{equation}
  \langle j | K_\rho^f(A) | k \rangle = \delta_{jk} \langle j | A | j \rangle f(1) p_j.
\end{equation}
We will assume the function $f(z)$ to be normalized such that $f(1)=1$, which leads for operators $A$ with $[A,\rho]=0$ to the property
\begin{equation}
  K_\rho^f (A) = A \rho = \rho A. 
\end{equation}

\subsection{Generalized quantum covariance}

At this point we make the tentative definition for a generalized quantum covariance or correlation function, based on a function $f(z)$ defined for $z\geq 0$,
\begin{equation}
  \mathcal{C}_\rho^f(A,B) = \text{Tr} \{ A K_\rho^f(B) \} = \text{Tr}\{ A  f\big(L_\rho R^{-1}_\rho\big) R_\rho B \}.
\label{eq:GeneralizedCovarianceDefCAB}
\end{equation}

The limit of classical statistics corresponds to a situation where all operators, $\rho = \sum_j a_j |j\rangle \langle j |$, $A= \sum_j a_j |j\rangle \langle j |$ and $B = \sum_j b_j |j \rangle \langle j |$ are simultaneously diagonal. In that case one has
\begin{equation}
  \mathcal{C}_\rho^f(A,B) = f(1) \sum_j p_j a_j b_j.
\end{equation}
It is therefore convenient so assume the normalization condition $f(1)=1$ to have the standard classical definition of a correlation function in that limit.

Working now with Heisenberg picture operators and with $\rho=\rho_\text{i}$ we write a general quantum two-point correlation function as
\begin{equation}
  \Delta^f_{mn}(t-t^\prime) = \text{Tr}\left\{ \phi_m^\text{H}(t) K_\rho^f \phi_n^\text{H}(t^\prime) \right\} = \text{Tr} \left\{ \phi_m^\text{H}(t) f\big(L_\rho R_\rho^{-1}\big) R_\rho \phi_n^\text{H}(t^\prime) \right\}.
\label{eq:defGeneralizedCovarianceCorrFunction} 
\end{equation}

The question arises whether the function $f(z)$ needs to be restricted further to have a sensible definition of quantum correlation functions. That is indeed the case, but the matter is non-trivial. We therefore first discuss a few examples, before we discuss further restrictions on $f(z)$. 

\subsubsection{Transposed correlation function and dual function}
The definition in eq.\ \eqref{eq:defGeneralizedCovarianceCorrFunction} is not necessarily symmetric. However, one may write
\begin{equation}
  f\big( L_\rho R_\rho^{-1} \big) R_\rho = \tilde f\big(R_\rho L^{-1}_\rho \big) L_\rho,
\end{equation}
where 
\begin{equation}
\tilde f(z) = z f(1/z),
\end{equation}
is a conjugate function, which can for example be checked with eq.\ \eqref{eq:explicitRepKf}.

It is then straight-forward to show
\begin{equation}
  \Delta^f_{mn}(t-t^\prime) = \text{Tr} \left\{ \phi_n^\text{H}(t^\prime) \tilde f\big( L_\rho R^{-1}_\rho \big) R_\rho \phi_m^\text{H}(t) \right\} = \Delta^{\tilde f}_{nm}(t^\prime-t).
\end{equation}
In other words, the generalized correlation function associated to the dual function $\tilde f(z)$ is the transpose of the correlation function associated to $f(z)$. In the special case where $f(z)$ is self-dual, $\tilde f(z) = f(z)$, one obtains a symmetric correlation functions, $\Delta^f_{mn}(t-t^\prime) = \Delta^f_{nm}(t^\prime-t)$. 

\subsection{Some examples}
Let us discuss a few examples:
\begin{enumerate}
\item For the constant function $f(z)=1$ the generalized covariance in eq.\ \eqref{eq:defGeneralizedCovarianceCorrFunction} becomes
\begin{equation}
  \Delta^f_{mn}(t-t^\prime) = \text{Tr}\{ \phi_m^\text{H}(t) R_\rho \phi_n^\text{H}(t^\prime) \} = \text{Tr}\{ \rho \phi_m^\text{H}(t) \phi_n^\text{H}(t^\prime) \} = \Delta^>_{mn}(t-t^\prime) = \Delta^+_{mn}(t-t^\prime).
\label{eq:WightmanProp} 
\end{equation}
This correlation function, with fixed order of the operators, is also known as Wightman propagator, or positive frequency propagator in quantum field theory, although at finite temperature its Fourier transform has support at negative frequencies, as well.
\item The linear function $f(z)=z$ is the dual to $f(z)=1$. The associated covariance matrix is accordingly the transpose of eq.\ \eqref{eq:WightmanProp},
\begin{equation}
  \Delta^f_{mn}(t-t^\prime) = \text{Tr}\{ \phi_m^\text{H}(t) L_\rho \phi_n^\text{H}(t^\prime) \} = \text{Tr}\{ \rho \phi_n^\text{H}(t^\prime) \phi_m^\text{H}(t) \} = \Delta^<_{mn}(t-t^\prime) = \Delta^+_{nm}(t^\prime-t). 
\end{equation}
One may call this the negative frequency propagator.
\item The symmetric combination $f(z) = (1 + z)/2$ is self-dual and leads to the symmetric correlation function
\begin{equation}
\begin{split}
  \Delta^f_{mn}(t-t^\prime) & = \text{Tr}\{ \phi_m^\text{H}(t) \frac{1}{2}\left[ R_\rho + L_\rho\right] \phi_n^\text{H}(t^\prime) \} \\
  & = \text{Tr}\{ \rho \frac{1}{2} \left[ \phi_m^\text{H}(t) \phi_n^\text{H}(t^\prime) + \phi_n^\text{H}(t^\prime) \phi_m^\text{H}(t) \right] \} = \Delta^S_{mn}(t-t^\prime). 
\end{split}
\end{equation}
This is known as statistical correlation function and it is the symmetric convex combination of positive and negative frequency correlation functions.
\item For the function $f(z) = z^\gamma$, with conjugate $\tilde f(z) = z^{1-\gamma}$, the correlation function is
\begin{equation}
  \Delta^f_{mn}(t-t^\prime) = \text{Tr}\{ \phi_m^\text{H}(t) (L_\rho R_\rho^{-1})^\gamma R_\rho \phi_n^\text{H}(t^\prime) \} = \text{Tr}\{\phi_m^\text{H}(t) \rho^\gamma \phi_n^\text{H}(t^\prime) \rho^{1-\gamma} \}.
\label{eq:fzgammaCorrelationFunction} 
\end{equation}
Note that this is not symmetric, except for $\gamma=1/2$. We also note that $\rho$ does not need to be invertible, for eq.\ \eqref{eq:fzgammaCorrelationFunction} to be well defined, as long as $0\leq \gamma \leq 1$. On the other side, for $\gamma<0$ or $\gamma>1$ one feels intuitively that \eqref{eq:fzgammaCorrelationFunction} is not defined in a sensible way, for example because the limit where eigenvalues of $\rho$ approach zero seems singular. We will see below how this can be excluded by restricting the class of allowed functions $f(z)$.

\item The correlation function corresponding to
\begin{equation}
  f(z) = \frac{z-1}{\ln(z)} = \int_0^1 d\gamma z^\gamma.
\end{equation}
which is self-dual, is given by,
\begin{equation}
  \Delta^f_{mn}(t-t^\prime) = \int_0^1 d\gamma \, \text{Tr}\{\phi_m^\text{H}(t) \rho^\gamma \phi_n^\text{H}(t^\prime) \rho^{1-\gamma} \} = \Delta^B_{mn}(t-t^\prime).
\end{equation}
This is the time-dependent symmetric Bogoliubov-Kubo-Mori correlation function we have encountered previously.

\item As another example, take the self-dual function
\begin{equation}
  f(z) = \frac{(\sqrt{z}+1)^2}{4} = \frac{z + 2\sqrt{z} + 1}{4}.
\end{equation}
The associated correlation function is
\begin{equation}
  \Delta^f_{mn}(t-t^\prime) = \frac{1}{4} \text{Tr}\left\{ \left( \sqrt{\rho}\phi_m^\text{H}(t) + \phi_m^\text{H}(t) \sqrt{\rho} \right) \left(\sqrt{\rho} \phi_n^\text{H}(t^\prime) + \phi_n^\text{H}(t^\prime) \sqrt{\rho}\right)  \right\},
\end{equation}
which is obviously symmetric, and seems well defined.
\end{enumerate}

\subsection{Monotonicity property of quantum correlation functions}
Now our aim is to discuss how the function $f(z)$ entering the definition \eqref{eq:defGeneralizedCovarianceCorrFunction} needs to be restricted for a sensible definition of a quantum correlation function.

\subsubsection{Quantum evolution with completely positive trace-preserving maps}
Recall that a quantum system described by $\rho$ evolves in time with a linear map. The latter is unitary for evolution in isolation, $\rho \to U \rho U^\dagger$, but more generally it can be classified to be a \textit{completely positive, trace-preserving map}, $\rho \to \alpha(\rho)$. This is a linear map $\alpha$ such that $\mathbbm{1}_N \otimes \alpha$ is positive for an identity transformation $\mathbbm{1}_N$ on some auxiliary system of arbitrary dimension $N$, and where positive means that density operators with real and non-negative eigenvalues are mapped to density operators with again real and non-negative eigenvalues. Trace-preserving means $\text{Tr}\{ \alpha(\rho) \} = \text{Tr}\{ \rho\}$, and is obviously related to the conservation of probability. Because it maps (density) operators to (density) operators, one may call $\alpha$ a \textit{superoperator}. 

Completely positive maps can be represented with Kraus operators $K_n$ (albeit not uniquely),
\begin{equation}
  \alpha(A) = \sum_n K_n A K_n^\dagger,
\end{equation}
and the trace-preserving property follows for $\sum_n K^\dagger_n K_n = \mathbbm{1}$. Of course, the usual unitary evolution for isolated quantum systems follows for just a single non-zero Kraus operator $K=U$. Another example is a projective measurement in a basis $|n \rangle$. When the result is not recorded, the Kraus operators are the projectors $K_n = |n \rangle \langle n |$. 

\subsubsection{Evolution maps and adoint evolution maps}
Using the inner product \eqref{eq:HSInnerProduct} one can define an adjoint $\alpha^\dagger$ to the evolution map $\alpha$, so that for arbitrary $A$ and $B$, with the inner product \eqref{eq:HSInnerProduct},
\begin{equation}
  (A, \alpha(B)) = (\alpha^\dagger(A), B). 
\end{equation}
It is unital, $\alpha^\dagger(\mathbbm{1}) = \mathbbm{1}$, because $\alpha$ is trace-preserving. The conjugate map has the Kraus representation
\begin{equation}
  \alpha^\dagger(A) = \sum_n K_n^\dagger A K_n. 
\end{equation}

As an example, the quantum expectation value of some observable $A$ transforms under a quantum evolution map like 
\begin{equation}
  \text{Tr} \{ \rho A \} \to \text{Tr} \{ \alpha(\rho) A \} = \text{Tr} \{ \rho \alpha^\dagger(A) \}.
\label{eq:ExpValueEvAlpha}
\end{equation}
In other words, one can either evolve the density matrix $\rho$ with $\alpha$, or, equivalently, the observable with the adjoint $\alpha^\dagger$. This is the reminiscent of the relation between the Schrödinger and Heisenberg picture for isolated quantum evolution.

\subsubsection{Monotonicity property for generalized covariances}
For expectation values we saw in eq.\ \eqref{eq:ExpValueEvAlpha} that one can equivalently evolve the state $\rho$ with a map $\alpha$, or the operator with the conjugate map $\alpha^\dagger$. For unitary evolution, we are expecting that one can also implement this evolution equivalently either on states or operators for higher-oder correlation functions,
\begin{equation}
  %C_{\rho}(A,B) \to 
  \mathcal{C}_{U \rho U^\dagger}(A,B) = \mathcal{C}_{\rho}(U^\dagger A U, U^\dagger B U).
\end{equation}

But what happens for a more general quantum evolution described by a map $\alpha$? There it was proposed by Petz \cite{petz2002, 10.1109/TIT.2008.2008142} that a sensible definition of a general quantum covariance matrices should be defined such that it has the monotonicity property
\begin{equation}
  \mathcal{C}_{\rho}(\alpha^\dagger(A), \alpha^\dagger(A)) \leq  \mathcal{C}_{\alpha(\rho)}(A, A),
\label{eq:monotonicityCovQuantum}
\end{equation}
for completely positive (trace-preserving) maps $\alpha$ with conjugate $\alpha^\dagger$. 
In other words, a covariance of some operator is smaller (or equal), when the evolution is done on the two instances of the operator $A$, than when it is done on density matrix $\rho$. Equality is obtained for unitary evolution.

It was shown by Dénes Petz \cite{Petz:1999xrh, petz2002, 10.1109/TIT.2008.2008142}, building up on earlier work by Elena A.~ Morozowa and Nikolai N.~Chentsov \cite{Morozova1991}, that eq.\ \eqref{eq:monotonicityCovQuantum} is fulfilled with a definition of the form \eqref{eq:GeneralizedCovarianceDefCAB} when $f(z)$ is an \textit{operator monotone function}. The latter have been investigated and charcterized by Charles Loewner, see e.\ g.\ \cite{simon2019} for a mathematical exposition.

\subsection{Generalized correlation functions at higher order}
It is tempting to consider also generalized correlation functions at cubic and higher order.

%\subsection{Superoperators acting on two operators}
We start by a definition of \textit{superoperators}, similar to the definition in eq.\ \eqref{eq:defSuperOperatorsLR}, but now acting on \textit{two} operators,
\begin{equation}
  R_\rho(A, B) = AB \rho, \quad\quad\quad
  L_\rho(A,B) = \rho AB, \quad\quad\quad 
  C_\rho(A,B) = A \rho B. 
\label{eq:defSuperOperatorsLRC}
\end{equation}
Based on a function with two arguments $f(z_1, z_2)$ we define also the superoperator, again acting on two operators,
\begin{equation}
  K_\rho^f(A,B) = \left[ f\big( L_\rho R_\rho^{-1}, C_\rho R_\rho^{-1} \big) R_\rho \right](A,B).
\end{equation}
With this we can define a generalized three-point correlation function as
\begin{equation}
  \Delta^f_{mnk}(t,t^\prime, t^{\prime\prime}) = \text{Tr} \left\{ 
    \phi_m^\text{H}(t)  K_\rho^f\big( \phi_n^\text{H}(t^\prime), \phi_k^\text{H}(t^{\prime\prime}) \big). 
   \right\}.
\end{equation}
If the function $f(z_1, z_2)$ can be expanded in a two-dimensional Taylor series,
\begin{equation}
  f(z_1, z_2) = \sum_{\lambda_1, \lambda_2=0}^\infty \frac{f^{(\lambda_1, \lambda_2)}(0,0)}{\lambda_1! \lambda_2!} z_1^{\lambda_1} z_2^{\lambda_2},
\end{equation}
one has for the generalized correlation function a formal expansion of the form
\begin{equation}
  \Delta^f_{mnk}(t,t^\prime, t^{\prime\prime}) = \sum_{\lambda_1, \lambda_2=0}^\infty \frac{f^{(\lambda_1, \lambda_2)}(0,0)}{\lambda_1! \lambda_2!} \text{Tr} \left\{ \rho^{1-\lambda_1-\lambda_2} \phi_m^\text{H}(t) \rho^{\lambda_1} \phi_n^\text{H}(t^\prime) \rho^{\lambda_2} \phi_k^\text{H}(t^{\prime\prime}) \right\}.
\end{equation}
For contributions with $\lambda_1+\lambda_2>1$ this assumes implicitly that $\rho$ is invertible.

Another interesting class of functions is of the form
\begin{equation}
  f(z_1, z_2) = \int_0^1 d\lambda_1 \int_0^{1-\lambda_1} d\lambda_2 \, w(\lambda_1,\lambda_2) \, z_1^{\lambda_1} z_2^{\lambda_2},
\end{equation}
with a real and positive weight function $w(\lambda_1, \lambda_2)$. In that case one has
\begin{equation}
  \Delta^f_{mnk}(t,t^\prime, t^{\prime\prime}) = \int_0^1 d\lambda_1 \int_0^{1-\lambda_1} d\lambda_2 \, w(\lambda_1,\lambda_2) \, \text{Tr} \left\{ \rho^{1-\lambda_1-\lambda_2} \phi_m^\text{H}(t) \rho^{\lambda_1} \phi_n^\text{H}(t^\prime) \rho^{\lambda_2} \phi_k^\text{H}(t^{\prime\prime}) \right\}.
\end{equation}

We note that this does not assume $\rho$ to be invertible. This construction principle can obviously be extended to higher order correlation functions. The precise requirements on the function $f(z_1, z_2)$ and its generalizations remain to be investigated further.

\subsection{Fluctuation-dissipation relations}

We now provide a fluctuation dissipation relation for the generalized two-point correlation functions introduced in eq.\ \eqref{eq:defGeneralizedCovarianceCorrFunction}. This follows and generalizes ref.~\cite{shitara2016}.

We start with the definition in eq.\ \eqref{eq:defGeneralizedCovarianceCorrFunction}, and use a basis of energy eigenstates $|j\rangle$, to the time-independent Hamiltonian $H_\text{i}-\mu N$ such that $(H_\text{i} - \mu N) |j \rangle =  \omega_j | j \rangle$ and
\begin{equation}
  \mathbbm{1} = \sum_j | j \rangle \langle j |, \quad\quad\quad \langle j | k \rangle = \delta_{jk}.
\end{equation}
We also use the definition of time-dependent field operators in the interaction piecture in \eqref{eq:defHeisenbergOperator}, to arrive as
\begin{equation}
\begin{split}
  \Delta^f_{mn}(t-t^\prime) = & \frac{1}{Z} \sum_{jk} \text{Tr}\left\{ e^{-\beta(H_\text{i}-\mu N)} | j \rangle \langle j | e^{i (t-t_\text{i}) (H_\text{i}-\mu N)} \phi_m e^{-i (t-t_\text{i}) (H_\text{i}-\mu N)} \right.\\
  & \left. \quad\quad\quad\times f\big( L_\rho R_\rho^{-1} \big) | k \rangle \langle k |  e^{i (t-t_\text{i}) (H_\text{i}-\mu N)} \phi_n e^{-i (t-t_\text{i}) (H_\text{i}-\mu N)}\right\} \\
  = & \frac{1}{Z} \sum_{j,k} \langle j | \phi_m | k \rangle \langle k | \phi_n | j \rangle f\big( \exp(\beta (\omega_j - \omega_k)) \big) \exp(-\beta \omega_j - i(\omega_k - \omega_j)(t-t^\prime)).
\end{split}
\label{eq:SpectralDeltafGeneral}
\end{equation}
In the last equation we have used that the definition of the superoperators $L_\rho$ and $R_\rho$ in \eqref{eq:defSuperOperatorsLR} with $\rho = \rho_\text{i} = \exp(-\beta(H_\text{i}-\mu N))/Z$ and the properties of the energy eigenstates. For the Fourier transform, defined through
\begin{equation}
   \Delta^f_{mn}(t-t^\prime) =  \int \frac{d\omega}{2\pi} e^{-i\omega(t-t^\prime)} \Delta^f_{mn}(\omega),
\end{equation}
one finds then
\begin{equation}
  \Delta^f_{mn}(\omega) = \frac{1}{Z} \sum_{j,k} \langle j | \phi_m | k \rangle \langle k | \phi_n | j \rangle f\big( \exp(-\beta \omega) \big) \exp(-\beta \omega_j) \delta(\omega - \omega_k + \omega_j).
\end{equation}
What we have obtained here is a very general expression for all kind of two-pint correlation functions. This can be used for operator monotone functions $f(z)$, but also for their linear combinations. Also the spectral function defined in \eqref{eq:SpectralTwoPointFunctionDef} falls into this class, with the replacement $f(z) \to 1-z$. All functions ob this type are therefore related, and in particular one can express $\Delta^f_{mn}(\omega)$ for any $f(z)$ in terms of the spectral function $\Delta^\rho_{mn}(\omega)$,
\begin{equation}
  \Delta^f_{mn}(\omega) = \frac{f(\exp(-\beta \omega))}{1-\exp(-\beta \omega)} \Delta^\rho_{mn}(\omega).
\label{eq:GeneralizedFDRelation}
\end{equation}
This is a generalized form of the quantum fluctuation-dissipation relation obtained in ref.\ \cite{Callen:1951vq}, see also \cite{Kubo:1966fyg}.

\subsubsection{Classical limit}
For operator monotone functions $f(z)$ that are normalized to $f(1)=1$ one finds from \eqref{eq:GeneralizedFDRelation} in the classical limit $\beta\omega \ll 1$,
\begin{equation}
  \Delta^f_{mn}(\omega) = \frac{T}{\omega} \Delta^\rho_{mn}(\omega).
\label{eq:GeneralizedFDRelationClassicalLimit}
\end{equation}
This is the well-known form of the classical fluctuation-dissipation relation.

\subsubsection{Examples}
Let us discuss a few examples.
\begin{enumerate}
  \item For $f(z)=1$ one obtains a relation between the positive frequency Wightman function and the spectral function,
  \begin{equation}
    \Delta^+_{mn}(\omega) = [1+n_B(\omega)] \Delta^\rho_{mn}(\omega),    
  \end{equation}
  where $n_B(\omega) = 1/(\exp(\beta\omega)-1)$ is the Bose distribution function.
  \item Similarly one obtains with $f(z)=z$ for the negative frequency propagator
  \begin{equation}
    \Delta^-_{mn}(\omega) = n_B(\omega) \Delta^\rho_{mn}(\omega).
  \end{equation}
  \item For $f(z) = (1+z)/2$ one finds the original form of the fluctuation-dissipation relation \cite{Callen:1951vq},
  \begin{equation}
    \Delta^S_{mn}(\omega) = \left[\frac{1}{2} + n_B(\omega) \right] \Delta^\rho_{mn}(\omega).
  \end{equation}
  \item For $f(z)=(z-1)/\ln(z)$ one finds a fluctuation-dissipation relation for the BKM correlation function that has the same form as the classical limit,
  \begin{equation}
  \Delta^B_{mn}(\omega) = \frac{T}{\omega} \Delta^\rho_{mn}(\omega).
  \end{equation}
  This can also be easily translated to the time domain,
  \begin{equation}
  \Delta^\rho_{mn}(t-t^\prime) = - i \beta \frac{d}{dt^\prime} \Delta^{B}_{mn}(t- t^\prime) = i \beta  \frac{d}{dt} \Delta^{B}_{mn}(t- t^\prime).
\end{equation}
\end{enumerate}

\section{Spatial dependence of response functions}
We now discuss quantum field theories with one time and three space dimensions. (A generalization to other numbers of space dimensions is straight forward.) We expand around a thermal equilibrium state that is assumed to be spatially homogenous and governed by a constant temperature $T$, chemical potential $\mu$ and fluid velocity $u^\nu$. There is one reference frame (the fluid rest frame) where the spatial part of the fluid velocity vanishes. It is often convenient to work in that frame, but for some purposes it is better to keep the frame arbitray and to accept that the fuid velocity is non-vanishing. 

\subsubsection{Field theoetic notation and expansion of Hamiltonian}
We first adapt the discussion in section \ref{sec:HamiltoniansLagrangiansActions} to the field theoretical setup, in a notation where $x=(x^0,\mathbf{x})$ are spacetime coordinates. In the notation of section \ref{sec:HamiltoniansLagrangiansActions} the spation position $\mathbf{x}$ would be part of the index $n$ used there, but now we make it explicit. One can still allow different field components, as necessary for example for vector fields, and we will also use discrete indices $m$, $n$ etc.\ to label them.  

We assume that the expansion of the Hamiltonian in eq.\ \eqref{eq:ExpansionHamiltonianResponseOperators} reads now
\begin{equation}
\begin{split}
  H[j(t)] = & H(j_\text{i}) - \int d^3 x \sum_m  (j^m(t, \mathbf{x})-j^m_\text{i}) \phi_m(\mathbf{x})(j_\text{i}) \\
  & - \int d^3 x \frac{1}{2}\sum_{mn} (j^m(t, \mathbf{x})-j^m_\text{i}) (j^n(t, \mathbf{x})-j^n_\text{i}) \phi_{mn}(\mathbf{x})(j_\text{i}) \\
  & -  \int d^3x \frac{1}{3!} \sum_{mnk} (j^m(t, \mathbf{x})-j^m_\text{i}) (j^n(t, \mathbf{x})-j^n_\text{i}) (j^k(t, \mathbf{x})-j^k_\text{i}) \phi_{mnk}(\mathbf{x})(j_\text{i}) - \ldots,
\end{split}
\label{eq:ExpansionHamiltonianResponseOperatorsFieldTheory}
\end{equation}
where the fields $\phi_m(\mathbf{x})(j_\text{i})$, $\phi_{mn}(\mathbf{x})(j_\text{i})$, $\phi_{mnk}(\mathbf{x})(j_\text{i})$ and so on are here operators in the Schrödinger picture. We take them to be hermitean. We will also work with the Heisenberg, or interaction picture, where these operators also depend on time $t=x^0$.

\subsection{Volterra series in field theoretic notation}

We consider the response of a field expectation value $\Phi_m(x)$, or more specifically its deviation from the constant equilibrium value $\Phi_m^\text{i}$, to perturbations in source fields $j^n(x)-j^n_\text{i}$. The Volterra series in eq.\ \eqref{eq:Volterra} becomes, 
\begin{equation} 
\begin{split}
  \Phi_m(x) - \Phi_m^\text{i} = & \sum_n \int d^4 x^\prime \Delta_{mn}(x-x^\prime) (j^n(x^\prime) - j^n_\text{i}) \\
  & + \frac{1}{2} \sum_{n,k} \int d^4x^\prime d^4 x^{\prime\prime} \Delta^R_{mnk}(x-x^\prime, x-x^{\prime\prime}) (j^n(x^\prime)-j^n_\text{i}) (j^k(x^{\prime\prime})-j^k_\text{i}) + \ldots
\end{split}\label{eq:VolterraFieldTheory}
\end{equation}
This is so far very general and holds for many kinds of field expectation values $\Phi_m(x)$ and source fields $j^n(x)$, independent of whether they are elementary or composite, scalar, vector or tensor fields, relativistic or non-relativistic.

\subsubsection{Fourier representation}
In the field theoretic setup it is convenient to work with Fourier transforms in space and time, so that eq.\ \eqref{eq:LinearResponseFourierRep} generalizes to
\begin{equation}
\begin{split}
  \Phi_m(x) - \Phi_m^\text{i} = & \int \frac{d^4 p}{(2\pi)^4} e^{-ip^0 x^0 + i \mathbf{p} \mathbf{x}} \Phi_m(p), \\
  j^n(x^\prime) - j^n_\text{i} = & \int \frac{d^4 p}{(2\pi)^4} e^{-ip^0 x^{\prime0} + i \mathbf{p} \mathbf{x}^\prime} j^n(p),
\end{split}\label{eq:LinearResponseFourierRep}
\end{equation}
we write the linear response function now as
\begin{equation}
  \Delta^R_{mn}(x-x^\prime) = \int \frac{d^4 p}{(2\pi)^4} e^{-ip^0(x^0-x^{\prime 0})+i\mathbf{p}(\mathbf{x}-\mathbf{x}^\prime)} \Delta^R_{mn}(p),
\end{equation}
the quadratic response function as
\begin{equation}
  \Delta^R_{mnk}(x-x^\prime, x-x^{\prime\prime}) = \int \frac{d^4p^\prime d^4p^{\prime\prime}}{(2\pi)^8} e^{-ip^{\prime 0} (x^0-x^{\prime 0})+i\mathbf{p}^\prime (\mathbf{x}-\mathbf{x}^\prime)-i p^{\prime\prime 0}(x^0-x^{\prime\prime 0})+i\mathbf{p}^{\prime\prime}(\mathbf{x}-\mathbf{x}^{\prime\prime})} \Delta^R_{mnk}(p^\prime, p^{\prime\prime}).
\label{eq:retardedQuadraticFourierFieldTheory}
\end{equation}
and so on at higher order.

The Volterra series \eqref{eq:VolterraFieldTheory} becomes then generalization of eq.\ \eqref{eq:VolterraSeriesFrequencySpace}
\begin{equation}
\begin{split}
  \Phi_m(p) = & \Delta^R_{mn}(p) j^n(p) \\
  & + \frac{1}{2} \sum_{n,k} \int \frac{d^4 p^\prime d^4 p^{\prime\prime}}{(2\pi)^8} (2\pi)^4 \delta^{(4)}(p - p^\prime - p^{\prime\prime}) \Delta^R_{mnk}(p^\prime, p^{\prime\prime}) j^n(p^\prime) j^k(p^{\prime\prime}) + \ldots
\end{split}
\end{equation}
The linear response part has the same frequency $p^0$ and wave vector $\mathbf{p}$ as the perturbation. For the quadratic part one has a similar conservation of the wave numbers as implied by the Dirac delta. This generalizes to higher orders. 

\subsection{Causality}
The implications of causality differ somewhat between relativistic and non-relativistic theories. For the latter, one has simply $\Delta^R_{mn}(x-x^\prime) = 0$ for $x^0-x^{\prime 0} < 0$ at linear order, and similarly $\Delta^R_{mnk}(x-x^\prime, x-x^{\prime\prime})=0$ for $x^0-x^{\prime 0} < 0$ or $x^0-x^{\prime\prime 0} < 0$, and so on at higher orders. For non-relativistic theories these conditions are Galilei invariant, these conditions automatically hold in any Galilei reference frame.

For relativistic field theories, the implications of causality are even stronger, and one must have $\Delta^R_{mn}(x-x^\prime) = 0$ when $x^0-x^{\prime 0} < 0$ in \textit{some} reference frame. %One can also state this as $\Delta^R_{mn}(x-x^\prime) = 0$, except when $x$ is in the causal future of $x^\prime$, or, in other words, when $x^\prime$ is on or within the past light cone of $x$. Equivalently, one must have 
This means that $\Delta^R_{mn}(x-x^\prime) = 0$ when $x^0-x^{\prime 0} < 0$, but also when $x-x^\prime$ is spacelike, i.\ e.\ when $-(x^0-x^{\prime 0})+ (\mathbf{x}-\mathbf{x}^\prime)^2 >0$. Only when $x$ is in the causal future of $x^\prime$ can $\Delta^R_{mn}(x-x^\prime)$ be non-zero.

\subsection{Decomposition of response functions and operator expressions}
The decomposition in eq.\ \eqref{eq:DecompositionLinearRetardedInstDelayed} becomes now, with the Hamiltonian in eq.\ \eqref{eq:ExpansionHamiltonianResponseOperatorsFieldTheory}
\begin{equation}
  \Delta^R_{mn}(x-x^\prime) = \delta^{(4)}(x-x^\prime) \Delta^\infty_{mn} + \Delta^\mathscr{R}_{mn}(x-x^\prime),
\end{equation}
where we have now an instantaneous and local response part $\Delta^\infty_{mn}$ that has the operator expression
\begin{equation}
  \Delta^\infty_{mn} = \text{Tr}\{ \rho_\text{i} \phi^\text{H}_{mn}(x) \}
\label{eq:InstResponseExpValueFieldTheory}
\end{equation}
that generalizes eq.\ \eqref{eq:InstResponseExpValue}. Note that the right hand side of eq.\ \eqref{eq:InstResponseExpValueFieldTheory} is independent of the spacetime position $x=(x^0, \mathbf{x})$ as a consequence of translation symmetries. The delayed part of the response function becomes now, in generalization of eq.\ \eqref{eq:delayedTwoPointResponse}
\begin{equation}
  \Delta^\mathscr{R}_{mn}(x-x^\prime) = \frac{\theta(x^0-x^{\prime 0})}{i} \text{Tr}\left\{ \phi_m^\text{H}(x) \left[\rho_\text{i}, \phi_n^\text{H}(x^\prime)\right] \right\}.
\label{eq:delayedTwoPointResponseFieldTheory}
\end{equation}

Similar expressions hold for the quadratic response function. The decomposition in eq.\ \eqref{eq:DecompositionQuadraticRetardedInstDelayed} becomes now
\begin{equation}
\begin{split}
  \Delta^R_{mnk}(x-x^\prime, x-x^{\prime\prime}) = & \delta^{(4)}(x-x^\prime) \delta^{(4)}(x-x^{\prime\prime}) \Delta^{\infty}_{mnk} + \delta^{(4)}(x-x^\prime) \Delta^{\infty\mathscr{R}}_{mnk}(x-x^{\prime\prime}) \\
  & + \delta^{(4)}(x-x^{\prime\prime}) \Delta^{\infty\mathscr{R}}_{mkn}(x-x^{\prime}) + \Delta^{\mathscr{R}}_{mnk}(x-x^\prime, x-x^{\prime\prime}),
\end{split}\label{eq:DecompositionQuadraticRetardedInstDelayedFieldTheory}
\end{equation}
which has instantaneous local, as well as delayed parts. Eq.\ \eqref{eq:quadraticResponseInstPartOperator} generalizes to
\begin{equation}
  \Delta^{\infty}_{mnk} = \text{Tr}\{ \rho_\text{i} \phi_{mnk}^\text{H}(x)\},
\label{eq:quadraticResponseInstPartOperatorFieldTheory}
\end{equation}
the mixed instantaneous delayed term in eq.\ \eqref{eq:mixedInstDelayedQuadraticResponse} to
\begin{equation}
  \Delta^{\infty\mathscr{R}}_{mnk}(x-x^{\prime\prime}) = \frac{\theta(x^0-x^{\prime\prime 0})}{i} \text{Tr}\left\{ \phi_{mn}^\text{H}(x) [\rho_\text{i}, \phi^\text{H}_k(x^{\prime\prime})] \right\},
  \label{eq:mixedInstDelayedQuadraticResponseFieldTheory}
\end{equation}
and the fully delayed response part generalizes to
\begin{equation}
\begin{split}
  \Delta^\mathscr{R}_{mnk}(x-x^\prime, x-x^{\prime\prime}) = & \frac{\theta(x^0-x^{\prime 0})\theta(x^{\prime 0}-x^{\prime\prime 0})}{i^2} \text{Tr}\left\{ \phi_m^\text{H}(x) \left[ \left[\rho_\text{i}, \phi_n^\text{H}(x^\prime)\right], \phi_k^\text{H}(x^{\prime\prime}) \right] \right\} \\
  + &  \frac{\theta(x^0-x^{\prime\prime 0})\theta(x^{\prime\prime 0}-x^{\prime 0})}{i^2} \text{Tr}\left\{ \phi_m^\text{H}(x) \left[ \left[\rho_\text{i}, \phi_k^\text{H}(x^{\prime\prime})\right], \phi_n^\text{H}(x^{\prime}) \right] \right\}.
\end{split}\label{eq:DelayedPartResponseFunctionCommutatorsFieldTheory}
\end{equation}

\subsection{Spectral representations}
The spectral two-point function in eq.\ \eqref{eq:SpectralTwoPointFunctionDef} becomes in the field theoretic notation
\begin{equation}
\begin{split}
  \Delta^\rho_{mn}(x-x^\prime) = & - \text{Tr} \left\{ \phi_m^\text{H}(x) \left[\rho_\text{i} , \phi^\text{H}_n(x^\prime) \right] \right\} = \text{Tr} \left\{ \rho_\text{i} \left[ \phi_m^\text{H}(x), \phi_n^\text{H}(x^\prime) \right] \right\} \\ = & \int \frac{d^4p}{(2\pi)^4} e^{-ip^0(x^0-x^{\prime 0})+i\mathbf{p}(\mathbf{x}-\mathbf{x}^\prime)} \Delta^\rho_{mn}(p).
\end{split}\label{eq:SpectralTwoPointFunctionDefFieldTheory}
\end{equation}
Here one has $\Delta^\rho_{mn}(x-x^\prime) = - \Delta^\rho_{nm}(x^\prime-x)$  
and $\Delta^\rho_{mn}(p) = - \Delta^\rho_{nm}(-p)$. For hermitian operators $\phi^\text{H}_m(x)$ and $\phi^\text{H}_n(x^\prime)$ one has $\Delta^\rho_{mn}(x-x^\prime) = - \Delta^\rho_{mn}(x-x^\prime)^*$. In the Fourier domain this implies 
\begin{equation}
  \Delta^\rho_{mn}(p) = - \Delta^\rho_{mn}(-p)^*.
  \label{eq:RealityConstraintSpectralTwoPointFunctionFourierSpace}
\end{equation}

With these definitions one can write the delayed part of the response function in eq.\ \eqref{eq:mixedInstDelayedQuadraticResponseFieldTheory} as
\begin{equation}
  \Delta^\mathscr{R}_{mn}(x-x^\prime) = i \theta(x^0 - x^{\prime 0}) \Delta^\rho_{mn}(x-x^\prime).
\label{eq:DelayedPartResponseFunctionSpectralFunctionFieldTheory}
\end{equation}
Introducing also the complex-frequency two-point correlation function through
\begin{equation}
  G_{mn}(z, \mathbf{p}) = -\int_{\infty}^\infty \frac{d\omega}{2\pi} \frac{\Delta^\rho_{mn}(\omega, \mathbf{p})}{z-\omega},
\label{eq:SpectralRepFieldTheory}
\end{equation}
we can now write $\Delta^\mathscr{R}_{mn}(p^0, \mathbf{p}) = G_{mn}(p^0+i\epsilon, \mathbf{p})$. For every spatial momentum $\mathbf{p}$, this is analytic in the upper half of the complex frequency plane.

We also note here that the spectral function $\Delta^\rho_{mn}(\omega, \mathbf{p})$ depends on frequeny and momenta. One can often use reotational symmetry in the fluid rest frame to reduce to two independent arguments, but that is still one argument more than in vacuum, where one is used to spectral densities with only one argument.

For quadratic response, the field theoretic version of eq.\ \eqref{eq:GmnkDef} reads
\begin{equation}
\begin{split}
  & \Delta^\rho_{mnk}(x - x^\prime, x^\prime -x^{\prime\prime}) = \text{Tr} \left\{ 
  \phi_m^\text{H}(x) \left[ \left[\rho_\text{i}, \phi_n^\text{H}(x^\prime)\right], \phi_k^\text{H}(x^{\prime\prime}) \right]  
  %\rho_\text{i} \left[ \left[ \phi^\text{H}_m(t), \phi_n^\text{H}(t^\prime) \right], \phi_k^\text{H}(t^{\prime\prime}) \right] 
  \right\} \\ = & \int \frac{d^4p_1}{(2\pi)^4} \frac{dp_2}{(2\pi)^4} e^{- i p_1^0 (x^0-x^{\prime 0}) + i \mathbf{p}_1 (\mathbf{x}-\mathbf{x}^\prime) - i p_2^0 (x^{\prime 0}-x^{\prime\prime 0})+ i \mathbf{p}_2 (\mathbf{x}^\prime-\mathbf{x}^{\prime\prime})} \Delta^\rho_{mnk}(p_1, p_2).
\end{split}
\label{eq:GmnkDefFieldTheory}
\end{equation}
Here one has the symmetry $\Delta^\rho_{mnk}(x-x^\prime, x^\prime -x^{\prime\prime}) = - \Delta^\rho_{knm}(x^{\prime\prime}-x^\prime, x^\prime - x)$ or $\Delta^\rho_{mnk}(p_1, p_2) = - \Delta^\rho_{knm}(-p_2, -p_1)$ in Fourier space. For hermitian fields $\phi^\text{H}_m(x)$ the reality constraints are $\Delta^\rho_{mnk}(x - x^\prime, x^\prime - x^{\prime\prime}) = \Delta^\rho_{mnk}(x - x^\prime, x^\prime - x^{\prime\prime})^*$ and 
\begin{equation}
 \Delta^\rho_{mnk}(p_1, p_2) = \Delta^\rho_{mnk}(-p_1, -p_2)^*.
\label{eq:RealityConstraintSpectralThreePointFunctionFourierSpace}
\end{equation}

One can write the fully retarded part of the quadratic response function in real space as
\begin{equation}
\begin{split}
  \Delta^\mathscr{R}_{mnk}(x-x^\prime, x-x^{\prime\prime}) = & i^2 \theta(x^0-x^{\prime 0}) \theta(x^{\prime 0} - x^{\prime\prime 0})\Delta^\rho_{mnk}(x - x^\prime, x^\prime - x^{\prime\prime}) \\
  & + i^2 \theta(x^0-x^{\prime\prime 0}) \theta(x^{\prime\prime 0} - x^{\prime 0}) \Delta^\rho_{mkn}(x - x^{\prime\prime}, x^{\prime\prime} - x^{\prime}),
\end{split}\label{eq:retardedQuadraticResponseFunctionSpectral}
\end{equation}
and in Fourier space as
\begin{equation}
\begin{split}
  \Delta^\mathscr{R}_{mnk}(p,p^\prime) = & G_{mnk}(p^0+p^{\prime 0}+i\varepsilon, \mathbf{p}+\mathbf{p}^\prime, p^{\prime 0}+i\varepsilon, \mathbf{p}^\prime)\\
  & + G_{mnk}(p^0+p^{\prime 0}+i\varepsilon, \mathbf{p}+\mathbf{p}^\prime, p^{0}+i\varepsilon, \mathbf{p}),
\end{split}
\end{equation}
with the complex frequency correlation function defined by the spectral representation
\begin{equation}
  G_{mnk}(z_1, \mathbf{p}_1, z_2, \mathbf{p}_2) = (-1)^2 \int_{-\infty}^\infty \frac{d\omega_1 d\omega_2}{(2\pi)^2} \frac{\Delta^\rho(\omega_1, \mathbf{p}_1, \omega_2, \mathbf{p}_2)}{[z_1-\omega_1][z_2-\omega_2]}.
\end{equation}
Similar expressions hold for higher order response functions.

\subsubsection{Causality for relativistic theories}
The right hand side of eq.\ \eqref{eq:DelayedPartResponseFunctionSpectralFunctionFieldTheory} fulfills the condition $\Delta^\mathscr{R}_{mn}(x-x^\prime)=0$ for $x^0-x^{\prime 0} <0$. For non-relativistic field theories that is necessary and sufficient for causality. For relativistic theories, causality is stronger, however, and one must additionally have
\begin{equation}
  \Delta^\rho_{mn}(x-x^\prime) = 0 \quad\quad\quad \text{for} \quad\quad\quad -(x^0-x^{\prime 0})^2 + (\mathbf{x}-\mathbf{x}^\prime)^2 >0.
\end{equation}
The spectral two-point function needs to vanish for space-like seperation of the two arguments $x$ and $x^\prime$. From a microscopic point of view this is ensured that the commutator $\left[ \phi_m^\text{H}(x), \phi_n^\text{H}(x^\prime) \right]$ that enters eq.\ \eqref{eq:SpectralTwoPointFunctionDefFieldTheory} vanishes when $x$ and $x^\prime$ are space-like seperated.

Assume now that $x-x^\prime$ is space-like. With a Lorentz boost one can always reach a frame where $x^0=x^{\prime 0}+ \sigma$, where $\sigma$ is infinitesimal positive or negative, although this is in general different from the fluid rest frame. For the Fourier space representation of the spectral function defined by the second line of eq.\ \eqref{eq:SpectralTwoPointFunctionDefFieldTheory} this implies that for any non-zero $\mathbf{x}-\mathbf{x}^\prime$ one must have 
\begin{equation}
  \Delta^\rho_{mn}(\sigma, \mathbf{x}-\mathbf{x}^\prime) = \int \frac{d^3 p}{(2\pi)^3} e^{i\mathbf{p}(\mathbf{x}-\mathbf{x}^\prime)} \int_{-\infty}^\infty \frac{d p^0}{2\pi} e^{-i p^0 \sigma}\Delta^\rho_{mn}(p^0, \mathbf{p}) = 0.
\end{equation}
The only possibility is $ \Delta^\rho_{mn}(\sigma, \mathbf{x}-\mathbf{x}^\prime) =  c_{mn} \delta^{(3)}(\mathbf{x}-\mathbf{x}^\prime)$, which means that
\begin{equation}
  \int_{-\infty}^\infty \frac{d p^0}{2\pi} e^{-ip^0 \sigma}\Delta^\rho_{mn}(p^0, \mathbf{p}) = c_{mn},
\end{equation}
must be constants independent of $\mathbf{p}$. Dependeing on $\sigma$, the $p^0$ integral can be closed above or below the real axis. This condition must hold in any Lorentz frame. To get a condition that can be evaluated in the fluid rest frame we intoduce a time-like four-velocity $v^\mu$, normalized to $(v^0)^2 - \mathbf{v}^2=1$, and decompose $p^\mu = \omega v^\mu + \hat p^\mu$, with $v_\mu \hat p^\mu = 0$. One must then have
\begin{equation}
  \int_{-\infty}^\infty \frac{d \omega}{2\pi} e^{-i\omega \sigma}\Delta^\rho_{mn}(\omega v^\mu + \hat p^\mu) = c_{mn},
\end{equation}
where the right hand side is independent of $\hat p^\mu$. This condition must hold for any four-velocity $v^\mu$.

For the quadratic response function one has a similar relativistic causality condition: The expression in \eqref{eq:retardedQuadraticResponseFunctionSpectral} and therefore $\Delta^\rho_{mnk}(x - x^\prime, x^\prime - x^{\prime\prime})$ must vanish when either $x-x^\prime$ or $x-x^{\prime\prime} = x-x^\prime + x^\prime - x^{\prime\prime}$ are space-like. This can be translated to Fourier space similarly as the above argument for the two-point spectral function.

At this point one should mention that an alternative spectral representation for two-point functions in relativistic quantum theories was proposed in ref.\ \cite{Bros:1992ey, Bros:1994ofl, Bros:1996mw, Bros:2001zs} and recently discussed e.\ g.\ in refs.\ \cite{PhysRevD.106.045028, nair2025spectralfunctionsnonzerotemperature}. One writes the spectral function, defined in eq.\ \eqref{eq:SpectralTwoPointFunctionDefFieldTheory}, in the form
\begin{equation}
  \Delta^\rho_{mn}(p^0, \mathbf{p}) =  \int_0^\infty ds \int \frac{d^3 u}{(2\pi)^3} \text{sign}(p^0) \delta\left(s-(p^0)^2 + (\mathbf{p}-\mathbf{u})^2 \right) 2\pi D_{mn}(\mathbf{u}, s).
\label{eq:spectralFunctionBrosBuchholz}
\end{equation}
The construction is such that in the vacuum limit ($T\to 0$, $\mu\to 0$) one has
\begin{equation}
  D_{mn}(\mathbf{u}, s) \to (2\pi)^3 \delta^{(3)}(\mathbf{u}) \rho_{mn}(s),
\end{equation}
where $\rho_{mn}(s)$ is the spectral density that enters the well-known Källén-Lehmann representation. For example, for a stable massive particle with mass $m$ it would be of the form $\delta(s-m^2)$.

Relativistic causality is automatically obeyed by the right hand side of eq.\ \eqref{eq:spectralFunctionBrosBuchholz}, and it might be easier to reconstruct from numerical data (typically available in the Euclidean domain) \cite{PhysRevD.106.045028}. On the other side, $D_{mn}(\mathbf{u}, s)$ is in general still a complicated function, as much as $\Delta^\rho_{mn}(p^0, \mathbf{p})$.

\subsection{Time reversal}
The identities derived in section \ref{sec:TimeReversal} transfer immediatly to field theory. The time-reversed source configuration is now $j^n_\mathsf{T}(x^0, \mathbf{x}) = \varepsilon_n j^n(-x^0, \mathbf{x})$, and fields in the Heiselberg or interaction picture transform as
\begin{equation}
  \Theta \phi_n^\text{H}(x^0, \mathbf{x}) \Theta^\dagger = \varepsilon_n \phi_n^\text{H}(2t_\text{i}-x^0, \mathbf{x}){\big |}_{\varepsilon j_\text{i}}.
\end{equation}
For the spectral function the identity in eq.\ \eqref{eq:TimreReflectionSpectralFunction} generalizes to
\begin{equation}
  \Delta^\rho_{mn}(x^0-x^{\prime 0}, \mathbf{x}-\mathbf{x}^\prime) = - \varepsilon_m \varepsilon_n \Delta^\rho_{mn}(x^{\prime 0}-x^0, \mathbf{x}-\mathbf{x}^\prime){\big |}_{\varepsilon j_\text{i}}.
\end{equation}
In Fourier space this reads
\begin{equation}
  \Delta^\rho_{mn}(p^0, \mathbf{p}) = - \varepsilon_m \varepsilon_n \Delta^\rho_{mn}(-p^0, \mathbf{p}){\big |}_{\varepsilon j_\text{i}}, 
\end{equation}
and becomes, together with eq.\ \eqref{eq:RealityConstraintSpectralTwoPointFunctionFourierSpace},
\begin{equation}
  \Delta^\rho_{mn}(p^0, \mathbf{p}) = \varepsilon_m \varepsilon_n \Delta^\rho_{mn}(p^0, -\mathbf{p})^*{\big |}_{\varepsilon j_\text{i}}.
\end{equation}
Similarly, eq.\ \eqref{eq:TimeReflectionThreePointSpectralTimeDomain} generalizes to
\begin{equation}
  \Delta^\rho_{mnk}(x^0-x^{\prime 0}, \mathbf{x}-\mathbf{x}^\prime, x^{\prime 0}-x^{\prime\prime 0}, \mathbf{x}^\prime-\mathbf{x}^{\prime\prime}) = \varepsilon_m \varepsilon_n \varepsilon_k \Delta^\rho_{mnk}(x^{\prime 0}-x^0, \mathbf{x}-\mathbf{x}^\prime, x^{\prime\prime 0} -x^{\prime 0}, \mathbf{x}^\prime-\mathbf{x}^{\prime\prime}){\big |}_{\varepsilon j_\text{i}},
\end{equation}
or, in Fourier space,
\begin{equation}
  \Delta^\rho_{mnk}(p_1^0, \mathbf{p}_1, p_2^0, \mathbf{p}_2) = \varepsilon_m \varepsilon_n \varepsilon_k \Delta^\rho_{mnk}(-p_1^0, \mathbf{p}_1, -p_2^0, \mathbf{p}_2) {\big |}_{\varepsilon j_\text{i}}.
\end{equation}
Together with eq.\ \eqref{eq:RealityConstraintSpectralThreePointFunctionFourierSpace} this becomes
\begin{equation}
  \Delta^\rho_{mnk}(p_1^0, \mathbf{p}_1, p_2^0, \mathbf{p}_2) = \varepsilon_m \varepsilon_n \varepsilon_k \Delta^\rho_{mnk}(p_1^0, -\mathbf{p}_1, p_2^0, -\mathbf{p}_2)^* {\big |}_{\varepsilon j_\text{i}}.
\end{equation}
In situations with time reversal symmetry such that $j_\text{i}= \varepsilon j_\text{i}$ and additional symmetries such as spatial parity, one can often derive from these identities that the spectral functions $\rho^\rho_{mn}(p^0, \mathbf{p})$, $\Delta^\rho_{mnk}(p_1^0, \mathbf{p}_1, p_2^0, \mathbf{p}_2)$ and their higher-order analogues are purely real or imaginary.

\subsection{Quantum correlation functions and fluctuation-dissipation relations}
The discussion of quantum correlation functions in section \ref{sec:QuantumCorrelationFunctions} is also directly extendable to field theories. Let us just mention that the general quantum two-point correlation function in eq.\ \eqref{eq:defGeneralizedCovarianceCorrFunction}, defined for an operator monotone function $f(z)$, becomes now
\begin{equation}
\begin{split}
  \Delta^f_{mn}(x-x^\prime) & = \text{Tr}\left\{ \phi_m^\text{H}(x) K_\rho^f \phi_n^\text{H}(x^\prime) \right\} = \text{Tr} \left\{ \phi_m^\text{H}(x) f\big(L_\rho R_\rho^{-1}\big) R_\rho \phi_n^\text{H}(x^\prime) \right\} \\
  & = \int \frac{d^4 p}{(2\pi)^4} e^{ip(x-x^\prime)} \Delta^f_{mn}(p).
\end{split}
\label{eq:defGeneralizedCovarianceCorrFunctionFieldTheory} 
\end{equation}
We also give the fluctuation dissipatation relation, in an arbitrary Lorentz frame, where the fluid velocity is $u^\mu$, such that $\omega = -u^\mu p_\mu$ is the fluid velocity in the fluid rest frame. Eq.\ \eqref{eq:GeneralizedFDRelation} generalizes to
\begin{equation}
  \Delta^f_{mn}(p) = \frac{f(\exp(-\beta \omega))}{1-\exp(-\beta \omega)} \Delta^\rho_{mn}(p).
\label{eq:GeneralizedFDRelationFieldTheory}
\end{equation}
%It should be possible to generalize this to higher order correlation functions.

\section{Conserved currents and consequences of gauge symmetry}
We now consider a situation where the operators $\phi_n$ (see eq.\ \eqref{eq:HamiltonianComplexScalarWithExternalGaugeFields}) constitute the components of a conserved current, and the corresponding source field $j^n$ are gauge fields. This leads to a few special features, like Ward identities for response functions. We concentrate for simplicity on an Abelian gauge symmetry, which is most important for many condensed matter applications. For a recent discussion diffeomorphism symmetry as an example for a non-Abelian symmetry, see refs. \cite{Stoetzel:2025dcc, Jeon:2025kem}.

\subsection{Ward identities}
We take the source fields $j^n$ to be an Abliean gauge field $A_\mu(x)$ for which the gauge transformation has the form (with $x^\mu=(t, \mathbf{x})$ in units where the velocity of light is $c=1$)
\begin{equation}
  A_\mu(x) \to A_\mu(x) + \frac{\partial}{\partial x^\mu} \alpha(x).
\label{eq:gaugeTransformationA}
\end{equation}
It is taken to be coupled to a conserved current which is a composite operator in terms of elementary fields. The expectation value of this current is denoted as $N^\mu(x)$. The gauge symmetry implies the local conservation law
\begin{equation}
  \sum_{\mu=0}^3 \frac{\partial}{\partial x^\mu} N^\mu(x) = \frac{\partial}{\partial t} N^0(x) + \boldsymbol{\nabla} \cdot \mathbf{N}(x) = 0.
\label{eq:ConservationLawNmu}
\end{equation}

\subsubsection{Volterra series}
The analog of the Volterra series in eq.\ \eqref{eq:Volterra} becomes for the expectation value of the conserved current
\begin{equation}
\begin{split}
 N^\mu(x) - N^\mu_\text{i} = &  \sum_{\nu=0}^3 \int d^4 x^\prime \; \Delta_R^{\mu\nu}(x-x^\prime)   \left(A_\nu(x^\prime) - A_\nu^\text{i} \right) \\
 & + \frac{1}{2} \sum_{\nu,\rho=0}^3 \int d^4 x^\prime d^4x^{\prime\prime} \; \Delta_R^{\mu\nu\rho}(x-x^\prime, x-x^{\prime\prime})   \left( A_\nu(x^\prime) - A_\nu^\text{i} \right) \left( A_\rho(x^{\prime\prime}) - A_\rho^\text{i} \right) + \ldots
\end{split}\label{eq:VolterraConservedCurrentGaugeField}
\end{equation}

\subsubsection{Consequences of the conservation law}
The conservation law \eqref{eq:ConservationLawNmu} implies for the response functions identities of the form
\begin{equation}
  \sum_{\mu=0}^3 \frac{\partial}{\partial x^\mu} \Delta_R^{\mu\nu}(x-x^\prime) = 
  \sum_{\mu=0}^3 \frac{\partial}{\partial x^\mu} \Delta_R^{\mu\nu\rho}(x-x^\prime, x-x^{\prime\prime}) = \ldots = 0.
\end{equation}
Note that this is independent of the fact that the source fields are here gauge fields, and can be generalized accordingly to situations where also other source fields appear. It simply follows from the conservation law and its linearity. These relations can be easily extended to Fourier space using corresponding generalizations of eq.\ \eqref{eq:LinearResponseFourierRep} and \eqref{eq:retardedQuadraticFourier}. 

\subsubsection{Consequences of gauge symmetry}
Assuming now that the left hand side of eq.\ \eqref{eq:VolterraConservedCurrentGaugeField} is unaffected by a gauge transformation \eqref{eq:gaugeTransformationA} leads, after a partial integration, for the linear response function to an identity
\begin{equation}
  \sum_{\nu=0}^3 \frac{\partial}{\partial x^{\prime\nu}} \Delta_R^{\mu\nu}(x-x^\prime) = 0.
\end{equation}
Here we may note that a generalization is possible to situations where the left hand side of eq.\ \eqref{eq:VolterraConservedCurrentGaugeField} is not a conserved current. Again, a translation to Fourier space is straight forward.

For the quadratic response one obtains in a similar way
\begin{equation}   
  \sum_{\nu=0}^3 \frac{\partial}{\partial x^{\prime\nu}} \Delta_R^{\mu\nu\rho}(x-x^\prime, x-x^{\prime\prime}) = 0,
\end{equation}
where we have used the symmetry $\Delta_R^{\mu\nu\rho}(x-x^\prime, x-x^{\prime\prime}) = \Delta_R^{\mu\rho\nu}(x-x^{\prime\prime}, x-x^{\prime})$. These identities generalize to response functions of arbitrary order.

\subsection{Electromagnetic current response in the fluid dynamic regime}
We now consider as an example the response of a conserved U$(1)$ current $N^\mu(x)=(n(x), \mathbf{J}(x))$ to a change in the corresponding electromagnetic gauge field $A_\mu(x)$. Our aim ist to reconstruct the response function in the vicinity of a global equilibrium state where $n(x)=\bar n$ and $\mathbf{J}(x)=0$, from the fluid dynamic equations of motion. We take the background fluid velocity $u^\mu$ as given and work in the global fluid rest frame where $u^\mu=(1,0,0,0)$. We follow the discussion in ref.\ \cite{Floerchinger:2021xhb} where further details can be found.

We use the Fourier representations
\begin{equation}
\begin{split}
  N^\mu(x) - \bar n u^\mu = & \int \frac{d^d p}{(2\pi)^d} e^{ipx} N^\mu(p),\\
  A_\mu(x)  = & \int \frac{d^d p}{(2\pi)^d} e^{ipx} A_\mu(p).
\end{split}
\end{equation}
The conservation law $\partial_\mu N^\mu(x) = 0$ is represented in Fourier space as $p_\mu N^\mu(p)$. It allows to express the density perturbation in terms of the spatial current, $ N^0(p) = \mathbf{p}\cdot \mathbf{J}(p)/p^0$. Besides this conservation law we are using the linearized equation of motion
\begin{equation}
\mathbf{J}(x) + \tau \frac{\partial}{\partial t} \mathbf{J}(x) = \sigma \mathbf{E}(x) - D \boldsymbol{\nabla} n(x),  
\end{equation}
where $\tau$ is a relaxation time, $\sigma$ the electric conductivity, $\mathbf{E}(x)=\boldsymbol{\nabla}A_0(x) - \partial_t \boldsymbol{A}(x)$ the electric field, and $D$ the diffusion constant. In Fourier space this reads
\begin{equation}
  \sum_{k=1}^3 \left[(1-i\tau p^0) \delta_{jk} +i \frac{D}{p^0} p_j p_k \right] J^k(p) = i \sigma \left[p_j A_0(p) + p^0 A_j(p) \right].
\end{equation}
The matrix in square brackets on the left hand side can be inverted, which yields
\begin{equation}
  J^k(p) = \sum_{j=1}^3 \left[ \frac{\delta_{kj}}{1-i\tau p^0} - \frac{i D p_k p_j}{[p^0-i\tau (p^0)^2 + i D \mathbf{p}^2][1 - i \tau p^0]} \right] i\sigma \left[p_j A_0(p) + p^0 A_j(p) \right].
\end{equation}
In a final step we can use the conservation law to bring this into the form
\begin{equation}
  N^\mu(p) = \sum_{\nu=0}^3 G_\text{R}^{\mu\nu}(p) A_\nu(p),
\end{equation}
with the response function
\begin{equation}
\begin{split}
  G^{00}(p) = & \frac{i\sigma \mathbf{p}^2}{p^0-i\tau (p^0)^2 + i D \mathbf{p}^2}, \quad\quad\quad\quad G^{0j}(p) = G^{j0}(p) = \frac{i\sigma p^0 p^j}{p^0 - i \tau (p^0)^2 + i D \mathbf{p}^2},\\
  G^{jk}(p) = & \frac{i\sigma p^0 \delta^{jk}}{1-i\tau p^0}  + \frac{D\sigma p^0 p^j p^k}{[p^0 - i \tau (p^0)^2 + i D \mathbf{p}^2][1-i\tau p^0]}.
\end{split}
\end{equation}
One may check that the Ward identity $p_\mu G^{\mu\nu}(p) = p_\nu G^{\mu\nu}(p)=0$ is fulfilled. In the static limit $p^0=0$ one one finds $G^{00}(p) =\sigma/D = \chi$, while all other components vanish. Here $\chi=(\partial n/\partial\mu)$ is the static charge susceptibility and we have used the Einstein relation. One may also check that $G^{\mu\nu}(p)$ has all its poles in the lower half of the complex $p^0$ plane, as needed for causality. 

\section{Conclusions}

In summary, we have discussed here response theory as it can be applied to quantum fields that evolve in isolation, with an overall unitary time evolution. The general idea to expand the response in a functional Taylor series in perturbations of the Hamiltonian or Lagrangian is very useful, and together with general pronciples like causality it leads already to very many interesting consequences.

One can see response theory as a natural extension of thermodynamic equilibrium considerations, and indeed, dynamic susceptibilities are sometimes considered as thermal equilibrium properties in an extended sense. What is most interesting is that the setup allows to study time-dependence, or dynamics, but in a way that is still very universal. Thermal equilibrium states are still playing a role as initial conditions and expansion points.

Quantum theories are particularly interesting, because of the richness that comes with the non-commutative nature of density matrices and operators. Here one feels that the theory is still not yet complete, in particular in the non-linear regime. We have discussed that quantum theory allows to define a large class of correlation functions, and that a generalized fluctuation-dissipation relation allows to relate them to the spectral density.

It would be interesting so see in general which correlation functions are meaningful in a quantum theory, how they can be experimentally accessed, and how they are related to corresponding spectral functions for close-to-thermal-equlibrium situations. We have discussed that in some detail for a higher-order generalization of Bogoliubov-Kubo-Mori correlation functions, but there is more to be explored.

What would also be interesting in this context is to study the relation to quantum information theory, and especially quantum information geometry (see e.\ g.\ \cite{bengtsson2017} for an introduction) further. Of course this is a topic that has already been explored to some extend, see e.\ g.\ \cite{Goold_2016,PhysRevB.102.155407,Tsang_2025}. Still, intuitively one feels that more can be understood here, especially for relativistic quantum field theories where variants of quantum relative entropies can be defined locally in terms of modular theory.

A promissing direction could also be an extended use of functional techniques. We have already made extensive use of generating functionals for response functions, including the description of an initial thermal state and a measurement scheme. This can be extended to more complicated measurement schemes, also at intermediate times, and one could explore the use of Legendre transforms in that context. That could directly lead to effective equations of motion for expectation values of (composite) operators, which might be interesting for the construction of effective theories for macroscopic observables. 

Taken together, we believe that the approach of response theory to the correlation functions and dynamics of quantum fields has proven to be very useful already, and will likely remain so for the foreseeable future.

\section*{Acknowledgments}
The authors thanks Tim Stötzel for useful discussions.

\providecommand{\href}[2]{#2}\begingroup\raggedright\endgroup

\end{document}